\documentclass[letterpaper,12pt,titlepage,oneside,final]{book}
 
\newcommand{\href}[1]{#1} % does nothing, but defines the command so the
\usepackage{ifthen}
\newboolean{PrintVersion}
\setboolean{PrintVersion}{false} 
\usepackage{amsmath,amssymb,amstext} % Lots of math symbols and environments
\usepackage[pdftex]{graphicx} % For including graphics N.B. pdftex graphics driver 

\usepackage{rotating}
\usepackage{mathrsfs}
\usepackage[caption=false]{subfig}
\usepackage{enumitem}
\usepackage{color}
\usepackage[percent]{overpic}
\usepackage{bm}
\usepackage{bbm}
\usepackage{rotating}
\usepackage{cancel}
\usepackage[normalem]{ulem}
\usepackage{verbatim}
\usepackage{mathtools}
\usepackage{graphicx}
\usepackage{tensor}
\usepackage{pdflscape}
\usepackage{afterpage}
\usepackage{capt-of}
\usepackage{verbatim}
\usepackage{MnSymbol}
\usepackage{authblk}
\graphicspath{ {images/} }
\numberwithin{equation}{section}
\usepackage{braket}
\let\textacute\'
\let\textgrave\`
\newcommand{\ii}{\mathrm{i}}
\newcommand{\dd}{\mathrm{d}}
\newcommand{\be}{\begin{equation}}
\newcommand{\bel}[1]{\begin{equation}\label{#1}}
\newcommand{\ee}{\end{equation}}

\newcommand{\openone}{\mathbbm{1}}
\renewcommand{\vec}{\text{vec}}

\usepackage[pdftex,pagebackref=false]{hyperref} % with basic options
\hypersetup{
    plainpages=false,       % needed if Roman numbers in frontpages
    unicode=false,          % non-Latin characters in Acrobat’s bookmarks
    pdftoolbar=true,        % show Acrobat’s toolbar?
    pdfmenubar=true,        % show Acrobat’s menu?
    pdffitwindow=false,     % window fit to page when opened
    pdfstartview={FitH},    % fits the width of the page to the window
    pdftitle={Interpolated\ Collision\ Models},    % title: CHANGE THIS TEXT! Done
    pdfauthor={Daniel Grimmer},    % author: CHANGE THIS TEXT! and uncomment this line. Done
    pdfsubject={Interpolated Collision Models},  % subject: CHANGE THIS TEXT! and uncomment this line. Done
    pdfnewwindow=true,      % links in new window
    colorlinks=true,        % false: boxed links; true: colored links
    linkcolor=blue,         % color of internal links
    citecolor=green,        % color of links to bibliography
    filecolor=magenta,      % color of file links
    urlcolor=cyan           % color of external links
}
\ifthenelse{\boolean{PrintVersion}}{   % for improved print quality, change some hyperref options
\hypersetup{	% override some previously defined hyperref options
    citecolor=black,%
    filecolor=black,%
    linkcolor=black,%
    urlcolor=black}
}{} % end of ifthenelse (no else)

\let\origdoublepage\cleardoublepage
\newcommand{\clearemptydoublepage}{%
  \clearpage{\pagestyle{empty}\origdoublepage}}
\let\cleardoublepage\clearemptydoublepage

\begin{document}

% For a large document, it is a good idea to divide your thesis
% into several files, each one containing one chapter.
% To illustrate this idea, the "front pages" (i.e., title page,
% declaration, borrowers' page, abstract, acknowledgements,
% dedication, table of contents, list of tables, list of figures,
% nomenclature) are contained within the file "uw-ethesis-frontpgs.tex" which is
% included into the document by the following statement.
%----------------------------------------------------------------------
% FRONT MATERIAL
%----------------------------------------------------------------------
% T I T L E   P A G E
% -------------------
% Last updated June 14, 2017, by Stephen Carr, IST-Client Services
% The title page is counted as page `i' but we need to suppress the
% page number. Also, we don't want any headers or footers.
\pagestyle{empty}
\pagenumbering{roman}
\let\textacute\'
\let\textgrave\`

% The contents of the title page are specified in the "titlepage"
% environment.
\begin{titlepage}
        \begin{center}
        \vspace*{1.0cm}

        \Huge
        {\bf Interpolated Collision Model Formalism}

        \vspace*{1.0cm}

        \normalsize
        by \\

        \vspace*{1.0cm}

        \Large
        Daniel Grimmer\\

        \vspace*{3.0cm}

        \normalsize
        A thesis \\
        presented to the University of Waterloo \\ 
        in fulfillment of the \\
        thesis requirement for the degree of \\
        Doctor of Philosophy \\
        in \\
        Physics \\

        \vspace*{2.0cm}

        Waterloo, Ontario, Canada, 2020 \\

        \vspace*{1.0cm}

        \copyright\ Daniel Grimmer 2020\\
        \end{center}
\end{titlepage}

% The rest of the front pages should contain no headers and be numbered using Roman numerals starting with `ii'
\pagestyle{plain}
\setcounter{page}{2}

\cleardoublepage % Ends the current page and causes all figures and tables that have so far appeared in the input to be printed.
% In a two-sided printing style, it also makes the next page a right-hand (odd-numbered) page, producing a blank page if necessary.

% E X A M I N I N G   C O M M I T T E E (Required for Ph.D. theses only)
% Remove or comment out the lines below to remove this page
\begin{center}\textbf{Examining Committee Membership}\end{center}
  \noindent
The following served on the Examining Committee for this thesis. The decision of the Examining Committee is by majority vote.
  \bigskip
  
  \noindent
\begin{tabbing}
Internal-External Member: \=  \kill % using longest text to define tab length
External Examiner: \>  Daniel James \\ 
\> Professor, Dept. of Physics, University of Toronto \\
\end{tabbing} 
  \bigskip
  
  \noindent
\begin{tabbing}
Internal-External Member: \=  \kill % using longest text to define tab length
Supervisor(s): \> Robert Mann \\
\> Professor, Dept. of Physics, University of Waterloo \\
\> Eduardo Mart\textacute{i}n-Mart\textacute{i}nez\\
\> Professor, Dept. of Applied Mathematics, University of Waterloo \\
\end{tabbing}
  \bigskip
  
  \noindent
  \begin{tabbing}
Internal-External Member: \=  \kill % using longest text to define tab length
Internal Member: \> Adrian Lupascu\\
\> Professor, Dept. of Physics, University of Waterloo \\
Internal Member: \> Achim Kempf\\
\> Professor, Dept. of Applied Mathematics, University of Waterloo \\
\end{tabbing}
  \bigskip
  
  \noindent
\begin{tabbing}
Internal-External Member: \=  \kill % using longest text to define tab length
Internal-External Member: \> Florian Girelli \\
\> Professor, Dept. of Applied Mathematics, University of Waterloo \\
\end{tabbing}
%  \bigskip
%  
%  \noindent
%\begin{tabbing}
%Internal-External Member: \=  \kill % using longest text to define tab length
%Other Member(s): \> Leeping Fang \\
%\> Professor, Dept. of Fine Art, University of Waterloo \\
%\end{tabbing}

\cleardoublepage

% D E C L A R A T I O N   P A G E
% -------------------------------
  % The following is a sample Delaration Page as provided by the GSO
  % December 13th, 2006.  It is designed for an electronic thesis.
\begin{center}\textbf{Author’s Declaration}\end{center}
\noindent
This thesis consists of material all of which I authored or co-authored: see Statement of Contributions included in the thesis. This is a true copy of the thesis, including any required final revisions, as accepted by my examiners.

\noindent
I understand that my thesis may be made electronically available to the public.

\cleardoublepage

\begin{center}\textbf{Statement of Contributions}\end{center}
\noindent
Daniel Grimmer was the sole author this thesis, although portions of this thesis are direct adaptations of collaborative publications.

Chapters 1, 2, 3, 4, 8 and 9 are new content produced for this thesis by Daniel Grimmer alone.

Chapter 5 is adapted for this thesis from the publications \cite{Grimmer2017a} and \cite{Grimmer_2019} which were prepared in collaboration with Daniel's supervisors Robert B Mann and Eduardo Mart{\'{\i}}n-Mart{\'{\i}}nez. Daniel Grimmer conducted a majority of the research on these projects with some guidance/input from his supervisors. Daniel Grimmer wrote the draft manuscript alone which was then collaboratively edited.

Chapter 6 and 7 are adapted for this thesis from the publication \cite{Dan2018} and \cite{PhysRevA.97.052120} respectively, These publication were prepared in collaboration with Eric Brown and Achim Kempf and Robert B Mann and Eduardo Mart\'{\i}n-Mart\'{\i}nez. Daniel Grimmer conducted a majority of the research on these projects with some guidance/input from his collaborators. Daniel Grimmer wrote the draft manuscript alone which was then collaboratively edited.

\cleardoublepage

% A B S T R A C T
% ---------------

\begin{center}\textbf{Abstract}\end{center}
The dynamics of open quantum systems (i.e., of quantum systems interacting with an uncontrolled environment) forms the basis of numerous active areas of research from quantum thermodynamics to quantum computing. One approach to modeling open quantum systems is via a \textit{Collision Model}. For instance, one could model the environment as being composed of many small quantum systems (ancillas) which interact with the target system sequentially, in a series of ``collisions''.

In this thesis I will discuss a novel method for constructing a continuous-time master equation from the discrete-time dynamics given by any such collision model. This new approach works for any interaction duration, $\delta t$, by interpolating the dynamics between the time-points $t=n\,\delta t$. I will contrast this with previous methods which only work in the \textit{continuum limit} (as $\delta t \to 0$). Moreover, I will show that any continuum-limit-based approach will always yield unitary dynamics unless it is fine-tuned in some way. Given the central role of information flow between the system and environment plays in open quantum systems, unitary models are wholly insufficient. Thus continuum limit master equations must be fine-tuned to even function as valid models of open quantum systems. For instance, it is common to find non-unitary dynamics in the continuum limit by taking an (I will argue unphysical) divergence in the interaction strengths, $g$, such that $g^2 \delta t$ is constant as $\delta t\to0$.

In addition to overcoming the above limitations, the new interpolation-based approach allows for the straightforward treatment of essentially any representation of a quantum system (e.g., Hilbert space vector, density matrix, Bloch vector, probability vector, in addition to a Gaussian state's mean vector and covariance matrix). Examples of each of these representations will be given throughout this thesis.

Moreover, the new interpolation-based approach allows for an order-by-order analysis of the dynamics as a series in $\delta t$. This allows us to identify which types of dynamics are ``fast'' and which are ``slow'' as well as how this ``speed'' depends on the interaction Hamiltonian between the system and ancilla. For instance, we can (and will) investigate under what conditions we can see purification effects at first order in $\delta t$. As I will show the ``speed'' of the purification effects are tied to the complexity of the interaction; Purification at first order in $\delta t$ requires the interaction Hamiltonian to be at least Schmidt rank-2. A necessary condition for thermalization is also discussed.

In addition to this purification study, I will present a complete analysis of Gaussian dynamics regarding which types of dynamics appear at which orders in $\delta t$ under which Hamiltonians. Given a Hamiltonian (either designed or fixed by fundamental considerations e.g., the light-matter interaction) we can determine what dynamics are supported at what orders in $\delta t$. Conversely, given some dynamics (e.g., from experiments) we can determine what class of interaction Hamiltonians could support it.

\cleardoublepage

% A C K N O W L E D G E M E N T S
% -------------------------------

\begin{center}\textbf{Acknowledgements}\end{center}
I would like to thank everyone who helped make this thesis possible.  Firstly, I would like to thank my research supervisors, Robert B. Mann and Eduardo Mart\textacute{i}n-Mart\textacute{i}nez, for their guidance and support throughout my Ph.D. I would also like to thank a few professors whose courses I have particularly enjoyed: Achim Kempf, Doreen Fraser, and Jacqueline Feke.

I have had many co-authors throughout my Ph.D. including: David Layden, Eric Brown, Marvellous Onuma-Kalu, Stella Seah, Stefan Nimmrichter, Jader P. Santos, Valerio Scarani, Gabriel T. Landi and Irene Melgarejo-Lermas. Collaborating on research with these individuals has been a great pleasure. I joyfully anticipate our future collaborations. In addition to these formal collaborations my research has been shaped by many conversations with Aida Ahmadzadegan, Paulina Corona Ugalde, Kfir Dolev, Jack Davis, Laura Henderson, Robie Hennigar, and Richard Lopp,  Maria Papageorgiou, Jose de Ramon Rivera, Nayeli Rodríguez Briones, Petar Simidzija, Erickson Tjoa, and Silas Vriend. I hope that we can all remain in contact as our lives develop.  

In the latter half of my Ph.D. I joined the Board of Directors of the Graduate Student Association - University of Waterloo. My time on the board has been extremely rewarding politically, socially, and professionally.

Finally, I would like to thank my family for their continued love and support, especially in the recent COVID-19 induced isolation. In this regard, my brother Benjamin Grimmer deserves special thanks.

I acknowledge the financial support of the Natural Sciences and Engineering Research Council of Canada (NSERC) through a Vanier Scholarship.

I acknowledge that I have conducted my Ph.D. studies on the traditional territory of the Neutral, Anishnaabe, and Haudenosaunee peoples. I acknowledge that the University of Waterloo resides on land designated as part of the Haldimand Treaty tract, and that this land was acquired unfairly.

\cleardoublepage

% D E D I C A T I O N
% -------------------

%\begin{center}\textbf{Dedication}\end{center}

%This is dedicated to the one I love.\ToDo{Write Dedication}
%\cleardoublepage

% T A B L E   O F   C O N T E N T S
% ---------------------------------
\renewcommand\contentsname{Table of Contents}
\tableofcontents
\cleardoublepage
\phantomsection    % allows hyperref to link to the correct page

% L I S T   O F   F I G U R E S
% -----------------------------
\addcontentsline{toc}{chapter}{List of Figures}
\listoffigures
\cleardoublepage
\phantomsection		% allows hyperref to link to the correct page

% L I S T   O F   T A B L E S
% ---------------------------
\addcontentsline{toc}{chapter}{List of Tables}
\listoftables
\cleardoublepage
\phantomsection		% allows hyperref to link to the correct page

% GLOSSARIES (Lists of definitions, abbreviations, symbols, etc. provided by the glossaries-extra package)
% -----------------------------
%\printglossaries
%\cleardoublepage
%\phantomsection		% allows hyperref to link to the correct page

% Change page numbering back to Arabic numerals
\pagenumbering{arabic}

%----------------------------------------------------------------------
% MAIN BODY
%----------------------------------------------------------------------
% Because this is a short document, and to reduce the number of files
% needed for this template, the chapters are not separate
% documents as suggested above, but you get the idea. If they were
% separate documents, they would each start with the \chapter command, i.e, 
% do not contain \documentclass or \begin{document} and \end{document} commands.
%======================================================================

\chapter{Introduction}\label{Ch1}
Open quantum dynamics---the study of quantum systems interacting with an uncontrolled environment---is relevant to a host of different disciplines ranging from applied physics and engineering to the foundations of quantum theory. It is fundamental to the quantum measurement problem and to quantum thermodynamics. Moreover, it is essential to building error models for quantum information technologies, from quantum computing to quantum sensing.

One common way to construct a model of open quantum dynamics is by using a \textit{Collision Model} (also know as a \textit{Repeated Interaction System}) \cite{Attal:2006, Attal:2007, Attal:2007b, Vargas:2008, Giovannetti:2012,PhysRev.129.1880,PhysRevA.72.022110,PhysRevE.99.042103}. In such a model, the state of a quantum system, $\hat{\rho}$, is repeatedly updated by the process,
\begin{align}\label{IntroEQ1}
\hat{\rho}(n\,\delta t)
\to\hat{\rho}\left((n+1)\delta t\right)
&=\phi(\delta t)[\hat{\rho}(n\,\delta t)],
\end{align}
where $\phi(\delta t)$ is some completely positive trace preserving (CPTP) map and where $\delta t$ is the duration of each interaction. Given an initial state $\hat{\rho}(0)$, such an update scheme defines the system state $\hat{\rho}(t)$ at time $t=n\,\delta t$ for $n=0,1,2,\dots$. 

Collision models have been used to study a wide scope of problems including decoherence and related effects \cite{Scarani:2002, Ziman:2002, Ziman:2005, Ziman:2005b}, quantum thermodynamics \cite{Bruneau:2006, Bruneau:2008, Bruneau:2008b, Karevski:2009, Bruneau:2014, Bruneau:2014b, Hanson:2015,PhysRevLett.123.140601,PhysRevA.100.042107,kosloff2013quantum}, quantum metrology \cite{PhysRevLett.123.180602}, and even gravitational decoherence \cite{Kafri:2014zsa,Kafri:2015iha,ACMZ,Altamirano:2016hug}

An intuitive example of a collision model is what we will call in this thesis \textit{ancillary bombardment}. In this type of scenario the environment is modeled as an infinite collection of quantum systems (ancillas, $A$) which the system interacts with sequentially. One may imagine a quantum system, $S$, in a gas interacting with each of the constituents of the gas, $A$, one at a time in sequence.

In this thesis we will be making the following assumptions on for any ancillary bombardment scenarios: 1) the system never interacts with the same ancilla twice, 2) the ancillas do not interact with each other, 3) the ancillas are all initially uncorrelated with each other, 4) the ancillas are all in the same state, $\hat{\rho}_\text{A}$, at the beginning of their interaction with the system and 5) each system-ancilla interaction is identical (same Hamiltonian, same duration, etc.). Given these assumptions, the dynamics for the system is time-independent and Markovian\footnote{Some modified collision models introduce non-Markovianity by relaxing assumption 2. After $S$ interacts with the $n^\text{th}$ ancilla, $A_n$, there is an interaction between $A_n$ and $A_{n+1}$ before $S$ interacts with $A_{n+1}$. In this way information about previous system states can influence the dynamics. See for instance, \cite{PhysRevA.87.040103,PhysRevA.100.052113}.} (there is no mechanism for memory effects). That is, given the above assumptions, this ancillary bombardment scenario is described by the update scheme \eqref{IntroEQ1}; the same $\phi(\delta t)$ is used at every time-point and $\hat{\rho}\left((n+1)\delta t\right)$ only depends on $\hat{\rho}\left(n\,\delta t\right)$. If the system and ancilla evolve for a duration $\delta t$ under a joint Hamiltonian, $\hat{H}$, then the update map is given by,
\begin{align}\label{IntroEQ2}
\phi(\delta t)[\hat{\rho}_\text{S}]
=\text{Tr}_\text{A}\Big(\exp(-\ii \, \delta t \, \hat{H}/\hbar)(\hat{\rho}_\text{S}\otimes \hat{\rho}_\text{A})\exp(\ii \, \delta t \, \hat{H}/\hbar)\Big).
\end{align}
This is not the only type of update map which is possible in a generic collision model, many others possibilities will be considered in Chapter \ref{Ch2.1}. Regardless, let us continue the introduction with this form of update map in mind.

One may hope that the Markovian time-independent dynamics given by the update scheme \eqref{IntroEQ1} can be modeled by a Markovian time-independent master equation. That is,
\begin{align}\label{IntroEQ3}
\frac{\dd}{\dd t} 
\hat{\rho}(t)
=\mathcal{L}[\hat{\rho}(t)],
\end{align}
for some super-operator, $\mathcal{L}$, which generates the dynamics. If this were the case then we could leverage many of the tools developed for master equations to our benefit.

The obvious difficulty in trying to construct a differential equation from \eqref{IntroEQ1} is that it only defines the system state at times $t=n\,\delta t$, whereas \eqref{IntroEQ3} presupposes a system state is defined at all times. One obvious solution to this problem is to consider the \textit{continuum limit} where $\delta t\to0$. This approach is common in the literature \cite{CM,PhysRevLett.115.120403,PhysRevLett.108.040401,PhysRevA.98.062104,PhysRevA.97.053811,PhysRevX.7.021003,PhysRevA.96.032107,PhysRevA.95.053838,Giovannetti_2012,Altamirano_2017,Attal2007,doi:10.1063/1.4879240}, however it is not without its limitations. Indeed, one of the central aims of this thesis is to draw into sharp relief these limitations.

Firstly, the $\delta t\to0$ limit cannot be achieved experimentally; in actuality, all interactions take a finite amount of time. The $\delta t\to0$ limit does not describe a realistic scenario and is taken out of mathematical convenience (i.e., in order to get a master equation). Contrast this with lattice models where we take the continuum limit ($\delta x\to0$) to recover the continuum theory which we treat as being exact. It is important to stress that this is not the case for collision models; reality has $\delta t$ finite and taking $\delta t\to0$ is an approximation which must be justified. In order to justify this treatment one needs a way to analyze finite duration effects, if only to verify that they are negligible. 

Secondly, as we will discuss in Chapter \ref{Ch3}, for a wide range of update maps (including all those of the form \eqref{IntroEQ2}) the continuum limit dynamics is unitary; that is, the system does not become entangled or even correlated with its environment.  If we are hoping to use a continuum limit collision model to describe non-trivial open dynamics, we are in trouble. Indeed, this amounts to a complete failure\footnote{Not to say this phenomena is always detrimental. An application of this unitary-in-the-continuum-limit phenomenon for the purpose of quantum control was given in \cite{Layden:2015b}. Other early examples of this phenomena can be seen in Refs. \cite{Zanardi:2014, Zanardi:2015, Layden:2015b, Zanardi:2016}.} as a model of open quantum systems. As prevalent as this unitary-in-the-continuum-limit phenomenon is, it is not completely unavoidable. For example, a common trick \cite{Giovannetti:2012, CM,ACMZ,PhysRevLett.115.120403,PhysRevLett.108.040401,PhysRevA.98.062104,PhysRevA.97.053811,PhysRevX.7.021003,PhysRevA.96.032107,PhysRevA.95.053838,Giovannetti_2012,Altamirano_2017,Attal2007,doi:10.1063/1.4879240} to get non-unitary dynamics in the continuum limit is to take the interaction strength, $g$, to diverge as $g\sim\sqrt{\kappa/\delta t}$ as $\delta t\to0$ for some constant $\kappa$ such that $g^2\delta t=\kappa$ is constant as $\delta t\to0$.

In Chapter \ref{Ch3} I will cast serious doubt on the physicality of this $g^2\delta t=\kappa$ approach. Indeed, I have never seen this approach justified on physical terms. Instead it is discussed as a mathematical trick or simply what it takes to ``make the math work'' (i.e., to find non-unitary dynamics in the continuum limit). In Chapter \ref{Ch3} I will provide a necessary and sufficient condition for any update map, $\phi(\delta t)$, to give non-unitary dynamics in the continuum limit. As we will see, satisfying this condition requires the dynamics to be in some way fine-tuned (the $g^2\delta t$ trick being one such possibility). 

Thus in order to use collision models to model non-unitary dynamics we must either:
\begin{enumerate}
    \item work without a master equation,
    \item work with the continuum limit master equation and somehow justify both the $\delta t\to0$ approximation and the necessary fine-tuning (e.g., $g^2\delta t$ constant),
    \item somehow construct a master equation outside of the continuum limit.
\end{enumerate}
The primary goal of this thesis is to pursue and develop the third option as an alternative to the second. Indeed, the formalism developed in this thesis will allow us to easily work outside of the continuum limit, thus avoiding both of the above discussed issues.

The new approach is to construct a uniquely well-behaved interpolations scheme which 1) satisfies a differential equation of the form \eqref{IntroEQ3}, 2) exactly matches the discrete dynamics at each time-point given by \eqref{IntroEQ1}, and 3) is amenable to study in the continuum limit. This allows us to model the system's evolution with a master equation without taking $\delta t\to0$. Keeping $\delta t$ finite, we automatically include the finite duration effects which will be present in any actual experiment. Moreover, these finite duration effects are typically non-unitary such that we do not need any fine tuning.

In addition to solving the above issues the approach outlined in this thesis has two more significant benefits. Firstly, the new interpolation-based approach allows for the straightforward treatment of essentially any representation of a quantum system (e.g., Hilbert space vector, density matrix, Bloch vector, probability vector, in addition to a Gaussian state's mean vector and covariance matrix). Examples of each of these representations will be given in this thesis.

Secondly, the master equation which we construct can often be expanded as a series in the interaction duration, $\delta t$. This allows us to identify not only finite duration effects, but to associate a perturbative-order to any given type of dynamics. That is, we can differentiate ``fast'' dynamics/processes (present at low orders in $\delta t$) from ``slow'' dynamics/processes (present only at higher orders in $\delta t$). 

For instance, in Chapter \ref{Ch4} I  will do an order-by-order analysis of the dynamics arising from an ancillary bombardment scenario with a generic joint Hamiltonian,\\ 
\mbox{$\hat{H}=\hat{H}_\text{S}\otimes\hat{\openone}_\text{A}+\hat{\openone}_\text{S}\otimes \hat{H}_\text{A}+\hat{H}_\text{SA}$}. As I will show the ``speed'' of the purification effects are tied to to the complexity of the interaction; Purification at first order in $\delta t$ happens if and only if the interaction Hamiltonian, $\hat{H}_\text{SA}$, is at least Schmidt rank-2 and obeys a certain commutation relationship. 

In Chapter \ref{Ch4} I will also discuss under what Hamiltonians and at what orders in $\delta t$ the dynamics depends on the energy scales associated to $\hat{H}_\text{S}$ and $\hat{H}_\text{A}$ as well as why this is a necessary condition for the dynamics to describe thermalization. As we will see this dependence does not occur until second order in $\delta t$. This suggests that the dynamics necessary for thermalization are unavoidably related to finite-duration effects. Thus a microscopic understanding of thermalization through collision models cannot be found in the continuum limit.

In general each of the terms in the interpolating master equation can be written directly and simply in terms of the above system-ancilla interaction Hamiltonian. This allows us to pick our favorite type of dynamics (e.g., purification, thermalization, squeezing, etc) and see which types of interaction Hamiltonian can support it. Conversely if we know the form of the system-ancilla interaction Hamiltonian from first principles (e.g., the light matter interaction) then we can quickly see which types of dynamics it can facilitate through ancillary bombardment. This dictionary between dynamics and interaction Hamiltonians is worked out completely in Chapter \ref{Ch5} and \ref{InterpolateGQM} for Bosonic Gaussian systems. That is, I will present a complete classification of Gaussian dynamics along with an analysis of which types of dynamics appear at which orders in $\delta t$ under which Hamiltonians.

\chapter{Interpolated Collision Model Formalism}\label{Ch2}
\section{Introduction to Collision Models}\label{IntroToCollision}
Consider a quantum system described by a density matrix, $\hat{\rho}$, being repeatedly updated by a completely positive trace preserving (CPTP) map, $\phi(\delta t)$, of duration $\delta t$ as,
\begin{align}
\hat{\rho}(n\,\delta t)
\to\hat{\rho}\left((n+1)\delta t\right)
&=\phi(\delta t)[\hat{\rho}(n\,\delta t)],
\end{align}
for integers $n\geq0$. Solving this recurrence relation gives,
\begin{align}\label{DiscreteMasterEqs}
\hat{\rho}(n \, \delta t)
=\phi(\delta t)^n[\hat{\rho}(0)],
\end{align}
for some initial state $\hat{\rho}(0)$. 

One way to analyze this discrete-time dynamics would be to find the eigendecomposition of the update map $\phi(\delta t)$. This sort of analysis would identify any fixed points (if they exist) as well as the rates of convergence. Such an eigendecomposition can be done analytically in simple cases but more often will end up needing to be done numerically. Instead, and in the hopes of obtaining generic analytic results, I will seek to recast \eqref{DiscreteMasterEqs} as a differential equation. A priori this seems difficult since the above update scheme only specifies the system state $\hat{\rho}(t)$ at the times $t=n\,\delta t$.

If we wanted to know the system's state exactly at time $t\neq n\,\delta t$, we would need to know something about how the update map, $\phi(\delta t)$, is performed in time. That is, we would need to somehow know the ``interruption map'' map $\phi(\delta t, r)$ which describes the result of interrupting an update of duration $\delta t$ at a time $0\leq r\leq \delta t$.\footnote{We cannot always be guaranteed that such an interruption map exist. Consider the case where the interaction Hamiltonian is controlled by a switching function, $\chi(t)$, as $\hat{H}_\textsc{int}(t)=\chi(t)\hat{H}$. For instance, $\chi(t)$ may be a Gaussian. In such cases, interrupting may involve switching the interaction off suddenly, effectively causing a discontinuity $\chi(t)$. In some quantum field theory settings this may cause uncontrolled divergences.} For continuity we will assume that $\phi(\delta t,0)=\openone$ and $\phi(\delta t, \delta t)=\phi(\delta t)$. 

Given such an interruption map the system state at intermediate times is given by, 
\begin{align}\label{ExactIntermediate}
\hat{\rho}_\textsc{exact}(t)
=\phi(\delta t, r)[\phi(\delta t)^n[\hat{\rho}(0)]],
\end{align}
where $n$ and $r$ are the quotient and remainder of $t/\delta t$ respectively. Can we build a differential equation for the system dynamics given this interruption map? As we will see this poses some difficulties. If $\phi(\delta t, r)$ is differentiable with respect to $r$ and invertible then for $t\neq n\,\delta t$ the time derivative of $\hat{\rho}_\textsc{exact}(t)$ is linear and time-dependent as,
\begin{align}\label{ExactME}
\frac{\dd}{\dd t} 
\hat{\rho}_\textsc{exact}(t)
=\mathcal{L}_{\delta t}(r)[\hat{\rho}_\textsc{exact}(t)],
\qquad
t\neq n\,\delta t
\end{align}
where $\mathcal{L}_{\delta t}(r)=\left(\frac{\dd}{\dd r}\phi(\delta t, r)\right) \, \phi(\delta t, r)^{-1}$. However, at $t=n\,\delta t$ it is not to guaranteed that $\hat{\rho}_\textsc{exact}(t)$ is differentiable\footnote{We will see in Fig \ref{ThesisFigure1} in Sec. \ref{Framework} an example where $\hat{\rho}_\textsc{exact}(t)$ is not differentiable at $t=n\,\delta t$.}. The cyclic time-dependence in $\mathcal{L}_{\delta t}(r)$ and its potential non-smoothness at $t=n\,\delta t$ make an analysis of the exact dynamics in terms of a differential equation infeasible.

An alternate approach to obtaining a differential equation from \eqref{DiscreteMasterEqs} (indeed the standard approach) is to restrict our attention to the \textit{continuum limit}, in which $\delta t\to0$. As we will discuss throughout this thesis this approach is limited in scope, unphysical, and in practice requires fine-tuning to provide a useful model of open quantum systems. For an overview see the discussion in Chapter \ref{Ch1} following Eq. \eqref{IntroEQ3}.

Instead the one approach taken in this thesis is to construct a continuous-time interpolation scheme for the dynamics, defining a system state $\hat{\rho}_\textsc{int}(t)$ for all $t\geq0$ (not just $t=n\,\delta t$). As we will see, we can take this interpolation scheme to be generated by a time-independent linear differential equation, $\frac{\dd}{\dd t}\hat{\rho}_\textsc{int}(t)=\mathcal{L}_{\delta t}[\hat{\rho}_\textsc{int}(t)]$ by using the \textit{interpolated collision model formalism} described in this thesis.

\section{A Note on Representations of Quantum Systems}\label{AltRep}

Before going through the interpolated collision model formalism in detail, it is worth considering the various representations of quantum systems to which these techniques can be applied.

In the previous section we  took our quantum system to be represented by a density matrix, $\hat{\rho}$, and updated by a CPTP map, $\phi(\delta t)$. This update map (and evolution in general) is linear for quantum systems (at least when the system is initially uncorrelated with its environment). In addition to this standard representation there are often alternate or simplified ways to describe the state of a quantum system that are available and preferable in certain contexts. For instance:
\begin{enumerate}
    \item If the system is known to be in a pure state throughout its evolution, we can describe it by a vector in a Hilbert space, $\ket{\psi}$.
    \item The state of a two-level system can be represented exactly by its Bloch vector \mbox{$\bm{a}=\text{Tr}(\hat{\bm{\sigma}}\,\hat{\rho})$} where $\hat{\bm{\sigma}}=(\hat{\sigma}_x,\hat{\sigma}_y,\hat{\sigma}_z)$ is the vector of Pauli operators.
    \item If the density matrix is known to be diagonal in some fixed basis, $\{\ket{k}\}$, then it can be represented by a probability vector $\bm{p}$ with \mbox{$p_k=\text{Tr}(\ket{k}\!\bra{k} \, \hat{\rho})$}.
    \item More generally, if the state is known to be a convex combination of some fixed set of linearly independent states $\{\hat{\rho}_k\}$ as $\hat{\rho}=\sum_k p_k\,\hat{\rho}_k$ then the state can be uniquely represented by the probability vector $\bm{p}=U^{-1}\bm{q}$ where $U_{jk}=\text{Tr}(\hat{\rho}_j\hat{\rho}_k)$ is the matrix of inner products and $\bm{q}_k=\text{Tr}(\hat{\rho}_k\,\hat{\rho})$
    \item The state of a collection of Bosonic modes (e.g., harmonic oscillators) can be equivalently described by its Wigner function. If this Wigner function is Gaussian, then it can be represented by the first and second moments of its quadrature operators $\hat{\bm{X}}=(\hat{q}_1,\hat{p_1},\hat{q}_2,\hat{p}_2,\dots)$ as $\bm{X}=\langle\hat{\bm{X}}\rangle$ and 
    $\sigma_{jk}=\big\langle\{\hat{X}_j,\hat{X}_k\}
    \big\rangle
    -2\big\langle\hat{X}_j\big\rangle
    \big\langle\hat{X}_k\big\rangle$.
\end{enumerate}
In each of these cases, the alternate representation is a linear function of the density matrix $\hat{\rho}$ and therefore by linearity must also evolve by the application of some linear map\footnote{In actuality the density matrix $\hat{\rho}$ of a valid quantum state lives on an affine subspace (the hyperplane of states with $\text{Tr}(\hat\rho)=1$). The linear maps by which quantum systems evolve preserve this subspace. If we restrict our attention to this affine subspace, then our formerly linear transformations become linear-affine (i.e., $\bm{v}\to A\bm{v}+b$ instead of just $\bm{v}\to A\bm{v}$). Thus it is more appropriate to say that quantum systems evolve by linear-affine transformations. Indeed as we will see several of the above discussed representations of quantum systems evolve in a linear-affine way.}.

Quantum systems represented in all of the above ways undergoing repeated updates will be discussed in this thesis. Hilbert space vectors, $\ket{\psi}$, are discussed in Section \ref{UnitaryExamples}.
Density matrices, $\hat{\rho}$, are discussed in Chapters \ref{Ch3} and \ref{Ch4} as well as Sections \ref{MixedUnitaryExample} and \ref{AncillaryBombardmentExample}. Bloch vectors, $\bm{a}$, are discussed in Section \ref{BlochExample}. Probability vectors, $\bm{p}$, are discussed in Section \ref{ZenoEffectExample}. Bosonic Gaussian systems, $\bm{X}$ and $\sigma$, are discussed in Chapter \ref{InterpolateGQM}.

\section{Interpolated Collision Model Formalism}\label{Framework}
To lay the groundwork as generally as possible, covering all\footnote{As mentioned in the previous footnote, some of the above representations of quantum systems require linear-affine update maps. The relevant extensions for these cases will be discussed in Sec. \ref{BlochExample} and Chapter \ref{InterpolateGQM}.} of the above discussed representations, in this section we will adopt a context-neutral notation. We will consider a quantum state described by a vector $\bm{v}$ (e.g., $\ket{\psi}$ or $\bm{a}$ 
or $\bm{p}$ or $\bm{X}$ from Sec. \ref{AltRep}) being repeatedly updated by some linear map, $M(\delta t)$, of duration $\delta t$ as,
\begin{align}
\bm{v}(n\,\delta t)
\to\bm{v}((n+1)\delta t)
&=M(\delta t)\bm{v}(n\,\delta t),
\end{align}
for integer $n\geq0$. As before solving this recurrence relation gives,
\begin{align}\label{VMUpdateScheme}
\bm{v}(n \, \delta t)
=M(\delta t)^n \, \bm{v}(0),
\end{align}
for some initial state $\bm{v}(0)$. A generalization to allow for linear-affine update equations is discussed in Sec. \ref{BlochExample}. A generalization to allow for matrix update equations is discussed in Chapter \ref{InterpolateGQM}. 

In terms of $\bm{v}$ and $M(\delta t)$ our goal is to construct from these discrete time-points  a continuous-time interpolation scheme for the dynamics, defining a system state $\bm{v}_\textsc{int}(t)$ for all $t\geq0$ (not just $t=n\,\delta t$). We will attempt to define this interpolation scheme by a time-independent linear differential equation,
\begin{align}\label{VMDiffyQ}
\frac{\dd}{\dd t}\bm{v}_\textsc{int}(t)=L_{\delta t}[\bm{v}_\textsc{int}(t)],
\end{align}
where $L_{\delta t}$ is some linear operator on $\bm{v}$.  We will refer to this operator as the interpolation generator or the Liouvillian depending on context. This differential equation will be called the \textit{master equation} regardless of context. 

\begin{figure}
\includegraphics[width=0.95\textwidth]{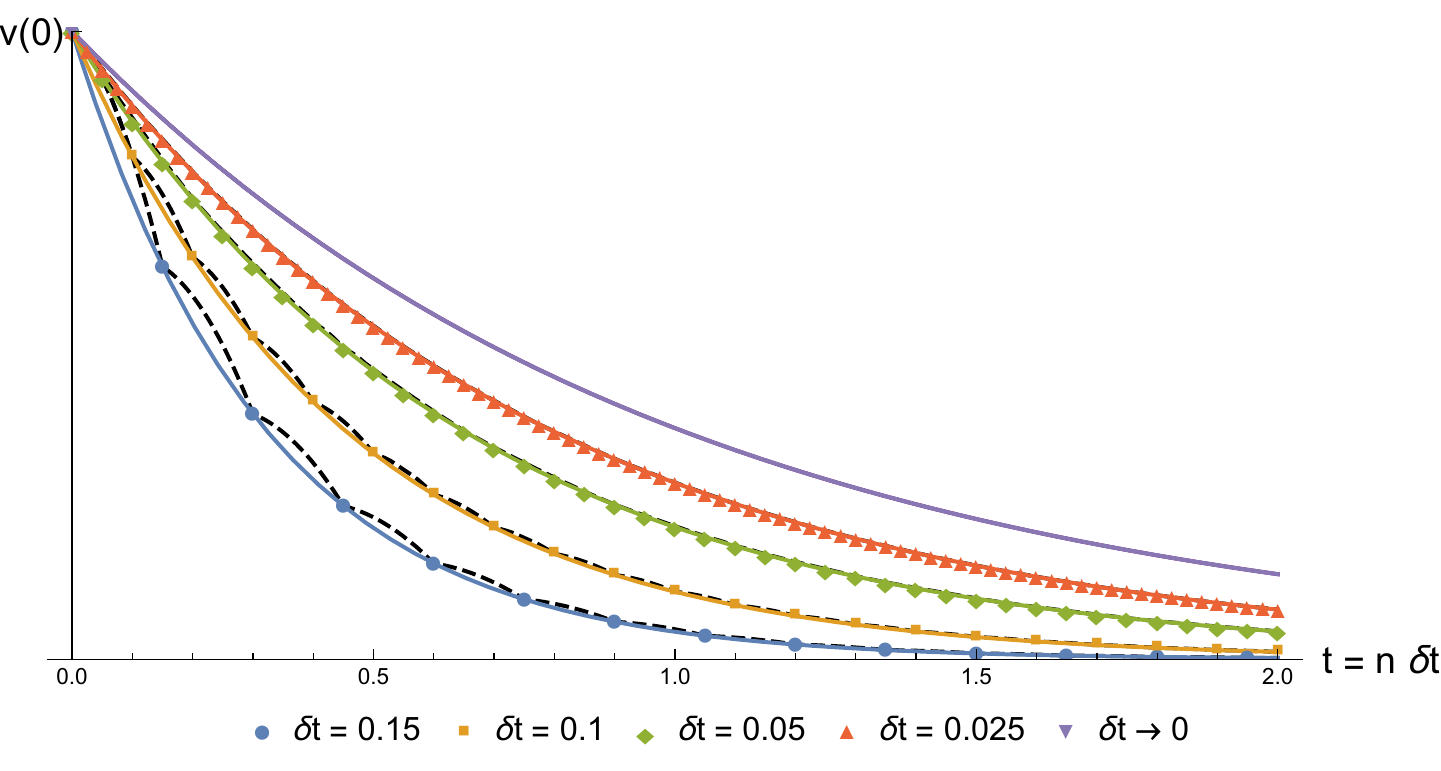}
\caption{The state, $\bm{v}(n\,\delta t)$, of a 1-dimensional system repeatedly updated by a linear map $M(\delta t)=1-b\,\delta t-a\,\delta t^2$ for various durations, $\delta t$, with $a=10$ and $b=1$. The exact dynamics (assuming an interruption map of \mbox{$M(\delta t,r)=M(r)$}) is shown as black dashed lines. The exact dynamics seems to ``bounce'' every $\delta t$ as the interaction is reset. This is an example of the non-smoothness discussed after Eq. \eqref{ExactME}. The interpolated dynamics (solid) is plotted for each duration. The purple solid line (above the others) shows the dynamics in the continuum limit, as $\delta t\to0$.}
\label{ThesisFigure1}
\end{figure}

To develop our intuition let us consider a very simple example scenario in which $\bm{v}$ is 1-dimensional (i.e., a real number), $M(\delta t)=1-b\,\delta t-a\,\delta t^2$ for some $a,b\geq0$, and $\bm{v}(0)=1$. We will return to this example throughout the rest of this section. The discrete-time dynamics of such a system is shown in  Fig. \ref{ThesisFigure1} (points) for a variety of interaction durations $\delta t$. The dashed lines in Fig \ref{ThesisFigure1} show the system dynamics at intermediate times, $t\neq n\,\delta t$, assuming the interruption map is $M(\delta t, r)=M(r)$. The solid lines in Fig \ref{ThesisFigure1} show the interpolated dynamics which will be constructed in this section. Note that in this example (and in general) two updates of half the duration has a different result than one update. That is $M(\delta t/2)^2\neq M(\delta t)$. Thus the interpolation scheme that we are seeking to construct (and the master equation \eqref{VMDiffyQ} which generates it) will generally depend on $\delta t$.

A straightforward approach to deriving continuous-time dynamics from the discrete update scheme \eqref{VMUpdateScheme} is to consider the \textit{continuum limit}, where $\delta t\to0$ with $t=n\,\delta t$ fixed. In this limit, the separation between each of the discrete time-points goes to zero such that they effectively form a continuous line, i.e., a continuum. Imagine the red triangles in Fig. \ref{ThesisFigure1} moving upwards and merging together into the purple (solid) line.

Within the continuum limit, we can define the state's time derivative as,
\begin{align}\label{MinumumRegularity}
v'(t)&\coloneqq
\lim_{\substack{\delta t\to0 \\ t=n\,\delta t}} \frac{\bm{v}((n+1)\delta t)-\bm{v}(n\,\delta t)}{\delta t}
=\left(\lim_{\delta t\to0}\frac{M(\delta t)-\openone}{\delta t}\right)\bm{v}(t)
=M'(0) \, \bm{v}(t),
\end{align}
where in the last step we are forced to make two assumptions: 
\begin{enumerate}
    \item $M(\delta t)\to\openone$ as $
    \delta t\to0$, where $\openone$ is the identity map. That is, nothing happens in no time.
    \item $M(\delta t)$ is differentiable at $\delta t=0$. That is, things happen at a finite rate.\footnote{Note that since, $M(\delta t)$ is only defined for $\delta t\geq0$ the limit which defines the derivative is ``one-sided''. The only way that $M(\delta t)$ could fail to be differentiable is if $(M(\delta t)-\openone)/\delta t$ diverges or oscillates wildly as $\delta t\to0^+$. In either case it is reasonable to say things are happening at an infinite rate.}
\end{enumerate}
Note that we will be using the notation $f'(x)$ to denote the derivative of a function $f$ at $x$. If the above assumptions hold then we can define the interpolation generator in the continuum limit as $L_0\coloneqq\lim_{\delta t\to0} L_{\delta t}=M'(0)$. The master equation \eqref{VMDiffyQ} can be easily solved in this limit yielding $\bm{v}(t)=\exp(M'(0)\,t)\,\bm{v}(0)$. In the simple 1-dimensional example discussed above $M'(0)=-b$ such that in the continuum limit we have $v(t)=\exp(-b\,t)\,v(0)$. Note that the continuum limit dynamics is independent of the $a$ parameter. This dynamics is plotted (purple solid) in Fig. \ref{ThesisFigure1}.  

Can we define an interpolation scheme outside of this limit? To do so we will need to fill in the gaps between the points $t=n\,\delta t$. In general, there are infinitely many interpolation schemes which match a given set of points. However, as first discussed in \cite{Grimmer2017a}, there is a unique interpolation scheme satisfying the following three assumptions: 
\begin{enumerate}
\item We assume the interpolated evolution is given by a linear, first-order, time-independent, time-local differential equation,
\begin{align}\label{AssumptionsMasterEquation}
\frac{\dd}{\dd t}\bm{v}_{\textsc{int}}(t)=L_{\delta t}[\bm{v}_{\textsc{int}}(t)],
\end{align}
for some linear map $L_{\delta t}$ (called the interpolation generator or the Liouvillian depending on context). As noted above, we will call this differential equation the \textit{master equation} regardless of the context. The master equation can be formally solved yielding,
\begin{align}\label{AssumptionsInterpolationDiffEQSol}
\bm{v}_{\textsc{int}}(t)=\exp(t \, L_{\delta t}) \, \bm{v}_{\textsc{int}}(0).
\end{align}

\item We assume the interpolated evolution exactly matches the discrete dynamics given by \eqref{DiscreteMasterEqs} at the end of every time step, that is at $t=n\,\delta t$,
\begin{align}
\bm{v}_{\textsc{int}}(n \, \delta t)
=\bm{v}(n\,\delta t),
\end{align}
for every integer $n\geq0$. From \eqref{VMUpdateScheme} and \eqref{AssumptionsInterpolationDiffEQSol} this means,
\begin{align}
\exp(n \, \delta t \, L_{\delta t})
=M(\delta t)^n,
\end{align}
for all $n\geq0$ or equivalently just,
\begin{align}\label{AssumptionsMatchingCondition}
\exp(\delta t \, L_{\delta t})
=M(\delta t).
\end{align}
Given the above assumptions about $M(\delta t)$ around $\delta t=0$, we have that for small enough $\delta t$, $M(\delta t)$ is ``near'' to $\openone$. Specifically, for small enough $\delta t$ we can assume that $M(\delta t)$ has a well defined logarithm\footnote{Throughout this thesis ``log'' and ``Log'' will both stand for the natural logarithm.} since eventually it will be in the radius of convergence of $\text{log}(x)$ around $x=1$. In this case \eqref{AssumptionsMatchingCondition} can be solved for the interpolation generator as,
\begin{align}\label{AssumptionsLDef0}
L_{\delta t}=\frac{1}{\delta t}\text{log}(M(\delta t)).
\end{align}
Note that this does not yet uniquely specify an interpolation scheme --  there is still ambiguity as to which branch cut of the logarithm we select.

\item Finally we assume that the interpolation generator converges in the limit $\delta t\to0$. This helps us resolve the ambiguity of the logarithm's branch cut. Taking $\delta t\to0$ in \eqref{AssumptionsLDef0} and demanding that $L_{\delta t}$ converges we are forced to take a branch cut with $\text{log}(\openone)=0$. We denote this choice by capitalizing the log function, writing,
\begin{align}\label{AssumptionsLDef1}
L_{\delta t}\coloneqq\frac{1}{\delta t}\text{Log}(M(\delta t)).
\end{align}
\end{enumerate}
These assumptions uniquely specify the interpolation scheme that is given by the master equation,
\begin{align}\label{MasterEquation}
\frac{\dd}{\dd t}\bm{v}_{\textsc{int}}(t)
=L_{\delta t}[\bm{v}_{\textsc{int}}(t)]
\quad\text{where}\quad
L_{\delta t}\coloneqq\frac{1}{\delta t}\text{Log}(M(\delta t)).
\end{align}
By construction this master equation is amenable to studying the \textit{continuum limit}, i.e. as $\delta t\to0$. By L'H\^opital's rule we have,
\begin{align}\label{AssumptionsL0Def}
L_0
\coloneqq\lim_{\delta t\to0}
\frac{\text{Log}(M(\delta t))}{\delta t}
=\frac{\dd}{\dd \, \delta t}
\bigg\vert_{\delta t=0}
\text{Log}(M(\delta t))
=M(0)^{-1}
M'(0)
=M'(0),
\end{align}
where in the last line we have made use of both of our earlier assumptions about $M(\delta t)$ (that nothing can happen in no time and that things happen at a finite rate). Thus interpolative approach produces the same continuum limit as the earlier straightforward approach whilst generalizing to outside this limit.

\begin{figure}
\includegraphics[width=0.95\textwidth]{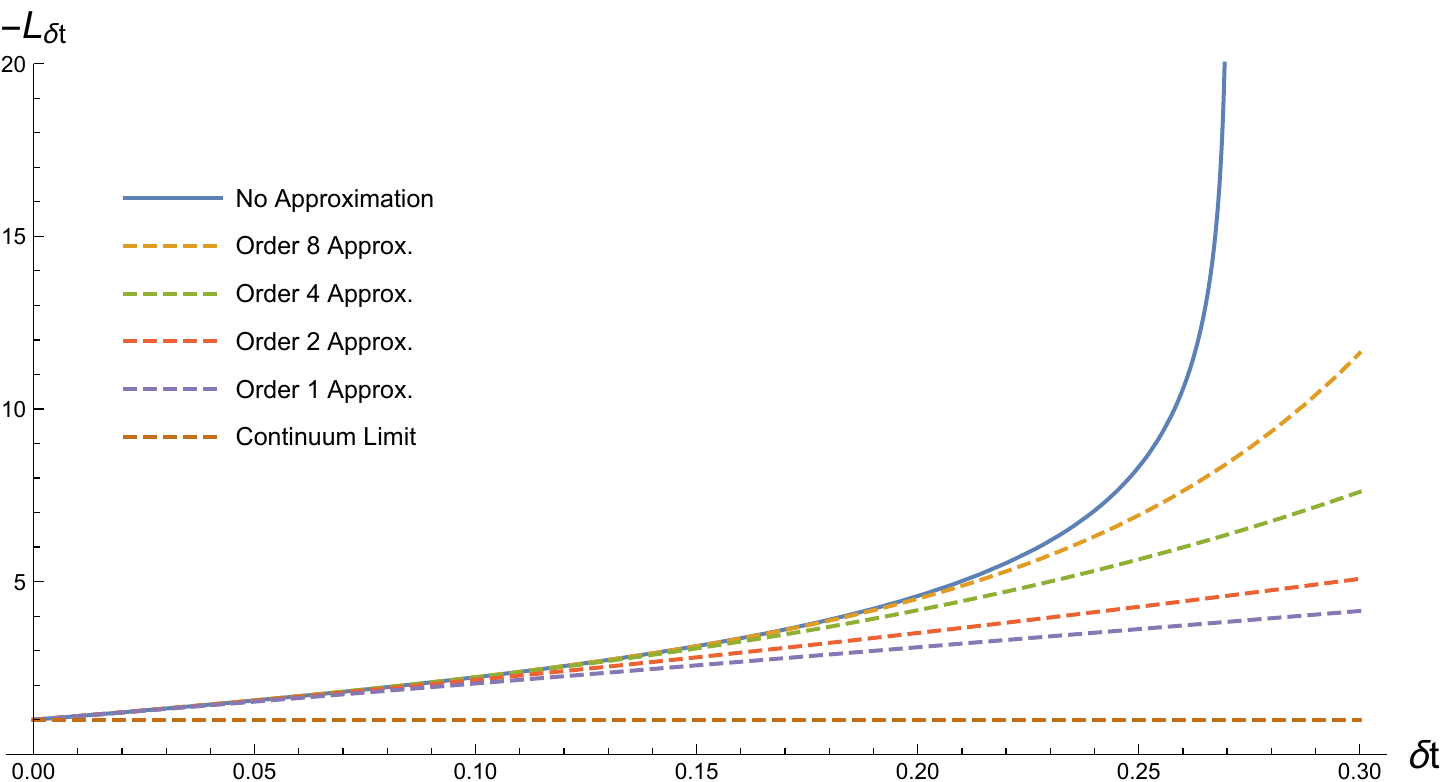}
\caption{The interpolation generator $L_{\delta t}$ for the update map $M(\delta t)=1-b\,\delta t-a\,\delta t^2$ with $a=10$ and $b=1$ is plotted (solid) as a function of $\delta t$. The interpolation generator is also plotted (dashed) at various levels of approximation, see Eq. \eqref{LdtSeries}.}
\label{ThesisFigure2}
\end{figure}

We can now complete our analysis of the simple example discussed above by providing an interpolation scheme for each duration $\delta t$. In this case, the solution to the interpolation equation \eqref{MasterEquation} is
\begin{align}
v_{\textsc{int}}(t)=\exp(t \, L_{\delta t}) \, v_{\textsc{int}}(0)
\quad\text{where}\quad
L_{\delta t}=\frac{1}{\delta t} \, \text{Log}(1-b\,\delta t-a\,\delta t^2).
\end{align}
That is, exponential decay at a rate of $-L_{\delta t}$. See the solid lines in Fig \ref{ThesisFigure1} that show the interpolation scheme for several interaction durations $\delta t$. Note that (by construction) the interpolated state exactly matches the discrete dynamics (and the exact dynamics as well) at each time point $t=n\,\delta t$. This rate of the exponential decay, $-L_{\delta t}$, is plotted (solid) as a function of $\delta t$ in Fig. \ref{ThesisFigure2}. Note that $L_{\delta t}$ diverges at $\delta t\approx0.27$, this corresponds to a root of $M(\delta t)=1-b\,t-a\,t^2=0$ with $a=10$ and $b=1$.

Unfortunately, in many scenarios of interest we will not be able to write down a clean analytic expression for $L_{\delta t}$ as we were able to in this simple example. In most cases of interest though we can expand the interpolation generator as a series around $\delta t=0$ and then explicitly characterize the first few low-order terms. The easiest way to do this is to first expand the update map $M(\delta t)$ as a series in $\delta t$ and then use \eqref{MasterEquation} to expand $L_{\delta t}$. For this approach to work, we need to make some additional assumptions about the update map. Recall that we have already assumed that $M(\delta t)\to\openone$ as $\delta t\to 0$ and that $M(\delta t)$ is differentiable at $\delta t=0$ in order for the interpolation generator to converge in the continuum limit. We now additionally assume that $M(\delta t)$ is analytic at $\delta t=0$ and can therefore be expanded as, 
\begin{align}\label{MSeries}
M(\delta t)
=\openone
+\delta t \, M_1
+\delta t^2 \, M_2
+\delta t^3 \, M_3
+\dots \, ,
\end{align}
for some linear maps $M_m$. From this we can then expand the interpolation generator, $L_{\delta t}$, as,
\begin{align}\label{LdtSeries}
L_{\delta t}
=L_0
+\delta t \, L_1
+\delta t^2 \, L_2
+\delta t^3 \, L_3
+\dots \, ,
\end{align}
for some linear maps $L_m$.

As was shown in an Appendix A of \cite{Grimmer2016a}, (and has been translated here in our new context-neutral notation) these coefficient maps can be computed recursively as,
\begin{align}
\label{L0def}
L_0
&\coloneqq M_1
&
&=M_1,\\
\label{L1def}
L_1
&\coloneqq M_2
-\frac{1}{2}L_0{}^2
&
&=M_2-\frac{1}{2}M_1{}^2,\\
\label{L2def}
L_2
&\coloneqq M_{3}
-\frac{1}{2}(L_0L_1+L_1L_0)
-\frac{1}{6}L_0{}^3
&
&=M_3
-\frac{1}{2}(M_1M_2+M_2M_1)
+\frac{1}{3}M_1{}^3.
\end{align}
In general the $m^{th}$-order term, $L_m$, is given by,
\begin{align}
L_{m} 
=M_{m+1}
-\sum_{n=1}^{m}\!\frac{1}{(n+1)!}
\sum_{\bm{\beta}\in\text{C}_\text{w}(m-n,n+1)}\prod_{i=1}^{n+1} L_{\beta_i},
\end{align}
where $\text{C}_\text{w}(M,N)$ are the weak compositions of $M$ of length $N$:  ordered lists of length $N$ of non-negative integers that sum to $M$. For example,
\begin{align}
\text{C}_\text{w}(3,2)
&=\{
(3,0),
(0,3),
(2,1),
(1,2)
\},\\
\text{C}_\text{w}(2,3)
&=\{
(2,0,0),
(0,2,0),
(0,0,2),
(1,1,0),
(1,0,1),
(0,1,1)
\}.
\end{align}
One can check using this formula that the next term, $L_3$, is given by,
\begin{align}
\label{L3def}
L_3
\coloneqq M_{4}
-\frac{1}{2}(L_0L_2+L_1{}^2+L_2L_0)
-\frac{1}{6}(L_0{}^2L_1+L_0L_1L_0+L_1L_0{}^2)
-\frac{1}{24}L_0{}^4.
\end{align}
Note that in general $L_n$ and $L_m$ will not commute.

For the simple 1-dimensional example discussed above we can easily compute $M_1=-b$ and $M_2=-a$ such that,
\begin{align}
L_0&= -b,\\
L_1&= -a -\frac{1}{2}b^2,\\
L_2&= -a\,b-\frac{1}{3}b^3,\\
L_3&= -a\,b^2-\frac{1}{2}a^2-\frac{1}{4}b^4.
\end{align}
A plot of $L_{\delta t}$ from our running example truncated at various orders in $\delta t$ can be found in Fig. \ref{ThesisFigure2} (dashed). Note how going even just to first order (past the zeroth order continuum limit approximation) improves the approximation significantly for small $\delta t$.

With the interpolation generator so expanded, the system's master equation can also be expanded as,
\begin{align}\label{MasterEqsSeries}
\frac{\dd}{\dd t}\bm{v}_\textsc{int}(t)
=L_0 \, \bm{v}_\textsc{int}(t)
+\delta t \, L_1 \, \bm{v}_\textsc{int}(t)
+\delta t^2 \, L_2 \, \bm{v}_\textsc{int}(t)
+\dots \, .
\end{align}
If $\delta t$ is small compared to the timescales set by the dynamics then we can study just the first few terms in this master equation to get a good approximation of the system's behavior.

\begin{figure}
\includegraphics[width=0.95\textwidth]{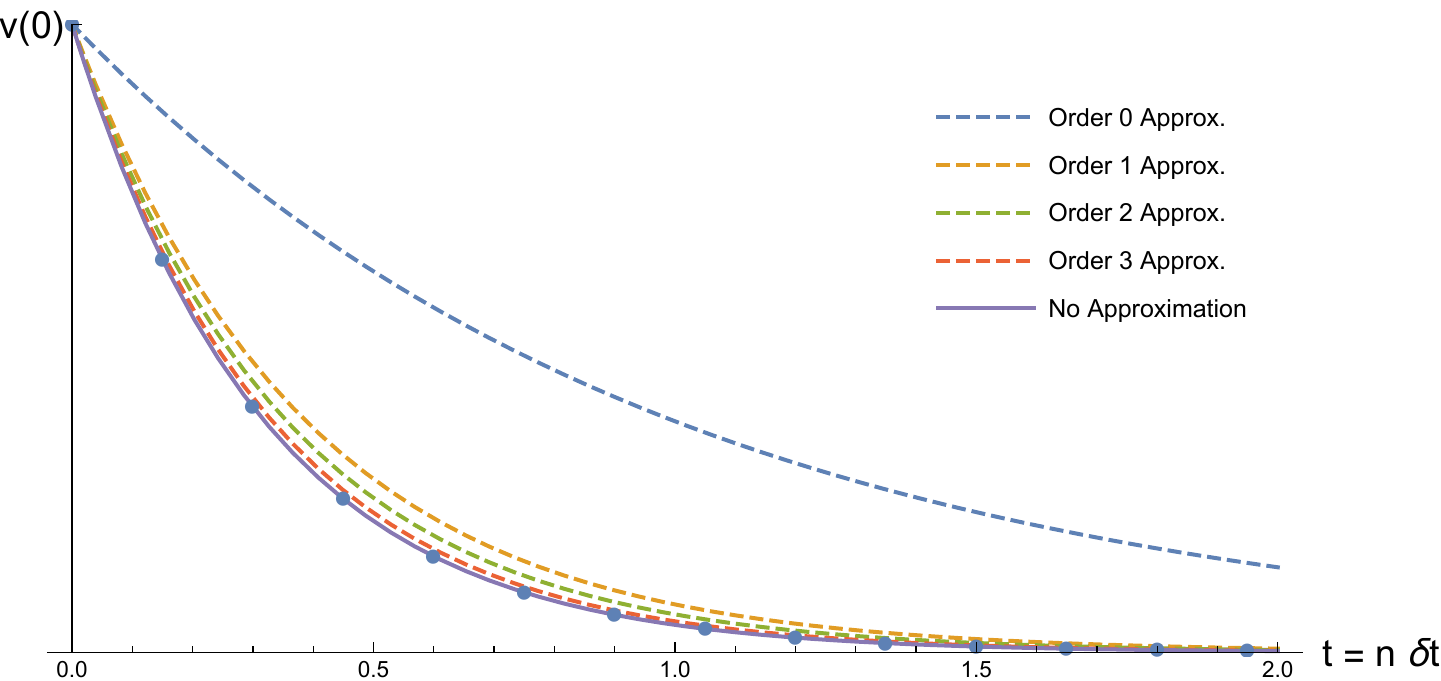}
\caption{The state, $\bm{v}(n\,\delta t)$, of a 1-dimensional system repeatedly updated by a linear map $M(\delta t)=1-b\,\delta t-a\,\delta t^2$ with $a=10$ and $b=1$ and $\delta t=0.15$ (blue circles). The full (untruncated) interpolation scheme is plotted (solid) going through these point. Truncating the interpolation at various orders results in the other lines (dashed).}
\label{ThesisFigure3}
\end{figure}

To get a sense of how this might work, let us once again consider our running example. Solving the truncated master equations is trivial; The result is exponential decay at a rate $-L_{\delta t}^{(m)}\geq0$ where \mbox{$L_{\delta t}^{(m)}=L_0+\dots+\delta t^m L_m$}, namely \mbox{$v_\text{int}(t)=\exp(L_{\delta t}^{(m)} \, t) \, v(0)$}. We plot the resulting truncated dynamics for $\delta t=0.15$ in Fig. \ref{ThesisFigure3}. Note that unlike the full (untruncated) interpolated dynamics, the truncated dynamics is not guaranteed to match the discrete dynamics at each time point. Note however that the truncated interpolation scheme does match the untruncated dynamics fairly well keeping only a few orders in $\delta t$.

While the application of the interpolative approach to this 1D example may seem relatively trivial in some respects, the application of this method to a variety of quantum scenarios will prove incredibly fruitful. We will consider a wide range of examples in the next chapter.

\chapter{Several Example Scenarios}\label{Ch2.1}
To build intuition and to develop some tools that will be useful later, we will now work through several basic example scenarios. Along the way we will develop some significant results.

\section{Three Unitary Examples (Branch Cuts and Time-dependent Hamiltonians)}\label{UnitaryExamples}
As a warm-up we will first look at three examples of unitary updates.

\subsubsection{Unitary Example 1: Branch Cuts}
Take the system to be represented by a vector, $\ket{\psi}$, in some Hilbert space that undergoes repeated updates by a unitary map as,
\begin{align}
\ket{\psi}
\to \hat{U}(\delta t)\ket{\psi},
\end{align}
where,
\begin{align}
\hat{U}(\delta t)
=\text{exp}(-\ii \, \hat{H} \, \delta t/\hbar),
\end{align}
for some Hamiltonian, $\hat{H}$. To match the notation of the previous Chapter: $\bm{v}\equiv\ket{\psi}$ and $M(\delta t)\equiv \hat{U}(\delta t)$.

Note that so far we have not said anything about the system's behavior between time points $t=n\,\delta t$. That is, we have not proscribed an interruption map, $\hat{U}(\delta t, r)$. One option (perhaps the most natural one) would be to take $\hat{U}(\delta t,r)=\hat{U}(r)$. In this case, interrupting an interaction of duration $\delta t$ after $r$ second is the same as a full interaction of duration $r$. Let us call this the intuited exact dynamics. 

However, this is not the only interrupted dynamics which is consistent with the update map $\hat{U}(\delta t)$. Another option is to take,
\begin{align}\label{Umod}
\hat{U}_\text{alt}(\delta t,r)
=\text{exp}\left(-\ii \, \hat{H}_\text{alt} \, r/\hbar\right)
\quad\text{where}\quad
\hat{H}_\text{alt}
=\hat{H}+\frac{2\pi\hbar\,z}{\delta t}\hat{\openone}
\end{align}
for some integers $z\in\mathbb{Z}$. Note that this interruption map has $\hat{U}_\text{alt}(\delta t,0)=\hat{\openone}$ and $\hat{U}_\text{alt}(\delta t,\delta t)=\hat{U}(\delta t)$. Under this dynamics the system states undergoes an extra $z$ complete rotations in phase between $t=0$ and $t=\delta t$.

As we have discussed in Sec \ref{IntroToCollision}, our interpolation scheme is constructed without knowledge or reference to such interruption maps, $\hat{U}(\delta t,r)$. Let us now investigate what our interpolation scheme gives between the timepoints $t=n\,\delta t$ and how this compares with the two possible interrupted dynamics discussed above.

To begin let us check that our update map has the minimum required regularity around $\delta t=0$ for $L_{\delta t}$ to converge in the continuum limit as $\delta t\to0$ (see the discussion around Equation \eqref{MinumumRegularity}). We can trivially confirm that $\hat{U}(0)=\hat{\openone}$ is the identity map and that $\hat{U}(\delta t)$ is differentiable at $\delta t=0$. Indeed, $\hat{U}'(0)=-\ii\hat{H}/\hbar$. These computations establish that in the continuum limit the interpolation generator is $L_0=\hat{U}'(0)=-\ii\hat{H}/\hbar$ such that the system's continuum limit master equation is,
\begin{align}
\frac{\dd}{\dd t}\ket{\psi(t)}
=L_0 \, \ket{\psi(t)}
=\hat{U}'(0)\,\ket{\psi}
=\frac{-\ii}{\hbar}\hat{H}\ket{\psi(t)}.
\end{align}
That is, the system evolves unitarily via the Hamiltonian $\hat{H}$. This is completely unsurprising, and matches the intuited exact dynamics between timepoints $t=n\,\delta t$.

In this example, we can also calculate the interpolation generator exactly outside of the continuum limit. Indeed we have,
\begin{align}\label{Unitary1}
L_{\delta t}
=\frac{1}{\delta t}\text{Log}(\hat{U}(\delta t))
=\frac{-\ii}{\delta t} \, \text{mod}_{[-\pi,\pi)}\Big(\frac{\hat{H} \, \delta t}{\hbar}\Big),
\end{align}
where $\text{mod}_{[-\pi,\pi)}(x)$ computes $x$ modulo $2\pi$ returning a value in the range $[-\pi,\pi)$. The action of this function on a Hermitian operator is to be understood as acting on its spectrum. This modulus arises from the branch cut that we have chosen for the logarithm.

To clarify the interpretation of this let us take for example $\hat{H}=0$. In this case $L_{\delta t}=0$, the interpolated evolution is no evolution, i.e. the state is stationary. Let us compare this with two possibilities for the interrupted dynamics. Firstly let us consider the intuited exact dynamics, i.e., $\hat{U}(\delta t,r)=U(r)$. In this case since $\hat{H}=0$ we have $\hat{U}(r)=\hat{\openone}$. That is, the intuited exact dynamics is also stationary. Secondly let us consider the modified interrupted dynamics, $\hat{U}_\text{alt}(\delta t, r)$ given by \eqref{Umod} with $z=3$. In this case the interpolated dynamics does not match the interrupted dynamics, it misses the rotations of the system's phases.

In general, if any of the system's phases have undergone more than a half rotation within the duration $\delta t$, then the interpolation scheme will take the more direct path to the final phase, not winding all the way around. As we have seen, this may produce a substantial departure from the underlying dynamics. However, if the duration of each update is short enough (or equivalently if the energy scale of the Hamiltonian is low enough) then this will not happen. Indeed if,
\begin{align}\label{SmallTimes}
-\pi\leq\frac{H_\text{min} \, \delta t}{\hbar}
\leq \frac{H_\text{max} \, \delta t}{\hbar}<\pi,
\end{align}
where $H_\text{max}$ and $H_\text{min}$ are the largest and smallest eigenvalues of $\hat{H}$ then $L_{\delta t}
=-\ii\hat{H}/\hbar$ such that,
\begin{align}
\frac{\dd}{\dd t}\ket{\psi(t)}
=L_{\delta t} \, \ket{\psi}
=\frac{-\ii}{\hbar} \, \hat{H}\ket{\psi}.
\end{align}
That is, if \eqref{SmallTimes} holds then the interpolation scheme will match the intuited exact dynamics between the timepoints $t=n\,\delta t$. Throughout this thesis we will typically assume that the duration of each update, $\delta t$, is small enough that this is the case.

\subsubsection{Unitary Example 2: Time-dependent Hamiltonian}
Take the system to again be represented by a vector, $\ket{\psi}$, in some Hilbert space, only now being updated by the map,
\begin{align}\label{UEx2}
\hat{U}(\delta t)
=\mathcal{T}\text{exp}\left(\frac{-\ii}{\hbar} \int_{0}^{\delta t}\hat{H}(\tau)\,\dd\tau\right),
\end{align}
in other words,  by the time-ordered exponential of some time-dependent Hamiltonian, $\hat{H}(t)$. Note that in \eqref{UEx2} the Hamiltonian is assumed to be independent of $\delta t$. This represents a case where $H(t)$ runs by a clock which gets rest to zero every $\delta t$ seconds. We will consider a case where $\hat{H}(t)$ depends on $\delta t$ in the next example.

It is again trivial to check that our update map has the required regularity around $\delta t=0$. Specifically we have $\hat{U}(0)=\hat{\openone}$ and that $\hat{U}(\delta t)$ is differentiable at $\delta t=0$. Indeed, we have $\hat{U}'(0)=-\ii\hat{H}(0)/\hbar$. From these simple computations we know that the interpolated dynamics converges in the continuum limit and is generated by $L_0=\hat{U}'(0)$. Thus the system's continuum limit master equation is,
\begin{align}
\frac{\dd}{\dd t}\ket{\psi(t)}
=L_0\ket{\psi(t)}
=\hat{U}'(0)\,\ket{\psi}
=\frac{-\ii}{\hbar}\hat{H}(0)\ket{\psi(t)}.
\end{align}
That is, the system evolves unitarily via the Hamiltonian $\hat{H}(0)$. This makes sense: if the Hamiltonian's clock is being reset constantly the Hamiltonian will never progress beyond $t=0$.

As in the previous example we can attempt to calculate the interpolation generator exactly outside of the continuum limit. Doing this we find,
\begin{align}
L_{\delta t}
=\frac{1}{\delta t}\text{Log}(\hat{U}(\delta t))
=\frac{-\ii}{\hbar} \, \hat{H}_{\delta t,\text{eff}}\ket{\psi},
\end{align}
where we have defined the the effective Hamiltonian, $\hat{H}_{\delta t,\text{eff}}$ to satisfy,
\begin{align}
\text{exp}\left(\frac{-\ii \, \delta t}{\hbar} \hat{H}_{\delta t,\text{eff}}\right)
=\hat{U}(\delta t)
=\mathcal{T}\text{exp}\left(\frac{-\ii}{\hbar} \int_{0}^{\delta t}\hat{H}(\tau)\dd \tau\right).
\end{align}
One may guess that the effective Hamiltonian will be the time averaged Hamiltonian,
\begin{align}
\hat{H}_{\delta t,\text{avg}}
=\frac{1}{\delta t}\int_0^{\delta t}\hat{H}(t) \, \dd t,
\end{align}
however this guess does not properly account for the Hamiltonian potentially not commuting with itself at different times, $[\hat{H}(t_1),\hat{H}(t_2)]\neq0$. 

To explore this we assume that we can expand the Hamiltonian around $t=0$ as,
\begin{align}
\hat{H}(t)
=H_0
+t \, H_1
+t^2 \, H_2
+t^3 \, H_3
+\dots \, .
\end{align}
Note that up to third order the time-averaged Hamiltonian is given by,
\begin{align}
\hat{H}_{\delta t,\text{avg}}
=\frac{1}{\delta t}\int_0^{\delta t}\hat{H}(t) \, \dd t
=H_0
+\delta t \, \frac{1}{2}H_1
+\delta t^2 \, \frac{1}{3}H_2
+\mathcal{O}(\delta t^3).
\end{align}

From the above series we can expand the unitary update map in a Dyson series as,  
\begin{align}
\hat{U}(\delta t)
=\hat{\openone}
+\delta t \, \hat{U}^{(1)}
+\delta t^2 \, \hat{U}^{(2)}
+\delta t^3 \, \hat{U}^{(3)}
+\dots . 
\end{align}
A straightforward calculation gives,
\begin{align}\label{UExpansion0}
\hat{U}^{(1)}&=\frac{-\ii}{\hbar}\hat{H}_0,\\
\hat{U}^{(2)}&=\frac{1}{2}\frac{-\ii \, }{\hbar}\hat{H}_1
+\frac{1}{2}\Big(\frac{-\ii}{\hbar}\Big)^2 \hat{H}_0{}^2,\\
\hat{U}^{(3)}&=\frac{1}{3}\frac{-\ii}{\hbar}\hat{H}_2
+\frac{1}{3}\Big(\frac{-\ii}{\hbar}\Big)^2
\big(\hat{H}_1\hat{H}_0
+\frac{1}{2}\hat{H}_0\hat{H}_1\big)
+\frac{1}{6}\Big(\frac{-\ii}{\hbar}\Big)^3
\hat{H}_0{}^3.
\end{align}
Using this we can next expand the interpolation generator as,
\begin{align}
L_{\delta t}
=L_0
+\delta t \, L_1
+\delta t^2 \, L_2
+\delta t^3 \, L_3
+\dots \, .
\end{align}
From this we can identify $H_{\delta t,\text{eff}}=\ii\hbar L_{\delta t}$.

Using equation \eqref{L0def}, \eqref{L1def} and \eqref{L2def} we then have
\begin{align}
L_0&=\frac{-\ii}{\hbar}\hat{H}_0,\\
L_1&=\frac{1}{2}\frac{-\ii \, }{\hbar}\hat{H}_1,\\
L_2&=\frac{1}{3}\frac{-\ii}{\hbar}\hat{H}_2
+\frac{1}{12}\left(\frac{-\ii}{\hbar}\right)^2
[\hat{H}_1,\hat{H}_0].
\end{align}
From these we can read off the effective Hamiltonian, $\hat{H}_{\delta t,\text{eff}}=\ii\hbar L_{\delta t}$, to third order as,
\begin{align}
\hat{H}_{\delta t,\text{eff}}
&=H_0
+\delta t \, \frac{1}{2}H_1
+\delta t^2 \, \left(\frac{1}{3}H_2
+\frac{1}{12}\frac{-\ii}{\hbar}
[\hat{H}_1,\hat{H}_0]\right)
+\mathcal{O}(\delta t^3)\\
&=\hat{H}_{\delta t,\text{avg}}
+\delta t^2\,\frac{1}{12}\frac{-\ii}{\hbar}
[\hat{H}_1,\hat{H}_0]
+\mathcal{O}(\delta t^3).    
\end{align}
The effective Hamiltonian differs from the time-averaged Hamiltonian starting at second order in $\delta t$. Note that this difference is written in terms of the non-commutation of the Hamiltonian with itself at different time points. Thus in general we can expect the interpolation scheme to pick up on such non-commutation-in-time effects.

\subsubsection{Unitary Example 3: Time-dependent Hamiltonian}
For our final unitary example, we once again take the system to be represented by a vector, $\ket{\psi}$, in some Hilbert space, only now being updated by the map,
\begin{align}
\hat{U}(\delta t)
=\mathcal{T}\text{exp}\left(\frac{-\ii}{\hbar} \int_{0}^{\delta t}\hat{H}_{\delta t}(\tau)\dd \tau\right)
\quad\text{where}\quad
\hat{H}_{\delta t}(t)=\hat{H}(t/\delta t),
\end{align}
or in other words by the time-ordered exponential of some time-dependent Hamiltonian, $\hat{H}_{\delta t}(t)$,  which also depends on the duration of the interaction. The duration dependence $\hat{H}_{\delta t}(t)=\hat{H}(t/\delta t)$ represents a case where time dependence of the Hamiltonian is based on its progress through the interaction. Imagine a spin qubit traveling through a region with a periodic position-dependent magnetic field. If the qubit travels through these magnetic domains twice as fast ($\delta t\to\delta t/2$) then it runs through the variation in magnetic field twice as fast as well ($\hat{H}(t)\to\hat{H}(2\,t)$). 

In this example the time-averaged Hamiltonian is,
\begin{align}
\hat{H}_\text{avg}
\coloneqq
\frac{1}{\delta t}\int_{0}^{\delta t}\hat{H}(\tau/\delta t) \, \dd \tau
=\int_{0}^{1}\hat{H}(\xi) \, \dd\xi,
\end{align}
where we have made a convenient change of variables to $\xi=\tau/\delta t$. Note that unlike in the previous example this time average is independent of $\delta t$. Taking this change of variables in the unitary map we have,
\begin{align}
\hat{U}(\delta t)
=\mathcal{T}\text{exp}\left(\frac{-\ii}{\hbar} \int_{0}^{\delta t}\hat{H}(\tau/\delta t) \, \dd \tau\right)
=\mathcal{T}\text{exp}\left(\frac{-\ii \, \delta t}{\hbar} \int_{0}^{1}\hat{H}(\xi) \, \dd\xi\right),
\end{align}
As always we begin by checking that $\hat{U}(0)=\hat{\openone}$ (it does) and computing the derivative at $\delta t=0$ as,
\begin{align}\label{Phi1TimeDepHam2}
\hat{U}'(0)=\frac{-\ii}{\hbar} \int_{0}^{1}\hat{H}(\xi) \, \dd\xi
=\frac{-\ii}{\hbar}\hat{H}_\text{avg}.
\end{align}
As always, $L_0=\hat{U}'(0)$ gives the system's master equation in the continuum limit,
\begin{align}
\frac{\dd}{\dd t}\ket{\psi(t)}
=L_0\,\ket{\psi(t)}
=\frac{-\ii}{\hbar}\hat{H}_\text{avg}\ket{\psi(t)}.
\end{align}
In this limit, the system evolves unitarily with respects to the time-averaged Hamiltonian.

To analyze the system's dynamics outside of this limit we again expand the time-ordered exponential $\hat{U}(\delta t)$ as a Dyson series in $\delta t$ as,
\begin{align}
\hat{U}(\delta t)
=\openone
+\delta t \, \hat{U}^{(1)}
+\delta t^2 \, \hat{U}^{(2)}
+\dots \, .
\end{align}
By a straightforward computation we find two terms to be,
\begin{align}
\hat{U}^{(1)}&=\frac{-\ii}{\hbar}\hat{H}_\text{avg}\\   
\hat{U}^{(2)}&=\frac{1}{2}\hat{U}^{(1)}{}^2 +\frac{-\ii}{\hbar}\hat{H}^{(1)},
\end{align}
where, 
\begin{align}
H^{(1)}=\frac{1}{2}
\frac{-\ii}{\hbar}\int_{0}^1 d\xi_1
\int_{0}^{\xi_1} d\xi_2 \, 
[\hat{H}(\xi_1),\hat{H}(\xi_2)],
\end{align}
captures how the Hamiltonian does not commute with itself at different times. Using these we can expand $L_{\delta t}$ as,
\begin{align}
L_{\delta t}
=L_0
+\delta t \, L_1
+\delta t^2 \, L_2
+\delta t^3 \, L_3
+\dots \, ,
\end{align}
where from \eqref{L0def} and \eqref{L1def} we have,
\begin{align}
L_0&=\frac{-\ii}{\hbar}\hat{H}_\text{avg},\\
L_1&=\frac{-\ii}{\hbar}\hat{H}^{(1)}.
\end{align}
In this case the non-commutation of the Hamiltonian at different times appears one order lower than the previous example.

\section{Mixed Unitary Example (Decoherence in Media)}\label{MixedUnitaryExample}
We can explore non-unitary dynamics by taking the system to be represented by a density matrix, $\hat{\rho}$, and updated by the mixed unitary map,
\begin{align}\label{PhiMixedUnitary}
\phi(\delta t)[\hat{\rho}]
=\sum_k p_k \exp(-\ii \, \delta t \, \hat{H}_k/\hbar) \, \hat{\rho} \, \exp(\ii \, \delta t \, \hat{H}_k/\hbar),
\end{align}
for some probabilities, $p_k$ and Hamiltonians $\hat{H}_k$. This update represents our system evolving for a time $\delta t$ under a random Hamiltonian $\hat{H}_k$ selected with probability $p_k$. One can imagine that the quantum system described by $\hat{\rho}$ is in some classical stochastic environment which determines the Hamiltonian, $\hat{H}_k$. We expect the uncertainty of this Hamiltonian will cause the state to become increasingly mixed over time. Thus we do not expect our interpolation scheme to give unitary evolution as it did in the previous examples.

Connecting with the notation of Sec. \ref{Framework} we have: $\bm{v}\equiv\hat{\rho}$ and $M(\delta t)\bm{v}\equiv \phi(\delta t)[\hat{\rho}]$. At first glance, this identification may seem odd, after all, $\bm{v}$ is a vector and $\hat{\rho}$ is a matrix. However, one must recall that matrices are themselves vectors (elements of a vector space) with extra structure added on. In Chapter \ref{InterpolateGQM} we work through an example where this is treated explicitly. For this example, it is sufficient to note that the update map \eqref{PhiMixedUnitary} still acts linearly on the state, $\hat{\rho}$, even though it does not act by ``left-multiplication'' as it did on $\bm{v}$. To capture this difference we change our notation for the interpolation generator as $L_{\delta t}\equiv\mathcal{L}_{\delta t}$.

It is easy to check that the minimum required regularity assumptions around $\delta t=0$ are satisfied in this scenario. Specifically, we have $\phi(0)=\openone$ and,
\begin{align}
\phi'(0)[\hat{\rho}]
=\sum_k p_k \, \frac{-\ii}{\hbar}[\hat{H}_k,\hat{\rho}]
=\frac{-\ii}{\hbar}[\langle\hat{H}\rangle,\hat{\rho}],
\end{align}
where $\langle\hat{H}\rangle=\sum_k p_k \,\hat{H}_k$ is the average Hamiltonian. Thus in the continuum limit (when the Hamiltonian is randomized very frequently) the system evolves as,
\begin{align}
\frac{\dd}{\dd t}\hat{\rho}(t)
=\mathcal{L}_0[\hat{\rho}(t)]
=\phi'(0)[\hat{\rho}(t)]
=\frac{-\ii}{\hbar}[\langle\hat{H}\rangle,\hat{\rho}(t)].
\end{align}
That is, perhaps surprisingly, the system evolves unitarily, with dynamics generated by the average Hamiltonian. The short duration of the interactions somehow insulates the system from the uncertainty of its environment. As we will see, however, this only happens in the limit $\delta t\to0$. 

To study this situation outside of the continuum limit, we expand the update map as,
\begin{align}
\phi(\delta t)
=\openone
+\delta t \, \phi_1
+\delta t^2 \, \phi_2
+\delta t^3 \, \phi_3
+\dots \, .
\end{align}
Since we have already computed $\phi_1=\phi'(0)$, we next compute $\phi_2$. A straightforward computation yields, 
\begin{align}
\phi_2[\hat{\rho}]
&=\frac{1}{2}\Big(\frac{-\ii}{\hbar}\Big)^2\sum_k p_k \, [\hat{H}_k,[\hat{H}_k,\hat{\rho}]].
\end{align}
Using \eqref{L1def}, we can compute the first correction to the dynamics accounting for finite interaction times $\mathcal{L}_1$ as,
\begin{align}\label{L1MixedUnitary}
\mathcal{L}_1[\hat{\rho}]
&=\left(\phi_2-\frac{1}{2}\mathcal{L}_0{}^2\right)[\hat{\rho}]
=\frac{1}{2}\Big(\frac{-\ii}{\hbar}\Big)^2\sum_{k\,\ell} (p_k \delta_{k\ell}-p_k p_\ell) \, [\hat{H}_k,[\hat{H}_\ell,\hat{\rho}]].
\end{align}

To help interpret this expression we can rewrite it in a standardized form, specifically in Lindblad form \cite{Lindblad}. For any Markovian time-independent master equation $\frac{\dd}{\dd t}\hat{\rho}(t)=\mathcal{L}[\hat{\rho}(t)]$ we can find operators $\hat{H}$, $\hat{F}_j$ and some scalars $\Gamma_j$ such that,
\begin{align}\label{LindbladForm}
\mathcal{L}[\hat{\rho}]
&=\frac{-\ii}{\hbar}[\hat{H},\hat{\rho}]
+\sum_j \Gamma_j \, \mathcal{D}(\hat{F}_j)[\hat{\rho}],
\end{align}
where $\hat{H}$ is self-adjoint and,
\begin{align}
\mathcal{D}(X)[\hat{\rho}]
&=X \hat{\rho} X^\dagger-\frac{1}{2} \{X^\dagger X,\hat{\rho}\}.
\end{align}
The dynamics generated by this master equation is CPTP if and only if the coefficients $\Gamma_j$ are real and non-negative \cite{Lindblad}. In this case $\hat{H}$ acts as the effective Hamiltonian of the dynamics and $\Gamma_j$ and $\hat{F}_j$ are the dynamics decoherence rates and decoherence modes respectively.

Using 1) the symmetry of the coefficients $p_k \delta_{kl}-p_k p_\ell$, and 2) the identity,
\begin{align}\label{ComIdentity}
\frac{1}{2}[\hat{A},[\hat{B},\hat{\rho}]]
+ \frac{1}{2}[\hat{B},[\hat{A},\hat{\rho}]]
= \frac{1}{2}\{\hat{A}\hat{B},\hat{\rho}\}
+ \frac{1}{2}\{\hat{B}\hat{A},\hat{\rho}\}
-\hat{A}\hat{\rho}\hat{B}
-\hat{B}\hat{\rho}\hat{A},
\end{align}
and 3) the fact that $\hat{H}_k=\hat{H}_k^\dagger$ we have,
\begin{align}\label{L1PsuedoLindblad}
\mathcal{L}_1[\hat{\rho}]
&=\frac{1}{\hbar^2}\sum_{k\ell} (p_k \delta_{k\,\ell}-p_k p_\ell)\big(\hat{H}_\ell\hat{\rho}\hat{H}_k^\dagger-\frac{1}{2} \{\hat{H}_\ell^\dagger\hat{H}_k,\hat{\rho}\}\big).
\end{align}
From here we can find the dynamics' decoherence modes and decoherence rates by diagonalizing the coefficient matrix $Q=[p_k \delta_{k\ell}-p_k p_\ell]$. 

To help us interpret this matrix we can think of a random variable which takes a value $x_k$ with probability $p_k$. The variance of this random variable is,
\begin{align}
\text{Var}(x)
&=\sum_k p_k x_k{}^2
-\big(\sum_k p_k x_k\big)
\big(\sum_\ell p_\ell x_\ell\big)\\
&\nonumber
=\sum_{k\,\ell} \, x_k(p_k \delta_{k\ell}
- p_k p_\ell)x_\ell\\
&\nonumber
=\bm{x}^\intercal Q\bm{x},
\end{align}
where $\bm{x}$ is a vector with entries $x_k$. Since the variance of a random variable is always non-negative we have $\bm{x}^\intercal Q\bm{x}\geq0$ for every $\bm{x}$. Thus, $Q$ must be positive semidefinite, $Q\geq0$, that is, its eigenvalues are real and non-negative. We can also bound the eigenvalues of $Q$ from above by noting that \mbox{$\text{Tr}(Q)=1-\vert\vert\bm{p}\vert\vert^2_2\leq1$}.

Let us assume that $Q$ has eigenvalues $\gamma_j\geq0$ and real eigenvectors\footnote{We can assume the eigenvectors of $Q$ are real valued because $Q$ is Hermitian and real-valued.} $v^{(j)}_k$ as \mbox{$Q_{k\ell}=\sum_j\gamma_j v^{(j)}_k v^{(j)}_\ell$}. From this we can ``diagonalize'' \eqref{L1PsuedoLindblad} as,
\begin{align}
\mathcal{L}_1[\hat{\rho}]
&=\sum_{j}\frac{\gamma_j}{\hbar^2}\big(\hat{A}_j^\dagger\hat{\rho}\hat{A}_j-\frac{1}{2} \{\hat{A}_j^\dagger\hat{A}_j,\hat{\rho}\}\big),
\end{align}
where $\hat{A}_j=\sum_{k} v^{(j)}_k \hat{H}_k$. We can identify $\hat{A}_j$ as the decoherence modes and $\delta t\,\gamma_j/\hbar^2$ as the decoherence rates. (Recall that $\mathcal{L}_1$ appears in the master equation multiplied by $\delta t$). Thus in general, at first order the dynamics can be non-unitary.

In fact the only way that the first order dynamics does not introduce decoherence is if all of the decoherence rate are zero. This would imply that each $\gamma_j=0$ such that $Q=0$. For this to happen we must have $\text{Tr}(Q)=1-\vert\vert\bm{p}\vert\vert^2_2=0$ such that $\vert\vert\bm{p}\vert\vert^2_2=1$. For normalized probability vectors, $\vert\vert\bm{p}\vert\vert_1=1$, such as $\bm{p}$ this only happens in the ``pure'' case where $\bm{p}$ picks out a single Hamiltonian with complete certainty.

As a concrete example, consider the spin degree of freedom of an electron traveling at a speed $v$ through a series of magnetic domains each of length $L$. The duration of the electron's interactions with each region is $\delta t=L/v$. Further assume that each domain has (with probability $p(B)$) a random magnetic field strength $B$ oriented vertically such that the spin evolves under a random Hamiltonian $\hat{H}_B=\mu \, B \, \hat{\sigma}_z/2$ where $\mu$ is the magnetic moment of the electron.

In the continuum limit (when the electron is traveling very fast or when the magnetric domains are very small) the spin evolves unitarily, by the average Hamiltonian given by the environment, $\langle\hat{H}\rangle
=\mu \, \langle B \rangle \, \hat{\sigma}_z/2$ where $\langle B \rangle=\int p(B) \, B  \, \dd B$ is the average magnetic field. Computing the first order dynamics we find,
\begin{align}
\mathcal{L}_1[\hat{\rho}]
&=\frac{\mu^2\,\Delta_B^2}{4\hbar^2}\left(\hat{\sigma}_z^\dagger\hat{\rho}\hat{\sigma}_z-\frac{1}{2} \{\hat{\sigma}_z^\dagger\hat{\sigma}_z,\hat{\rho}\}\right)
=-\frac{\mu^2\, \Delta_B^2}{8\hbar^2}[\hat{\sigma}_z,[\hat{\sigma}_z,\hat{\rho}]],
\end{align}
where $\Delta_B^2=\langle B^2\rangle-\langle B\rangle^2$ is the variance of the field strength. Note, we find a single decoherence mode, $\hat{\sigma}_z$. Recalling that $\mathcal{L}_1$ appears in the master equation with a factor of $\delta t$ we have a decoherence rate of $\Gamma=(\delta t \, \mu^2 \, \Delta_B^2)/(4\hbar^2)$. Note that the decoherence rate is proportional to the uncertainty of the magnetic field. Solving this dynamics we see that the spin undergoes phase damping at a rate of $\Gamma$. Ultimately the spin ends up losing all of its coherence in essentially undergoing a non-selective measurement in $\hat{\sigma}_z$.

\section{Bloch Sphere Extension and Example (Partial Swap)}\label{BlochExample}
In this example we will take the system to be a qubit, i.e. a two-level quantum system. The state of this two level system can be described by a $2\times2$ density matrix, $ \hat{\rho}$, with $ \hat{\rho}= \hat{\rho}^\dagger$ and $\text{Tr}( \hat{\rho})=1$ and $ \hat{\rho}\geq0$. Equivalently, we can describe this density matrix, $ \hat{\rho}$, by its projection on the $2\times2$ basis,
\be
\hat{\openone}_2
=\begin{pmatrix}
1 & 0 \\
0 & 1 \\
\end{pmatrix},
\ 
\hat{X}=\begin{pmatrix}
0 & 1 \\
1 & 0 \\
\end{pmatrix},
\ 
\hat{Y}=\begin{pmatrix}
0 & \ii \\
-\ii & 0 \\
\end{pmatrix},
\ 
\hat{Z}=\begin{pmatrix}
1 & 0 \\
0 & -1 \\
\end{pmatrix}
\ee
which span the set of valid $ \hat{\rho}$. That is, we can identify the state with the $4$-Bloch vector,
\begin{align}
\bm{u}
=(u_0, \ 
u_1, \ 
u_2, \ 
u_3)^\intercal
=(\text{Tr}(\hat{\openone}_2\, \hat{\rho}), \ 
\text{Tr}(\hat{X}\, \hat{\rho}), \ 
\text{Tr}(\hat{Y}\, \hat{\rho}), \ 
\text{Tr}(\hat{Z}\, \hat{\rho}))^\intercal.    
\end{align}
The above discussed conditions on $ \hat{\rho}$ translate here to $u_k\in\mathbb{R}$ and $u_0=1$ and $u_1^2+u_2^2+u_3^2\leq1$. The 4-Bloch vector, $\bm{u}$, thus lives in a 3D-sphere embedded in an affine subspace of $\mathbb{R}^4$. Typically this embedding in $\mathbb{R}^4$ is neglected and only the 3-Bloch vector $\bm{a}=(u_1,u_2,u_3)^\intercal$ is discussed. However, as we will see, thinking of the 4-Bloch vector is helpful in this case.

Any CPTP map, $\phi$, (e.g., our update map) will act linearly on $ \hat{\rho}$ as $ \hat{\rho}\to\phi[ \hat{\rho}]$. By linearity the update will also act linearly on $\bm{u}$ as $\bm{u}\to\Lambda \bm{u}$ for some $4\times4$ matrix $\Lambda $. (As will will see this is not true of the 3-Bloch vector). Indeed by linearity we have,
\begin{align}
\Lambda 
=\frac{1}{2}\begin{pmatrix}
\text{Tr}(\hat{\openone}_2 \, \phi [\hat{\openone}_2]) & 
\text{Tr}(\hat{\openone}_2 \, \phi [\hat{X}]) & 
\text{Tr}(\hat{\openone}_2 \, \phi [\hat{Y}]) & 
\text{Tr}(\hat{\openone}_2 \, \phi [\hat{Z}])\\
\text{Tr}(\hat{X} \, \phi [\hat{\openone}_2]) & 
\text{Tr}(\hat{X} \, \phi [\hat{X}]) & 
\text{Tr}(\hat{X} \, \phi [\hat{Y}]) & 
\text{Tr}(\hat{X} \, \phi [\hat{Z}])\\
\text{Tr}(\hat{Y} \, \phi [\hat{\openone}_2]) & 
\text{Tr}(\hat{Y} \, \phi [\hat{X}]) & 
\text{Tr}(\hat{Y} \, \phi [\hat{Y}]) & 
\text{Tr}(\hat{Y} \, \phi [\hat{Z}])\\
\text{Tr}(\hat{Z} \, \phi [\hat{\openone}_2]) & 
\text{Tr}(\hat{Z} \, \phi [\hat{X}]) & 
\text{Tr}(\hat{Z} \, \phi [\hat{Y}]) & 
\text{Tr}(\hat{Z} \, \phi [\hat{Z}])\\
\end{pmatrix}.
\end{align}
The factor $1/2$ arises because $\text{Tr}(\hat{\openone}_2^2)=\text{Tr}(\hat{X}^2)=\text{Tr}(\hat{Y}^2)=\text{Tr}(\hat{Z}^2)=2$. The condition that $\Lambda $ produces a valid state from any valid initial state implies that $\Lambda $ is real-valued and that its top row is $(1,0,0,0)$. Thus we can decompose $\Lambda $ as,
\begin{align}
\Lambda 
=\begin{pmatrix}
1 & \bm{0}^\intercal \\
\bm{d}  & T  \\ 
\end{pmatrix},
\end{align}
for some $\bm{d} \in\mathbb{R}^3$ and some $3\times 3$ real-valued matrix $T$. From this we can read off the update map for the 3-Bloch vector, $\bm{a}$, as, 
\bel{BlochUpdate1}
\bm{a}\to T \bm{a}+\bm{d} .
\ee
Note that $\bm{a}$ undergoes a linear-affine update not a linear one. The valid-in-valid-out condition now tells us that we must have,
\begin{align}\label{FiniteBlochCPCond}
\vert\vert T \bm{a}+\bm{d} \vert\vert_2^2\leq1 
\quad\text{for all}\quad \bm{a}
\quad\text{with}\quad \vert\vert\bm{a}\vert\vert_2^2\leq1.
\end{align}
That is, the dynamic must map all vectors within the Bloch sphere back into the Bloch sphere.

We next consider generic differential evolution by taking $T=\openone_3+\dd t \,A$ and $\bm{d}=\dd t\,\bm{b}$ for some $\bm{b}\in\mathbb{R}^3$ and some $3\times3$ real-valued matrix $A$. This gives the most general form of a Bloch sphere master equation:
\be\label{BlochME}
\frac{\dd}{\dd t}\bm{a}(t)= A\,\bm{a}(t)+\bm{b}.
\ee
The differential version of \eqref{FiniteBlochCPCond} is,
\begin{align}\label{DiffBlochCPCond}
\bm{a}^\intercal A\bm{a}+\bm{a}^\intercal\bm{b}\leq0 
\quad\text{for all}\quad \bm{a}
\quad\text{with}\quad \vert\vert\bm{a}\vert\vert_2^2=1.
\end{align}

Before applying the interpolated collision model formalism to this example, let us briefly characterize the dynamics which \eqref{BlochME} can produce. First note that the affine part of the dynamics, $\bm{b}$, can be thought of as the dynamics effect on the maximally mixed state $\bm{a}=0$. Thus the dynamics is unital (mapping $\hat{\rho}=\openone_2/2$ to  $\hat{\rho}=\openone_2/2$ or equivalently $\bm{a}=0$ to $\bm{a}=0$) if and only if $\bm{b}=0$.

To understand the dynamics arising from the linear part of the dynamics, $A$, we can decompose $A$ into its symmetric and antisymmetric parts as \mbox{$A_s=(A+A^\intercal)/2$} and \mbox{$A_a=(A-A^\intercal)/2$}.  

Since $A$ is real, its symmetric part is Hermitian. Moreover \eqref{DiffBlochCPCond} implies that $A_s$ is negative semi-definite, $A_s\leq0$. Consider the master equation \eqref{BlochME} with $A_a=0$ and $\bm{b}=0$ written in the basis in which $A_s$ is diagonal. In this basis, the components of $\bm{a}$ evolve independently each by exponential decay. Indeed we can think of $A_s$ as mapping the Bloch sphere onto progressively shrinking ellipsoids with axes depending on the eigensystem of $A_s$.

The antisymmetric part of $A$ can be be understood using the fact that for any real-valued antisymmetric $3\times3$ matrix, $\Omega$, there is a vector $\bm{\omega}\in\mathbb{R}^3$ such that $\Omega\bm{v}=\bm{\omega}\times\bm{v}$ where $\times$ here is the standard cross product. Viewed this way
\eqref{BlochME} with $A_s=0$ and $\bm{b}=0$ describes rotation around the axis given by the vector $\bm{h}=(h_1,h_2,h_3)$ corresponding $A_a$. This evolution is unitary with repsects to a Hamiltonian $\hat{H}\propto h_1\hat{X}+h_2\hat{Y}+h_3\hat{Z}$.

The question now arises: Can we apply the interpolated collision model formalism to the 3-Bloch vector's dynamics? Our update map will be of the form,
\be\label{BlochUpdate2}
\bm{a}\to T(\delta t)\,\bm{a}+\bm{d}(\delta t),
\ee
that is, linear-affine and not linear as we assumed in Sec. \ref{Framework}. Moreover, the master equation that we would hope to produce would be of the form,
\be\label{BlochMEGoal}
\frac{\dd}{\dd t}\bm{a}= A_{\delta t}\,\bm{a}+\bm{b}_{\delta t}.
\ee
This interpolating master equation is linear-affine, not linear as we assumed in Sec. \ref{Framework}.

The solution of course is to apply the interpolated collision model formalism to the 4-Bloch vector, $\bm{u}$, which is update by the matrix,
\begin{align}
\Lambda(\delta t) 
=\begin{pmatrix}
1 & \bm{0}^\intercal \\
\bm{d}(\delta t)  & T(\delta t)  \\ 
\end{pmatrix}
\end{align}
and the interpret whatever results in terms of the 3-Bloch vector, $\bm{a}$.

To work this out we will need the matrix identity,
\begin{align}
\text{Log}\left(\begin{pmatrix}
1 & \bm{0}^\intercal \\
\bm{m} & M  \\ 
\end{pmatrix}\right)
=\begin{pmatrix}
1 & \bm{0}^\intercal \\
\frac{\text{Log}(M)}{M-\openone}\bm{m} & \text{Log}(M)  \\ 
\end{pmatrix},
\end{align}
where $\text{Log}(M)/(M-\openone)$ is to be understood via the series,
\be
\frac{\text{Log}(M)}{M-\openone}
=\sum_{k=0}^\infty\frac{(-1)^k}{k+1}(M-\openone)^k.
\ee
Using this we have the interpolation generator for the 4-Bloch vector,
\be
L_{\delta t}=\frac{1}{\delta t}\begin{pmatrix}
1 & \bm{0}^\intercal \\
\frac{\text{Log}(T(\delta t))}{T(\delta t)-\openone}\bm{d}(\delta t) & \text{Log}(T(\delta t))  \\ 
\end{pmatrix},
\ee
and so the master equation,
\be
\frac{\dd}{\dd t}\bm{u}(t)
=\frac{1}{\delta t}\begin{pmatrix}
1 & \bm{0}^\intercal \\
\frac{\text{Log}(T(\delta t))}{T(\delta t)-\openone}\bm{d}(\delta t) & \text{Log}(T(\delta t))  \\ 
\end{pmatrix}
\bm{u}(t).
\ee
Noting that $\bm{u}=(1,\bm{a})^\intercal$ we can rewrite this as a master equation of the form \eqref{BlochMEGoal} with
\begin{align}
\label{BlochAdef}
A_{\delta t}
&=\frac{1}{\delta t}\text{Log}(T(\delta t)),\\
\label{Blochbdef}
\bm{b}_{\delta t}
&=\frac{1}{\delta t}
\frac{\text{Log}(T(\delta t))}{T(\delta t)-\openone_{2N}}\bm{d}(\delta t).
\end{align}

We can use the above technique to extend the interpolated collision model formalism presented in Sec. \ref{Framework} to any system with a linear-affine update equation; Any linear-affine equation can be ``linearized'' by embedding the vector in an affine subspace. We will see another example of this in Chapter \ref{InterpolateGQM}.

Let us now look at one of the most commonly used interactions in collision models, specifically the partial swap interaction \cite{PhysRevLett.113.100603,PhysRevA.75.052110,PhysRevE.97.022111,1367-2630-16-9-095003,PhysRevA.76.062307} first discussed in \cite{Scarani2002}. This interaction consists of a system, $S$, interacting with an ancilla, $A$, via the Hamiltonian, $\hat{H}_\text{sw}=\hbar \, \omega \, U_\text{sw}$, where $U_\text{sw}$ is the unitary matrix which swaps the states of $S$ and $A$ as \mbox{$U_\text{sw}(\ket{S}\otimes\ket{A})=\ket{A}\otimes\ket{S}$}. Let us take $S$ and $A$ to be qubits in which case, then \mbox{$\hat{H}_\text{sw}=\hbar \, \omega(\openone_2\otimes\openone_2
+X\otimes X
+Y\otimes Y
+Z\otimes Z)/2$}. 
That is, the isotropic spin coupling. For simplicity we will neglect the systems' free dynamics.

Evolution under $\hat{H}_\text{sw}$ for a time $\delta t$ is described by the partial swap unitary,
\begin{align}
U(\delta t) 
&=\exp(-\ii \, \hat{H}_\text{sw} \, \delta t/\hbar)
=\cos(\omega\,\delta t) \, \hat{\openone}_\textsc{sa}
-\ii \, \sin(\omega\,\delta t) \, U_\text{sw}.
\end{align}
The reduced state of the system after the partial swap is given by the update map,
\begin{align}
\hat{\rho}_\text{S}\,\to\,
\phi(\delta t)[\hat{\rho}_\text{S}]=\text{Tr}_\text{A}(U(\delta t)(\hat{\rho}_\text{S}\otimes\hat{\rho}_\text{A})U(\delta t)^\dagger),
\end{align}
where $\hat{\rho}_\text{A}$ is the ancilla's initial state. 

Without loss of generality we can take the ancilla to initially be in the state $\hat{\rho}_\text{A}=(\openone+r\,\hat{Z})/2$ for some $r\in[0,1]$. Note that $r=0$ corresponds to the maximally mixed state whereas $r=1$ is a pure state. The system's 3-Bloch vector is updated as \eqref{BlochUpdate2} with,
\begin{align}
T(\delta t)&=
\begin{pmatrix}
\text{cos}(\omega\,\delta t)^2 & r\,\text{cos}(\omega\,\delta t)\,\text{sin}(\omega\,\delta t) & 0\\
-r\,\text{cos}(\omega\,\delta t)\,\text{sin}(\omega\,\delta t) & \text{cos}(\omega\,\delta t)^2 & 0\\
0 & 0 & \text{cos}(\omega\,\delta t)^2\\
\end{pmatrix}\\
\nonumber
\bm{d}(\delta t)&=
\begin{pmatrix}
0\\
0\\
r\,\text{sin}(\omega\,\delta t)^2.\\
\end{pmatrix}
\end{align}
From these we can compute $A_{\delta t}$ and $\bm{b}_{\delta t}$ non-perturbatively using \eqref{BlochAdef}. 

In the regime where $\omega\,\delta t\ll 1$ we can make the expansion $A_{\delta t}=A_0+\delta t \, A_1+\delta t^2 \, A_2+\dots $ where,
\begin{align}
A_0
&=\begin{pmatrix}
0 & \omega\,r & 0\\
-\omega\,r & 0 & 0\\
0 & 0 & 0\\
\end{pmatrix},\\
A_1
&=\begin{pmatrix}
-\omega^2\,(1 - r^2/2) & 0 & 0\\
0 & -\omega^2\,(1 - r^2/2) & 0\\
0 & 0 & -\omega^2\\
\end{pmatrix},\\
A_2
&=\begin{pmatrix}
0 & \omega^3 r (1 - r^2)/3 & 0\\
-\omega^3 r (1 - r^2)/3 & 0 & 0\\
0 & 0 & 0\\
\end{pmatrix}.
\end{align}
Similarly we can expand $\bm{b}_{\delta t}=\bm{b}_0+\delta t \,\bm{b}_1+\delta t^2 \,\bm{b}_2+\dots $ where,
\begin{align}
\bm{b}_0=0,
\quad
\bm{b}_1
&=(0,0,\omega^2\,r)^\intercal,
\quad
\bm{b}_2=0.
\end{align}

Thus in the continuum limit we have rotation about the $z$-axis at a rate $\omega\,r$. Note that this is unitary dynamics. At first order we have dephasing in the $x$ and $y$ directions at a rate of $\delta t \, \omega^2\,(1 - r^2/2)$. At first order in the $z$ direction we approach a fixed point of $\bm{a}=(0,0,r)$ at a rate $\delta t\,\omega^2$. Note that this fixed point is the ancilla's 3-Bloch vector prior to their interaction with the system. At second order we find another unitary contribution to the dynamics. Specifically, the rate of rotation we found in the continuum limit, $\omega\,r$ is at second order corrected by an amount $\delta t^2\,\omega^3 r (1 - r^2)/3$.

\section{Repeated Measurement Example (Zeno Effect)}\label{ZenoEffectExample}
In this example we show that the update map can include (non-selective) projective measurements. We take the quantum system to be represented by a $D$ dimensional density matrix, $\hat{\rho}$. We take the system to be updated by a CPTP map,
\begin{align}
\phi(\delta t)[\hat{\rho}]
&=\sum_{k=1}^D\hat{\Pi}_k \exp(-\ii\hat{H}\delta t/\hbar) \,
\hat{\rho} \, \exp(\ii\hat{H}\delta t/\hbar)\hat{\Pi}_k\\
&=\sum_{k=1}^D\left(\bra{k} \exp(-\ii\hat{H}\delta t/\hbar) \, 
\hat{\rho} \, \exp(\ii\hat{H}\delta t/\hbar)\ket{k}\right)\ket{k}\!\bra{k}\\
&=\sum_{k=1}^D\,q_k \ \ket{k}\!\bra{k}\\,
\end{align}
where $\{\hat{\Pi}_k\}=\{\ket{k}\!\bra{k}\}$ are a complete set of rank-one orthogonal projectors, i.e.,
\begin{align}
\hat{\Pi}_n\hat{\Pi}_m=\delta_{nm}\Pi_n,
\quad
\sum_{k=1}^D\Pi_k=\hat{\openone}, 
\quad\text{and}\quad 
\text{Tr}(\Pi_k)=1
\end{align}
and
\begin{align}
q_k
=\bra{k} \exp(-\ii\hat{H}\delta t/\hbar) \, 
\hat{\rho} \, \exp(\ii\hat{H}\delta t/\hbar)\ket{k}.
\end{align}
This update represents the system evolving unitarily under a Hamiltonian $\hat{H}$ for a time $\delta t$ and then undergoing an instantaneous projective measurement without postselection.

Unfortunately we cannot directly apply the interpolated collision model formalism to this scenario since,
\begin{align}
\phi(0)[\hat{\rho}]=\sum_{k=1}^D\hat{\Pi}_k \, \hat{\rho} \, \hat{\Pi}_k,
\end{align}
is not the identity map; something happens in no time. This is due to the instantaneous nature of our measurements. 

We can circumvent this problem by noting that as shown above after each update the system is always in a state of the form, $\hat{\rho}(n\,\delta t)=\sum_k p_k\hat{\Pi}_k=\sum_k p_k\ket{k}\!\bra{k}$ for some $p_k$. If we assume that the initial state $\hat{\rho}(0)$ can also be written in this form then we can represent the state $\hat{\rho}(n\,\delta t)$ at each discrete time point by the probability vector $\bm{p}(n\,\delta t)$ with entries,
\begin{align}
p_k(n\,\delta t)=\text{Tr}(\hat{\Pi}_k\,\hat{\rho}(n\,\delta t)).    
\end{align}

This does not mean that state can always be written in this form. Indeed if take the interruption map (see Equation \eqref{ExactIntermediate}) to be,
\begin{align}
\phi(\delta t,r)[\hat{\rho}]
=\exp(-\ii\hat{H}\,r/\hbar)
\,\hat{\rho}\,
\exp(\ii\hat{H}\,r/\hbar),
\end{align}
for $0\leq r < \delta t$ then the intermediate states will not generally be of this form. The dynamics generated by $\hat{H}$ will generally create coherences in the $\{\ket{k}\}$ basis. These coherences are removed at the end of the update by the measurement. Despite this we can still restrict our attention to the subspace of states which are incoherent in the basis $\ket{k}$ and construct our interpolation scheme there.

We can identify the update map for $\bm{p}(\delta t)$ through the following computation,
\begin{align}\label{PhiTildeDefZeno}
p_k((n+1)\,\delta t)
&=\text{Tr}\left(\hat{\Pi}_k \, \phi(\delta t)[\hat{\rho}(n\,\delta t)]\right)\\
&=\text{Tr}(\hat{\Pi}_k \, \phi(\delta t)[\sum_\ell p_\ell(n\,\delta t) \hat{\Pi}_\ell])\\
&=\sum_\ell\text{Tr}(\hat{\Pi}_k \, \phi(\delta t)[\hat{\Pi}_\ell]) \ p_\ell(n\,\delta t)\\
&=\sum_\ell\,\Lambda(\delta t)_{k\ell} \ p_\ell(n\,\delta t).
\end{align}
From this we can identify the update map $\Lambda(\delta t)$ as having entries,
\begin{align}
\Lambda(\delta t)_{k\ell}=\text{Tr}(\hat{\Pi}_k \, \phi(\delta t)[\hat{\Pi}_\ell])
=\bra{k}\phi(\delta t)[\ket{\ell}\!\bra{\ell}]\ket{k}.
\end{align}
Connecting with the notion of Sec. \ref{Framework} we now have $\bm{v}\equiv\bm{p}$ and $M(\delta t)\equiv \Lambda(\delta t)$.

We can write the update map, $\Lambda(\delta t)$, in terms of the Hamiltonian as,
\begin{align}
\Lambda(\delta t)_{k\ell}
&=\text{Tr}(\hat{\Pi}_k \, \phi(\delta t)[\hat{\Pi}_\ell])\\
&=\text{Tr}(\hat{\Pi}_k \, \sum_{n=1}^D\hat{\Pi}_n \exp(-\ii\hat{H}\delta t/\hbar) \, 
\hat{\Pi}_\ell \, \exp(\ii\hat{H}\delta t/\hbar)\hat{\Pi}_n)\\
&=\text{Tr}(\hat{\Pi}_k \exp(-\ii\hat{H}\delta t/\hbar) \, 
\hat{\Pi}_\ell \, \exp(\ii\hat{H}\delta t/\hbar))\\
&=\text{Tr}(\ket{k}\!\bra{k} \exp(-\ii\hat{H}\delta t/\hbar) \, 
\ket{\ell}\!\bra{\ell} \, \exp(\ii\hat{H}\delta t/\hbar))\\
&=\vert\bra{k} \exp(-\ii\hat{H}\delta t/\hbar) \, 
\ket{\ell}\vert^2.
\end{align}
We can identify the entries of this matrix as the probability that time evolution by $\hat{H}$ takes $\ket{\ell}\to\ket{k}$. 

We can now apply the interpolated collision model formalism since this new update map now has the minimum required regularity around $\delta t=0$. Indeed $\Lambda(0)=\openone$,
and computing the derivative at zero we find,
\begin{align}\label{Lambda1Zeno}
\Lambda'(0)_{k\ell}
=\frac{-\ii}{\hbar}\text{Tr}(\hat{\Pi}_k \, [\hat{H},\hat{\Pi}_\ell])\\
=\frac{-\ii}{\hbar}\text{Tr}(\hat{H} [\hat{\Pi}_k,\hat{\Pi}_\ell])
=0,
\end{align}
where we have used the commutator identity, \mbox{$\text{Tr}([\hat{A},\hat{B}]\,\hat{C})=\text{Tr}(\hat{A}\,[\hat{B},\hat{C}])$} and the fact that \mbox{$[\hat{\Pi}_k,\hat{\Pi}_\ell]=0$} for orthogonal projectors. This means that the continuum limit dynamics vanishes,
\begin{align}
\frac{\dd}{\dd t}\bm{p}(t)
=L_0 \ \bm{p}(t)
=\Lambda'(0) \ \bm{p}(t)
=0.
\end{align}
We can understand this as an example of the Zeno effect \cite{Degasperis1974,QuantumZenoEffect}. By rapidly and repeatedly measuring the system we can freeze out any dynamics happening in between our measurements. The ``watched'' system is frozen. However, as we will now see, this only happens in the continuum limit, i.e. when the measurements are happening infinitely often. 

We can expand \eqref{PhiTildeDefZeno} as a series around $\delta t=0$ as,
\begin{align}
\Lambda(\delta t)
&=\openone
+\delta t \, \Lambda_1
+\delta t^2 \, \Lambda_2
+\delta t^3 \, \Lambda_3
+\dots \, ,
\end{align}
where $\Lambda_1=0$ and,
\begin{align}
\Lambda_2{}_{k\ell}
&=\frac{1}{2}\Big(\frac{-\ii}{\hbar}\Big)^2\text{Tr}(\hat{\Pi}_k \, [\hat{H},[\hat{H},\hat{\Pi}_\ell]])
=\frac{1}{2}\Big(\frac{\ii}{\hbar}\Big)^2\text{Tr}([\hat{H},\hat{\Pi}_k]^\dagger[\hat{H},\hat{\Pi}_\ell]).
\end{align}
We can interpret \mbox{$\frac{\ii}{\hbar}[\hat{H},\hat{\Pi}_\ell]$} as the derivative of the projector $\hat{\Pi}_\ell$ in the Heisenberg picture and the trace of a product as an inner product. In this light $\Lambda_2$ captures the degree to which the projectors rotate into each other.

From this we can compute the first order interpolation generator \mbox{$L_1=\Lambda_2-\Lambda_1^2/2=\Lambda_2$}. This gives us the first order master equation,
\begin{align}
\frac{\dd}{\dd t}p_k(t)
&=\frac{\delta t}{2}\Big(\frac{\ii}{\hbar}\Big)^2\sum_{n=1}^D\text{Tr}([\hat{H},\hat{\Pi}_k]^\dagger[\hat{H},\hat{\Pi}_\ell]) \, p_\ell(t).
\end{align}
Thus outside of the continuum limit we see that the system's dynamics is not fixed by the repeated measurements.

For a concrete example consider the scenario where we rapidly measure a qubit with free dynamics, \mbox{$\hat{H}=\hbar\omega\hat{\sigma}_z/2$} in the $\hat{\Pi}_+=\ket{+}\!\bra{+}$, $\hat{\Pi}_-=\ket{-}\!\bra{-}$ basis. If we assume that $\delta t \, \omega\ll 1$ (such that we can neglect second order terms) we find,
\begin{align}
\frac{\dd}{\dd t}p_+(t)
&=\frac{\delta t}{2}\Big(\frac{\ii}{\hbar}\Big)^2\sum_{n=\pm}\text{Tr}([\hat{H},\hat{\Pi}_+]^\dagger[\hat{H},\hat{\Pi}_n]) \, p_n(t)\\
\nonumber
&=-\frac{\delta t \, \omega^2}{8}\sum_{n=\pm}\text{Tr}([\hat{\sigma}_z,\hat{\Pi}_+]^\dagger[\hat{\sigma}_z,\hat{\Pi}_n]) \, p_n(t)\\
\nonumber
&=-\frac{\delta t \, \omega^2}{4}\big(p_+(t)-p_-(t)\big),
\end{align}
and similarly,
\begin{align}
\frac{\dd}{\dd t}p_-(t)
=-\frac{\delta t \, \omega^2}{4}\big(p_-(t)-p_+(t)\big).
\end{align}
Note that as expected the sum of the probabilities is fixed \mbox{$\frac{\dd}{\dd t}(p_+(t)+p_-(t))=0$}. The dynamics of the difference, \mbox{$\Delta p(t)\coloneqq p_+(t)-p_-(t)$}, is,
\begin{align}
\frac{\dd}{\dd t}\Delta p(t)
=-\frac{\delta t \, \omega^2}{2}\Delta p(t).
\end{align}
Thus the dynamics induced by the rapid repeated measurement ultimately drives the system to the maximally mixed state ($\Delta p=0$ and therefore $p_+=1/2$ and $p_-=1/2$) at a rate $\Gamma=\delta t \, \omega^2/2$.

\section{Main Example: Ancillary Bombardment}\label{AncillaryBombardmentExample}
Let us next turn our attention to the main scenario considered in this thesis. Consider a quantum system which is represented by its density matrix, $\hat{\rho}_\text{S}(t)$, and which is updated by the CPTP map,
\begin{align}\label{PhiAncillaryBombardment1}
\phi(\delta t)[\hat{\rho}_\text{S}]
=\text{Tr}_\text{A}\Big(\exp(-\ii \, \delta t \, \hat{H}/\hbar)(\hat{\rho}_\text{S}\otimes \hat{\rho}_\text{A})\exp(\ii \, \delta t \, \hat{H}/\hbar)\Big).
\end{align}
This represents our system engaging with an ancillary system, A, (initially uncorrelated with our system and with reduced state $\hat{\rho}_\text{A}$) and interacting with it for a time $\delta t$ under a joint time-independent Hamiltonian, 
\begin{align}
\hat{H}
=\hat{H}_\text{S}\otimes\openone_\text{A}
+\openone_\text{S}\otimes\hat{H}_\text{A}
+\hat{H}_\text{SA},
\end{align}
before finally decoupling from ancilla, which is discarded. This scenario, which we term \textit{ancillary bombardment}, will be considered extensively throughout the remainder of this thesis.

We first confirm that the update map satisfies the required regularity assumptions around $\delta t=0$. Specifically we have, $\phi(0)=\openone$ and,
\begin{align}
\phi'(0)[\hat{\rho}_\text{S}]
\nonumber
&=\text{Tr}_\text{A}\big(\frac{-\ii}{\hbar}[\hat{H},\hat{\rho}_\text{S}\otimes \hat{\rho}_\text{A}]\big)\\
\nonumber
&=\frac{-\ii}{\hbar}\big[\text{Tr}_\text{A}(\hat{H}\hat{\rho}_\text{A}),\hat{\rho}_\text{S}\big]\\
&=\frac{-\ii}{\hbar}\big[\hat{H}^{(0)},\hat{\rho}_\text{S}\big],
\end{align}
where $\hat{H}^{(0)}=\text{Tr}_\text{A}(\hat{H}\hat{\rho}_\text{A})$ is the ``average'' Hamiltonian with respects to $\hat{\rho}_\text{A}$. Thus under very rapid bombardment (in the continuum limit) the system evolves as,
\begin{align}\label{ABZerothMasterEq1}
\frac{\dd}{\dd t}\hat{\rho}_\text{S}(t)
&=\mathcal{L}_0[\hat{\rho}_\text{S}(t)]\\
\nonumber
&=\phi'(0)[\hat{\rho}_\text{S}(t)]\\
\nonumber
&=\frac{-\ii}{\hbar}\big[\hat{H}^{(0)},\hat{\rho}_\text{S}(t)\big].
\end{align}
That is, in the continuum limit the system evolves unitarily with a Hamiltonian,
\begin{align}\label{2482}
\hat{H}^{(0)}
=\hat{H}_\text{S}
+\text{Tr}_\text{A}(\hat{H}_\text{SA}\hat{\rho}_\text{A})
=\hat{H}_\text{S}+\hat{H}_\textsc{ind}^{(0)},
\end{align}
where $\hat{H}_\textsc{ind}^{(0)}=\text{Tr}_\text{A}(\hat{H}_\text{SA}\hat{\rho}_\text{A})$ is a new Hamiltonian induced by the rapid bombardment. This phenomena of unitary evolution under rapid bombardment was first discussed by David Layden in \cite{Layden:2015b}. He gives the interpretation that the system is ``pushed'' be the rapid stream of ancillas (accounting for the modification of the system's dynamics) but that it does not have time to ``talk'' with them (there is quantum information flow between them, i.e. no possibility for entanglement to be generated). Moreover, in \cite{Layden:2015b}, it was shown how this phenomenon could yield rapid quantum control.

We can explore dynamics outside of the continuum limit by expanding the interaction map as,
\begin{align}
\phi(\delta t)
=\openone
+\delta t \, \phi_1
+\delta t^2 \, \phi_2
+\delta t^3 \, \phi_3
+\dots \, ,
\end{align}
where,
\begin{align}
\label{TimelessPhi1}
&\phi_1[\hat{\rho}_\text{S}]
=\frac{-\ii}{\hbar} \, 
\text{Tr}_\text{A}\Big(
[\hat{H},\hat{\rho}_\text{S}\otimes \hat{\rho}_\text{A}]\Big)\\
\label{TimelessPhi2}
&\phi_2[\hat{\rho}_\text{S}]
=\frac{1}{2!}\Big(\frac{-\ii}{\hbar}\Big)^2
\text{Tr}_\text{A}\Big(
[\hat{H},[\hat{H},\hat{\rho}_\text{S}\otimes \hat{\rho}_\text{A}]]\Big)\\
\label{TimelessPhi3}
&\phi_3[\hat{\rho}_\text{S}]
=\frac{1}{3!}\Big(\frac{-\ii}{\hbar}\Big)^3
\text{Tr}_\text{A}\Big(
[\hat{H},[\hat{H},[\hat{H},\hat{\rho}_\text{S}\otimes \hat{\rho}_\text{A}]]]\Big)\\
\label{TimelessPhi4}
&\phi_4[\hat{\rho}_\text{S}]
=\frac{1}{4!}\Big(\frac{-\ii}{\hbar}\Big)^4
\text{Tr}_\text{A}\Big(
[\hat{H},[\hat{H},[\hat{H},[\hat{H},\hat{\rho}_\text{S}\otimes \hat{\rho}_\text{A}]]]]\Big),
\end{align}
with higher order terms following the same pattern. From these we can compute the corrections to the master equation \eqref{ABZerothMasterEq1} accounting for the finite duration of each interaction.  

The first order term in the master equation is given by $\mathcal{L}_1
=\phi_2-\frac{1}{2}\phi_1{}^2$. In order to compute this it is useful to write the full Hamiltonian in the general form,
\begin{align}\label{HamK1}
\hat{H}=\sum_{k\in K} \hat{Q}_k\otimes\hat{R}_k,
\end{align}
where $\hat{Q}_k$ and $\hat{R}_k$ are Hermitian operators on the system and ancilla respectively. Note that the system and ancilla's free Hamiltonians can be included in this representation by taking \mbox{$k\in K=\{\text{S},\text{A},1,2,3,\dots,N\}$} with
$\hat{Q}_\text{S}=\hat{H}_\text{S}$, $\hat{R}_\text{S}=\hat{\openone}_\text{A}$, $\hat{Q}_\text{A}=\hat{\openone}_\text{S}$, and $\hat{R}_\text{A}=\hat{H}_\text{A}$.
Separating the free Hamiltonians out of the sum we have the interaction Hamiltonian,
\begin{align}
\hat{H}_\text{SA}=\sum_{k=1}^N \hat{Q}_k\otimes\hat{R}_k.
\end{align}
Plugging this expression into \eqref{TimelessPhi1} and rearranging terms we find,
\begin{align}\label{Phi1AB}
\phi_1[\hat{\rho}_\text{S}]
&=\frac{-\ii}{\hbar}
\text{Tr}_\text{A}\Big([\hat{H},\hat{\rho}_\text{S}\otimes \hat{\rho}_\text{A}]\Big)\\
&=\frac{-\ii}{\hbar}\sum_{k\in K}
\langle \hat{R}_k\rangle \ [\hat{Q}_k,\hat{\rho}_\text{S}],
\end{align}
where $\langle\hat{R}_n\rangle = \text{Tr}_\text{A}(\hat{R}_k\,\hat{\rho}_\text{A})$. From this we find,
\begin{align}\label{HalfPhi1Squared}
\frac{1}{2}\phi_1{}^2[\hat{\rho}_\text{S}]
&=\frac{1}{2}\left(\frac{-\ii}{\hbar}\right)^2\sum_{n,m\in K}
\langle\hat{R}_n\rangle\,\langle\hat{R}_m\rangle \  [\hat{Q}_n,[\hat{Q}_m,\hat{\rho}_\text{S}]]\\
&=\frac{1}{\hbar^2}\sum_{n,m\in K}
\langle\hat{R}_n^\dagger\rangle\,\langle\hat{R}_m\rangle \big(
\hat{Q}_n\hat{\rho}_\text{S}\hat{Q}_m^\dagger-\frac{1}{2}\{\hat{Q}_m^\dagger\hat{Q}_n,\hat{\rho}_\text{S}\}\big),
\end{align}
where we have used
1) the symmetry of the coefficients $\langle\hat{R}_n\rangle\,\langle\hat{R}_m\rangle$, and 2) the identity \eqref{ComIdentity}, and 3) the fact that $\hat{Q}_k=\hat{Q}_k^\dagger$.

We can similarly compute $\phi_2[\hat{\rho}]$ as follows. Using \eqref{HamK1} we have,
\begin{align}\label{Phi2AB}
\phi_2[\hat{\rho}_\text{S}]
&=\frac{1}{2!}\Big(\frac{-\ii}{\hbar}\Big)^2
\text{Tr}_\text{A}\Big(
[\hat{H},[\hat{H},\hat{\rho}_\text{S}\otimes \hat{\rho}_\text{A}]]\Big)\\
&=\frac{1}{2}\left(\frac{-\ii}{\hbar}\right)^2\sum_{n,m\in K}
\text{Tr}_\text{A}\left([\hat{Q}_n\otimes\hat{R}_n,[\hat{Q}_m\otimes\hat{R}_m,\hat{\rho}_\text{S}\otimes\hat{\rho}_\text{A}]]\right),
\end{align}
Expanding the sum we have,
\begin{align}
&\sum_{n,m\in K}
\text{Tr}_\text{A}\left([\hat{Q}_n\otimes\hat{R}_n,[\hat{Q}_m\otimes\hat{R}_m,\hat{\rho}_\text{S}\otimes\hat{\rho}_\text{A}]]\right)\\
&=\sum_{n,m\in K}
\text{Tr}_\text{A}(\hat{R}_n\hat{R}_m\hat{\rho}_\text{A})\hat{Q}_n\hat{Q}_m\hat{\rho}_\text{S}
+\text{Tr}_\text{A}(\hat{\rho}_\text{A}\hat{R}_m\hat{R}_n)\hat{\rho}_\text{S}\hat{Q}_m\hat{Q}_n\\
&-\sum_{n,m\in K}
\text{Tr}_\text{A}(\hat{R}_m\hat{\rho}_\text{A}\hat{R}_n)\hat{Q}_m\hat{\rho}_\text{S}\hat{Q}_n
+\text{Tr}_\text{A}(\hat{R}_n\hat{\rho}_\text{A}\hat{R}_m)\hat{Q}_n\hat{\rho}_\text{S}\hat{Q}_m\\
&=\sum_{n,m\in K}
\langle\hat{R}_n\hat{R}_m\rangle \hat{Q}_n\hat{Q}_m\hat{\rho}_\text{S}
+\langle\hat{R}_m\hat{R}_n\rangle
\hat{\rho}_\text{S}\hat{Q}_m\hat{Q}_n
-\langle\hat{R}_n\hat{R}_m\rangle\hat{Q}_m\hat{\rho}_\text{S}\hat{Q}_n
-\langle\hat{R}_m\hat{R}_n\rangle \hat{Q}_n\hat{\rho}_\text{S}\hat{Q}_m,
\end{align}
Exchanging the indices $m$ and $n$ in the second and fourth terms as,
\begin{align}
\sum_{n,m\in K}
\langle\hat{R}_m\hat{R}_n\rangle
\hat{\rho}_\text{S}\hat{Q}_m\hat{Q}_n
&=\sum_{n,m\in K}
\langle\hat{R}_n\hat{R}_m\rangle
\hat{\rho}_\text{S}\hat{Q}_n\hat{Q}_m,
\end{align}
and,
\begin{align}
\sum_{n,m\in K}\langle\hat{R}_m\hat{R}_n\rangle \hat{Q}_n\hat{\rho}_\text{S}\hat{Q}_m
&=\sum_{n,m\in K}\langle\hat{R}_n\hat{R}_m\rangle \hat{Q}_m\hat{\rho}_\text{S}\hat{Q}_n,
\end{align}
we have,
\begin{align}
&\sum_{n,m\in K}
\text{Tr}_\text{A}\left([\hat{Q}_n\otimes\hat{R}_n,[\hat{Q}_m\otimes\hat{R}_m,\hat{\rho}_\text{S}\otimes\hat{\rho}_\text{A}]]\right)\\
&=\sum_{n,m\in K}
\langle\hat{R}_n\hat{R}_m\rangle\left( \hat{Q}_n\hat{Q}_m\hat{\rho}_\text{S}
+\hat{\rho}_\text{S}\hat{Q}_n\hat{Q}_m
-2\hat{Q}_m\hat{\rho}_\text{S}\hat{Q}_n\right)\\
&=-2\sum_{n,m\in K}
\langle\hat{R}_n\hat{R}_m\rangle\left(
\hat{Q}_m\hat{\rho}_\text{S}\hat{Q}_n
-\frac{1}{2}\{\hat{Q}_n\hat{Q}_m,\hat{\rho}_\text{S}\}\right).
\end{align}
Thus in total we have,
\begin{align}
\phi_2[\hat{\rho}_\text{S}]
=\frac{1}{\hbar^2}\sum_{n,m\in K}
\langle\hat{R}_n^\dagger\hat{R}_m\rangle \, \big(\hat{Q}_n\hat{\rho}_\text{S}\hat{Q}_m^\dagger-\frac{1}{2}\{\hat{Q}_m^\dagger\hat{Q}_n,\hat{\rho}_\text{S}\}\big),
\end{align}
where we have used that $\hat{R}_k$ and $\hat{Q}_k$ are Hermitian to insert daggers. From this and \eqref{HalfPhi1Squared} we have,
\begin{align}
\mathcal{L}_1[\hat{\rho}_\text{S}]
=\frac{1}{\hbar^2}\sum_{n,m\in K}
\big(\langle\hat{R}_n^\dagger\hat{R}_m\rangle-\langle\hat{R}_n^\dagger\rangle\langle\hat{R}_m\rangle\big) \, \big(
\hat{Q}_n\hat{\rho}_\text{S}\hat{Q}_m^\dagger-\frac{1}{2}\{\hat{Q}_m^\dagger\hat{Q}_n,\hat{\rho}_\text{S}\}\big).
\end{align}
This is almost in Linblad form and is reminiscent of \eqref{L1MixedUnitary} from our earlier mixed unitary example. In this case however, the coefficient matrix, $D_{nm}\coloneqq\langle\hat{R}_n^\dagger\hat{R}_m\rangle-\langle\hat{R}_n^\dagger\rangle\langle\hat{R}_m\rangle$, is not necessarily symmetric since $\hat{R}_n^\dagger$ and $\hat{R}_m$ may not commute. Moreover the coefficients are not necessarily real since $\hat{R}_n^\dagger\hat{R}_m$ may not be Hermitian. Nonetheless we can see that this coefficient matrix is positive semi-definite by the following argument. Contracting the coefficient matrix with any complex vector $v_k$ we have
\begin{align}
\bm{v}^\dagger D \bm{v}
&=\sum_{n,m\in K} v_n^* \big(\langle\hat{R}_n^\dagger\hat{R}_m\rangle-\langle\hat{R}_n^\dagger\rangle\langle\hat{R}_m\rangle\big)v_m\\
&=\langle (\sum_{n\in K} \hat{R}_n v_n)^\dagger
(\sum_{m\in K} \hat{R}_m v_m)\rangle
-\langle\sum_{n\in K} \hat{R}_n v_n\rangle^*\langle\sum_{m\in K} \hat{R}_m v_m\rangle\\
&=\langle \hat{B}^\dagger\hat{B}\rangle
-\langle\hat{B}^\dagger\rangle\langle\hat{B}\rangle\\
&\geq0,
\end{align}
where $\hat{B}=\sum_{k\in K} \hat{R}_k v_k$. Thus $D\geq0$. This shows that the dynamics generated by $\mathcal{L}_1$ is CPTP.

As discussed above we can separate out the terms dealing with the free Hamiltonians by isolating the parts of the sum at least one of $n$ or $m$ being in $\{\text{S},\text{A}\}$. 
First consider the terms with $n=\text{S}$ and $m$ free. Recall that $\hat{R}_\text{S}=\openone_\text{A}$ and $\hat{Q}_\text{S}=\hat{H}_\text{S}$ such that,
\begin{align}
\frac{1}{\hbar^2}\sum_{m\in K}
\big(\langle\hat{R}_m\rangle-\langle\hat{R}_m\rangle\big) \, \big(\hat{H}_\text{S}
\hat{\rho}_\text{S}\hat{Q}_m^\dagger-\frac{1}{2}\{\hat{Q}_m^\dagger\hat{H}_\text{S},\hat{\rho}_\text{S}\}\big)
=0.
\end{align}
These terms vanish. Similarly all the terms with $m=\text{S}$ and $n$ free also vanish. All the other terms under consideration have either $n=\text{A}$ and $m\in[1,N]$ or vice versa. Taken together they give (after significant manipulation),
\begin{align}\label{24112}
\frac{-\ii}{\hbar}
[H^{(1)},\hat{\rho}_\text{S}]
\quad\text{where}\quad
H^{(1)}
=\frac{1}{2}\text{Tr}_\text{A}\big(\frac{\ii}{\hbar}[\hat{H}_\text{SA},\hat{H}_\text{A}]\hat{\rho}_\text{A}\big).
\end{align}
We can interpret this term as the free Hamiltonian of the ancilla evolved forward in time (as in the interaction picture) by the interaction Hamiltonian.
\begin{comment}
\begin{align}
&\frac{1}{\hbar^2}\sum_{m=1}^N
\big(\langle \hat{H}_\text{A}\hat{R}_m\rangle-\langle \hat{H}_\text{A}\rangle\langle\hat{R}_m\rangle\big) \, \big(
\hat{\rho}_\text{S}\hat{Q}_m^\dagger-\frac{1}{2}\{\hat{Q}_m^\dagger,\hat{\rho}_\text{S}\}\big)\\
&+\frac{1}{\hbar^2}\sum_{m=1}^N
\big(\langle\hat{R}_m\hat{H}_\text{A}\rangle-\langle \hat{H}_\text{A}\rangle\langle\hat{R}_m\rangle\big) \, \big(
\hat{Q}_m\hat{\rho}_\text{S}-\frac{1}{2}\{\hat{Q}_m,\hat{\rho}_\text{S}\}\big)\\
&=\frac{1}{2}\frac{1}{\hbar^2}\sum_{m=1}^N
\big(\langle \hat{H}_\text{A}\hat{R}_m\rangle-\langle \hat{H}_\text{A}\rangle\langle\hat{R}_m\rangle\big) \, [\hat{\rho}_\text{S},\hat{Q}_m]\\
&+\frac{1}{2}\frac{1}{\hbar^2}\sum_{m=1}^N
\big(\langle\hat{R}_m \hat{H}_\text{A}\rangle-\langle \hat{H}_\text{A}\rangle\langle\hat{R}_m\rangle\big) \, [\hat{Q}_m,\hat{\rho}_\text{S}]\\
&=\frac{1}{2}\frac{1}{\hbar^2}\sum_{m=1}^N
\langle\hat{R}_m\hat{H}_\text{A}-\hat{H}_\text{A}\hat{R}_m\rangle \, [\hat{Q}_m,\hat{\rho}_\text{S}]\\
&=\frac{1}{2}\frac{1}{\hbar^2}\sum_{m=1}^N
\langle[\hat{R}_m,\hat{H}_\text{A}]\rangle \, [\hat{Q}_m,\hat{\rho}_\text{S}]\\
&=\dots\\
&=\frac{1}{2}\Big(\frac{-\ii}{\hbar}\Big)\Big(\frac{\ii}{\hbar}\Big)
[\langle[\hat{H}_\text{SA},\hat{H}_\text{A}]\rangle,\hat{\rho}_\text{S}]
\end{align}
\end{comment}

Thus in total we have the first order dynamics,
\begin{align}\label{ABL1}
\mathcal{L}_1[\hat{\rho}_\text{S}]
=\frac{-\ii}{\hbar}
[H^{(1)},\hat{\rho}_\text{S}]
+\frac{1}{\hbar^2}\sum_{n,m=1}^N
\big(\langle\hat{R}_n^\dagger\hat{R}_m\rangle-\langle\hat{R}_n^\dagger\rangle\langle\hat{R}_m\rangle\big) \, \big(
\hat{Q}_n\hat{\rho}_\text{S}\hat{Q}_m^\dagger-\frac{1}{2}\{\hat{Q}_m^\dagger\hat{Q}_n,\hat{\rho}_\text{S}\}\big).
\end{align}
That is, the first order dynamics consists of a unitary term coming from the non-commutation of $\hat{H}_\text{SA}$ and $\hat{H}_\text{A}$ as well as some new dissipative dynamics. The decoherence modes and decoherence rates can be found by diagonalizing the coefficient matrix $D$ as in Sec \ref{MixedUnitaryExample}.

\begin{comment}
It is worth noting that the dissipative dynamics in \eqref{ABL1} is of a completely generic form. That is, if we would like the dynamics to have decoherence rates $\gamma_k$ and decoherence modes $\hat{F}_k$ as
\begin{align}
D[\hat{\rho}_\text{S}]
=\sum_{k=1}^N
\gamma_k 
\big(\hat{F}_k\hat{\rho}_\text{S}\hat{F}_k^\dagger-\frac{1}{2}\{\hat{F}_k^\dagger\hat{F}_k,\hat{\rho}_\text{S}\}\big)    
\end{align}
we can find an interaction Hamiltonian, $\hat{H}_\text{SA}$ and ancilla state, $\hat{\rho}_\text{A}$ which produces them. To see this one example of how this can be done, take $\{\hat{Q}_k\}$ to be a basis of Hermitian operators such that $\hat{F}_k=\sum_n c_{k,n}\hat{Q}_n$ for some complex coefficients $c^{k,n}$. Thus we have
\begin{align}
D[\hat{\rho}_\text{S}]
=\sum_{k=1}^N
\gamma_k c_{k,m}^* c_{k,n}  \ 
\big(\hat{Q}_n\hat{\rho}_\text{S}\hat{Q}_m^\dagger-\frac{1}{2}\{\hat{Q}_m^\dagger\hat{Q}_n,\hat{\rho}_\text{S}\}\big).    
\end{align}
We can find operators $\hat{R}_k$ and state $\hat{\rho}_\text{A}$ which give these desired coefficients as follows. Take $\hat{\rho}_\text{A}=\sum_k\gamma_k\ket{k}\!\bra{k}/\mathcal{N}$ for some normalization constant $\mathcal{N}=\sum_k \gamma_k$ and $\hat{R}_m=\sum_k c_{k,m} \ket{k}\bra{m}$
such that 
\begin{align}
\langle\hat{R}_m\rangle    
=\text{Tr}(\hat{\rho}_\text{A}\hat{R}_m)
=\frac{1}{\mathcal{N}}\sum_k\gamma_k 
\bra{k}\hat{R}_m\ket{k}
=\frac{1}{\mathcal{N}}\sum_k\gamma_k 
c_{k,k}
\end{align}
\end{comment}

As the examples considered in this Chapter have shown, the interpolated collision model formalism can be applied to a wide variety of scenarios and representations of quantum systems. It is noteworthy that in every example so far discussed the continuum limit dynamics has been unitary, even when the update map for $\delta t\neq0$ is clearly non-unitary. In the following chapter we will investigate in detail this unitary-in-the-continuum-limit phenomena. 

\chapter{Non-Unitary Dynamics in the Continuum Limit}\label{Ch3}
In all of the examples discussed in the previous chapter the continuum limit dynamics was unitary. Is this a generic phenomena? To address this question we will first show a simple (albeit fine-tuned) example in which the continuum limit dynamics is not unitary. We will then work out under what conditions in general the continuum limit dynamics is non-unitary. This discussion will show that fine-tuning along the lines of our simple example is always necessary to see non-unitary dynamics in the continuum limit. Finally, we will discuss in what ways this fine-tuning is often going to be unnatural.

In the final section of this Chapter we will discuss the implications of these results for anyone wishing to use master equations derived from collision models to describe open quantum systems. As was argued in the Chapter \ref{Ch1}, the $\delta t\to0$ limit is already a departure from reality (realistic interactions have finite duration). Moreover, as this Chapter shows, in addition to the $\delta t\to0$ approximation, the continuum limit master equation approach will require some fine tuning in order to yield non-unitary dynamics. As we will discuss, the required sort of fine tuning seems difficult to explain.

For these reasons I believe that the continuum limit master equation approach should be largely abandoned in favor of the interpolative approach developed in this thesis. As we have seen already in Chapter \ref{Ch2} and \ref{Ch2.1} the interpolative approach generalizes the continuum limit approach (we can take $\delta t\to0$ anytime) and can be applied to a wide variety of interesting scenarios. In later Chapters we will see further application of the interpolative approach.

\section{A simple fine-tuned example of non-unitary dynamics in the continuum limit}\label{CavesMilburn}

Let us consider a continuous variable quantum system, S, with a position operator, $\hat{q}_\text{S}$, and canonically conjugate momentum, $\hat{p}_\text{S}$. Let us assume that within each update the system interacts for a duration $\delta t$ with an ancilla with position and momentum operators, $\hat{q}_\text{A}$ and $\hat{p}_\text{A}$. Let us take each ancilla to be in the state $\hat{\rho}_\text{A}$ at the beginning of its interaction and for the interaction to be generated by the Hamiltonian,
\be
\hat{H}
=\hat{H}_\text{S}\otimes\hat{\openone}_\text{A}
+\hat{\openone}_\text{S}\otimes\hat{H}_\text{A}
+g\,\hat{q}_\text{S}\otimes\hat{p}_\text{A},
\ee
for some interaction strength, $g$. This is an example of \textit{ancillary bombardment} which is discussed in general in Chapter \ref{AncillaryBombardmentExample}.

This choice of Hamiltonian is inspired by the continuous-position measurement model introduced by Caves and Milburn in 1987 \cite{CM}. This sort of interaction has in recent years been applied to models gravitational decoherence \cite{Altamirano:2016hug,ACMZ} and is often used in simple pointer measurements scenarios.

From equations \eqref{ABZerothMasterEq1} and \eqref{2482} we have the continuum limit dynamics,
\begin{align}\label{ABZerothMasterEq2}
\frac{\dd}{\dd t}\hat{\rho}_\text{S}(t)
&=\mathcal{L}_0[\hat{\rho}_\text{S}(t)]\\
\nonumber
&=\frac{-\ii}{\hbar}\big[\hat{H}_\text{S}+\hat{H}^{(0)}_\text{ind},\hat{\rho}_\text{S}(t)\big],
\end{align}
where $\hat{H}_\text{ind}^{(0)}=g\,\langle\hat{p}_\text{A}\rangle\,\hat{q}_\text{S}$ is the induced dynamics at zeroth order. Note that at this point in the continuum limit the dynamics is unitary. From equation \eqref{24112} and \eqref{ABL1}, the first order dynamics is given by,
\begin{align}
\mathcal{L}_1[\hat{\rho}_\text{S}]
&=\frac{-\ii}{\hbar}
[\hat{H}^{(1)},\hat{\rho}_\text{S}]
+\frac{1}{\hbar^2}
\big(\langle\hat{p}_\text{A}^2\rangle-\langle\hat{p}_\text{A}\rangle^2\big) \, \big(
\hat{q}_\text{S}\hat{\rho}_\text{S}\hat{q}_\text{S}^\dagger-\frac{1}{2}\{\hat{q}_\text{S}^\dagger\,\hat{q}_\text{S},\hat{\rho}_\text{S}\}\big)\\
&=\frac{-\ii}{\hbar}
[\hat{H}^{(1)},\hat{\rho}_\text{S}]
-\frac{g^2}{2\hbar^2}\big(\langle\hat{p}_\text{A}^2\rangle-\langle\hat{p}_\text{A}\rangle^2\big)[\hat{q}_\text{S},[\hat{q}_\text{S},\hat{\rho}_\text{S}]],
\end{align}
where,
\be
\hat{H}^{(1)}
=\frac{g}{2}\,\big\langle\frac{\ii}{\hbar}[\hat{p}_\text{A},\hat{H}_\text{A}]\big\rangle\,\hat{q}_\text{S}.
\ee
Thus to first order the dynamics is,
\begin{align}
\frac{\dd}{\dd t}\hat{\rho}_\text{S}(t)
&=(\mathcal{L}_0+\delta t\mathcal{L}_1)[\hat{\rho}_\text{S}(t)]\\
\nonumber
&=\frac{-\ii}{\hbar}
[\hat{H}_\text{S}
+g\,\langle\hat{p}_\text{A}\rangle\,\hat{q}_\text{S}+\frac{g\,\delta t}{2}\,\big\langle\frac{\ii}{\hbar}[\hat{p}_\text{A},\hat{H}_\text{A}]\big\rangle\,\hat{q}_\text{S},\hat{\rho}_\text{S}]
-\frac{g^2\delta t}{2\hbar^2}\big(\langle\hat{p}_\text{A}^2\rangle-\langle\hat{p}_\text{A}\rangle^2\big)[\hat{q}_\text{S},[\hat{q}_\text{S},\hat{\rho}_\text{S}]].
\end{align}
Note that the first order dynamics will in general be non-unitary. The non-unitary term is proportional to $g^2 \delta t \Delta_p^2$ where  $\Delta_p^2\coloneqq\langle\hat{p}_\text{A}^2\rangle-\langle\hat{p}_\text{A}\rangle^2$ is variance of the ancilla momentum. As such, this term will vanish in the continuum limit $\delta t\to 0$. We can ``promote'' this non-unitary term into the continuum limit by either the interaction strength, $g$, or the variance of the ancilla momentum, $\Delta_p^2$, depend on the interaction duration $\delta t$ diverging as $\delta t\to0$.

For instance we could take the ancillas to have increasingly uncertain momentum as the interactions shorten as $\Delta_p^2\sim \kappa_1/\delta t$ such that $\Delta_p^2\,\delta t\to\kappa_1$ a constant. This is the approach taken in \cite{CM}. In this modified continuum limit we have,
\begin{align}
\frac{\dd}{\dd t}\hat{\rho}_\text{S}(t)
&=\frac{-\ii}{\hbar}
[\hat{H}_\text{S}
+g\,\langle\hat{p}_\text{A}\rangle\,\hat{q}_\text{S},\hat{\rho}_\text{S}]
-\frac{g^2\,\kappa_1}{2\hbar^2}[\hat{q}_\text{S},[\hat{q}_\text{S},\hat{\rho}_\text{S}]].
\end{align}
It is worth noting that taking $\Delta_p^2\to\infty$ as $\delta t\to0$ does not imply that the ancillas are becoming more mixed. This effect can be achieved by taking the ancillas to be pure and to have increasingly well defined positions which by the uncertainty principle this guarantees a divergence in momentum variance.

Alternatively we could take the interaction strength $g\sim\sqrt{\kappa_2/\delta t}$ to diverge as $\delta t\to0$ such that $g^2\delta t\to \kappa_2$ a constant. In this limit, the $g\,\delta t\sim\sqrt{\kappa_2\,\delta t}$ term in the effective Hamiltonian goes to zero. However, there is an issue in the zeroth order dynamics, the unitary term proportional to $g\sim\sqrt{\kappa_2/\delta t}$ will diverge unless we also take $\langle p_\text{A}\rangle\to0$ fast enough that $\langle p_\text{A}\rangle/\sqrt{\delta t}\to0$. In this finely balanced limit we will have non-unitary dynamics in the continuum limit,
\begin{align}
\frac{\dd}{\dd t}\hat{\rho}_\text{S}(t)
&=\frac{-\ii}{\hbar}
[\hat{H}_\text{S},\hat{\rho}_\text{S}]
-\frac{\kappa_2\,\Delta_p^2}{2\hbar^2}[\hat{q}_\text{S},[\hat{q}_\text{S},\hat{\rho}_\text{S}]].
\end{align}
This $g^2\delta t\to\kappa_2$ trick is very common in collisional models \cite{CM,PhysRevLett.115.120403,PhysRevLett.108.040401,PhysRevA.98.062104,PhysRevA.97.053811,PhysRevX.7.021003,PhysRevA.96.032107,PhysRevA.95.053838,Giovannetti_2012,Altamirano_2017,Attal2007,doi:10.1063/1.4879240}.

In any case, both of these methods to achieve non-unitary dynamics in the continuum have required that we take some property of the interaction (either the Hamiltonian or the ancilla state) to diverge in a specific way as $\delta t\to0$. As we will see in the next section this is always the case, some degree of fine-tuning is always necessary to see non-unitary dynamics in the continuum limit. 

\section{Continuum Limit of a General Update Scheme (Kraus Representation)}\label{L0isUnitary}
Consider a quantum system represented by its density matrix, $\hat{\rho}$, which is repeatedly updated by a generic CPTP map, $\phi(\delta t)$, which is analytic at $\delta t=0$ and has $\phi(0)=\openone$. Recall that any CPTP map can be written in Kraus form as,
\bel{KrausUpdateMap2}
\phi(\delta t)[\hat{\rho}]
=\sum_{k=1}^N \hat{A}_k(\delta t)\,\hat{\rho}\,\hat{A}_k(\delta t)^\dagger,
\ee
where $\hat{A}_k(\delta t)$ are operators (called Kraus operators) satisfying the trace preserving condition, 
\bel{TPCond1}
\sum_{k=1}^N \hat{A}_k(\delta t)^\dagger \hat{A}_k(\delta t)=\hat{\openone}.
\ee
Given such a general update map, can we find a necessary and sufficient condition on the Kraus operators such that the continuum limit dynamics will be non-unitary? Indeed we can. To do this we will first need to transfer our assumptions about the regularity of $\phi(\delta t)$ at $\delta t=0$ (i.e., being analytic at $\delta t=0$ with $\phi(0)=\openone$) onto the Kraus operators. 

Let us begin with the assumption that $\phi(\delta t)$ is analytic at $\delta t=0$. Does this imply that $\hat{A}_k(\delta t)$ is also analytic at $\delta t=0$? First let us, consider a simplified problem. Consider a function $f(x)\geq0$ defined for $x\geq0$. Does the fact that $f(x)^2$ is analytic at $x=0$ imply that its square root $f(x)$ is also analytic at $x=0$? 

It does not. Consider the function $f(x)^2=x$ and note that $f(x)=\sqrt{x}$ is not differentiable at $x=0$. However, as can be seen in Appendix \ref{KrausApp}, this is the only sort of non-analytic behavior that $f(x)$ can have at $x=0$. Specifically, if $f(x)^2$ is analytic at $x=0$ then either $f(x)$ or $f(x)/\sqrt{x}$ is analytic at $x=0$. Equivalently, if $f(x)^2$ is analytic at $x=0$ then $f(x)$ is either analytic at $x=0$ or is $\sqrt{x}$ times something analytic at $x=0$. 

Moreover in Appendix \ref{KrausApp} it is shown that something similar is true for Kraus operators. Specifically, $\phi(\delta t)$ being analytic at $\delta t=0$ implies that each of its Kraus operators are one of the following two types:
\begin{enumerate}
\item A Kraus operator of the first kind is analytic at $\delta t=0$. These operators can be expanded as,
\bel{KrausType1}
A^{(1)}_m(\delta t)
=\hat{A}_{m,0}
+\delta t \, \hat{A}_{m,1}
+\delta t^2 \, \hat{A}_{m,2}
+\dots.
\ee
We mark Kraus operators of the first kind with a superscript $(1)$ and index them with $m\in[1,N^{(1)}]$. Note the subscript following the $m$ index is always an integer.

\item A Kraus operator of the second kind is not analytic at $\delta t=0$. These operators can be expanded as,
\bel{KrausType2}
A^{(2)}_n(\delta t)
=\sqrt{\delta t} \, \big(
\hat{A}_{n,1/2}
+\delta t \, \hat{A}_{n,3/2}
+\delta t^2 \, \hat{A}_{n,5/2}
+\dots\big).
\ee
We mark Kraus operators of the second kind with a superscript $(2)$ and index them with $n\in[1,N^{(2)}]$. Note the subscript following the $n$ index is always a half integer. 
\end{enumerate}
Dividing the Kraus operators into these two types we have,
\begin{align}\label{KrausDecompIn2}
\phi(\delta t)[\hat{\rho}]
&=\sum_{m=1}^{N^{(1)}} \hat{A}_m^{(1)}(\delta t)\,\hat{\rho} \,\hat{A}_m^{(1)}(\delta t)^\dagger
+\sum_{n=1}^{N^{(2)}} \hat{A}_n^{(2)}(\delta t)\,\hat{\rho}\, \hat{A}_n^{(2)}(\delta t)^\dagger.
\end{align}

Next we can transfer the assumption that ``Nothing happens in no time'', i.e. $\phi(0)=\openone$ to the Kraus operators. Decomposing the Kraus operators as discussed above we have at $\delta t=0$,
\begin{align}
\hat{\rho}
=\phi(0)[\hat\rho]
&=\sum_{m=1}^{N^{(1)}} \hat{A}_m^{(1)}(0)\,\hat{\rho}\, \hat{A}_m^{(1)}(0)^\dagger    
\end{align}
since all Kraus operators of the second kind vanish at $\delta t=0$. Thus, the Kraus operators of the first kind evaluated at $\delta t=0$ should form a Kraus representation of the identity channel. The most general Kraus representation of the identity channel is,
\be
\hat{A}^{(1)}_m(0)=\sqrt{p_m} \, e^{\ii\theta_m} \, \hat{\openone},
\ee
for some probabilities $p_m\geq0$ and phases $\theta_m\in[-\pi,\pi)$. The trace preserving condition at $\delta t=0$ requires that $\sum_k p_k = 1$. Noting that the channel $\phi(\delta t)$ is unaffected by a phase rotation of each of its Kraus operators, $\hat{A}_k(\delta t)\to \hat{A}_k(\delta t) e^{\ii\phi_k}$, we can take $\theta_m=0$ without loss of generality. Thus the condition that $\phi(0)=\openone$ implies that $\hat{A}_m^{(1)}(0)=\hat{A}_{m,0}=\sqrt{p_m}\openone$.

To summarize any CPTP map $\phi(\delta t)$ which is analytic at $\delta t=0$ and has $\phi(0)=\openone$ has a Kraus decomposition of the form \eqref{KrausDecompIn2} with $\hat{A}_m^{(1)}(0)=\hat{A}_{m,0}=\sqrt{p_m}$ for some probabilities $p_m$.

Now that we have determined which Kraus representations correspond to our analytic update maps, we can proceed to calculate the dynamics in the continuum limit for such a general update map. Taking the derivative at $\delta t=0$ of \eqref{KrausDecompIn2} and using $\hat{A}_{m,0}=\sqrt{p_m}\,\hat\openone$ we have,
\begin{align}\label{L0Kraus1}
\frac{\dd}{\dd t}\hat{\rho}(t)
&=\mathcal{L}_0[\hat{\rho}(t)]
=\phi'(0)[\hat{\rho}]
=\sum_{m=1}^{N^{(1)}} \sqrt{p_m} \, \big(\hat{A}_{m,1}\hat{\rho}
+\hat{\rho} \hat{A}_{m,1}^\dagger\big)
+\sum_{n=1}^{N^{(2)}} \hat{A}_{n,1/2} \, \hat{\rho} \, \hat{A}_{n,1/2}^\dagger.
\end{align}
In order to determine if this dynamics is unitary or non-unitary we will now write it in Lindblad form \cite{Lindblad}. To do this we first divide $\sum_{m=1}^{N^{(1)}} \sqrt{p_m} \, \hat{A}_{m,1}$ into its Hermitian and anti-Hermitian parts as $\hat{G}+\frac{-\ii}{\hbar}\hat{H}$ with,
\begin{align}
\hat{G}
&\coloneqq\frac{1}{2}\sum_{m=1}^{N^{(1)}} \sqrt{p_m} \, (\hat{A}_{m,1}+\hat{A}_{m,1}^\dagger)\\
\hat{H}
&\coloneqq\frac{\ii\hbar}{2}\sum_{m=1}^{N^{(1)}} \sqrt{p_m} \, (\hat{A}_{m,1}-\hat{A}_{m,1}^\dagger),
\end{align}
being Hermitian operators. Using this decomposition we have,
\be
\mathcal{L}_0[\hat{\rho}]
=\frac{-\ii}{\hbar}[\hat{H},\hat{\rho}]
+\{\hat{G},\hat{\rho}\}
+\sum_{n=1}^{N^{(2)}} \hat{A}_{n,1/2} \, \hat{\rho} \, \hat{A}_{n,1/2}^\dagger.
\ee
This can be simplified using the trace preserving condition \eqref{TPCond1}. To first order in $\delta t$ this is,
\begin{align}
0&=\sum_{m=1}^{N^{(1)}} \sqrt{p_m} \, \big(\hat{A}_{m,1}
+ \hat{A}_{m,1}^\dagger\big)
+\sum_{n=1}^{N^{(2)}} \hat{A}_{n,1/2}^\dagger \hat{A}_{n,1/2}\\
&=2\,\hat{G}+\sum_{n=1}^{N^{(2)}} \hat{A}_{n,1/2}^\dagger \hat{A}_{n,1/2},
\end{align}
such that,
\begin{align}
\hat{G}
=\frac{-1}{2}\sum_{n=1}^{N^{(2)}} \hat{A}_{n,1/2}^\dagger \hat{A}_{n,1/2}.
\end{align}
Thus we find the continuum limit dynamics in Lindblad form as,
\begin{align}\label{L0Kraus2}
\frac{\dd}{\dd t}\hat{\rho}(t)
=\mathcal{L}_0[\hat{\rho}(t)]
=\frac{-\ii}{\hbar}[\hat{H},\hat{\rho}]
+\sum_{n=1}^{N^{(2)}} \, \hat{A}_{n,1/2} \, \hat{\rho} \, \hat{A}_{n,1/2}^\dagger
-\frac{1}{2} \, \{\hat{A}_{n,1/2}^\dagger \, \hat{A}_{n,1/2},\hat{\rho}\}.
\end{align}
Thus in general the continuum dynamics can be non-unitary as the example in the previous subsection had suggested. The decoherence modes of the continuum limit dynamics are given by the set of operators $\{\hat{A}_{n,1/2}\}$. 

That is, the decoherence modes are entirely determined by the Kraus operators of the second kind. If there are no Kraus operators of the second kind then there are no decoherence modes and the dynamics is unitary.

Recall that the Kraus operators of the second kind are exactly those Kraus operators  which are non-analytic at $\delta t=0$. Thus if $\phi(\delta t)$ has only analytic Kraus operators then the continuum limit dynamics will be unitary.

Thus the continuum limit dynamics is non-unitary if and only if  1) the update map has Kraus operators which are non-analytic at $\delta t=0$ and therefore can be expanded as \eqref{KrausType2} and 2) at least one of these non-analytic Kraus operators has $\hat{A}_{n,1/2}\neq0$.

\section{Non-Unitary Continuum Limit for Ancillary Bombardment}
As we have seen in the previous section, an generic update map will yield non-unitary dynamics in the continuum limit only if it has non-analytic Kraus operators. We will now investigate how difficult it is to build an update map with this property. Specifically we will investigate what modifications to the ancillary bombardment example in Sec. \ref{AncillaryBombardmentExample} are necessary to produce non-analytic Kraus operators.

Suppose that evolution of the system and ancilla is generated by a time-independent\footnote{Note that time-independence is assumed here for simplicity, the same conclusions hold for a time-dependent Hamiltonian. Note that time-dependence and duration-dependence are strictly independent concepts.} Hamiltonian $\hat{H}$ such that we have the update map, 
\be
\phi(\delta t)[\hat{\rho}_\text{S}]
=\text{Tr}_\text{A}\Big(\exp(-\ii \, \delta t \, \hat{H}/\hbar)(\hat{\rho}_\text{S}\otimes \hat{\rho}_\text{A})\exp(\ii \,\delta t \, \hat{H}/\hbar)\Big).
\ee
Without loss of generality we may  assume that $\hat{\rho}_\text{A}=\sum_\ell q_\ell \ket{\ell}\!\bra{\ell}$ for some probabilities $q_\ell$ and orthonormal vectors $\{\ket{\ell}\}$. We can find Kraus operators for this update map (indexed by $k$ and $\ell$) as,
\be
\hat{A}_{k\ell}(\delta t)=\sqrt{q_\ell} \bra{k}\exp(-\ii \, \delta t \, \hat{H}/\hbar)\ket{\ell}.
\ee
If none of $\hat{H}$, $q_\ell$, or $\ket{\ell}$ depend on the interaction duration $\delta t$ then all of these Kraus operators would be analytic at $\delta t=0$; The matrix exponential is an analytic function. In particular we would have,
\be
\hat{A}_{k\ell}(\delta t)
=\sqrt{q_\ell} \ \delta_{k\ell}
+\frac{-\ii \, \delta t}{\hbar}\bra{k}\hat{H}\ket{\ell}
+\frac{1}{2}\left(\frac{-\ii \, \delta t}{\hbar}\right)^2\bra{k}\hat{H}^2\ket{\ell}
+\dots.
\ee
Thus, in order to have some of these Kraus operators be non-analytic at $\delta t=0$ we must take at least one of $\hat{H}$, $q_\ell$, or $\ket{\ell}$ to depend on the interaction duration.

Compare this with types of modifications that we were able to come up with in Sec. \ref{CavesMilburn} to achieve non-unitary dynamics in the continuum limit. We found before that we could either change the ancilla state, $\hat{\rho}_\text{A}$, as we changed $\delta t$ (by either keeping it pure or making it more mixed) or we could change the Hamiltonian (particularly the coupling strength, $g$). As we can now see, in general one of these two is necessary to see non-unitary dynamics in the continuum limit.

It is worth delving into the $\delta t$-dependent Hamiltonian option in some detail. Let us take $\hat{H}=\hat{H}_0+g\,\hat{H}_\text{int}$ for some coupling strength $g$ which may depend on $\delta t$. For simplicity we can take $\hat{\rho}_\text{A}=\ket{0}\bra{0}$ independent of $\delta t$. In this case we have an update map
\be
\phi(\delta t)[\hat{\rho}_\text{S}]
=\text{Tr}_\text{A}\Big(\exp(-\ii \, \delta t \, (\hat{H}_0+g\,\hat{H}_\text{int})/\hbar)(\hat{\rho}_\text{S}\otimes \ket{0}\bra{0})\exp(\ii \,\delta t \, (\hat{H}_0+g\,\hat{H}_\text{int})/\hbar)\Big).
\ee
First note that to ensure that nothing happens in no time ($\phi(0)=\openone$) we can take \mbox{$\lim_{\delta t\to0} g \, \delta t=0$}. From this update map we can construct the Kraus operators,
\be
\hat{A}_{k}(\delta t)=\bra{k}\exp(-\ii \, \delta t \, (\hat{H}_0+g\hat{H})/\hbar)\ket{0}.
\ee
Since $\lim_{\delta t\to0} g \, \delta t=0$ we can expand the exponential in each Kraus operator as a series for small $\delta t$,
\begin{align}
\hat{A}_k(\delta t)
&=\bra{k}\Big(\hat{\openone}_\text{SA}
-\frac{\ii\,\delta t}{\hbar} \big(\hat{H}_0+g\,\hat{H}_\text{int}\big)
+\dots\Big)\ket{0}\\
&=\hat{\openone}_\text{S} \, \delta_{k0}
-\frac{\ii\,\delta t}{\hbar}
\bra{k}\hat{H}_0\ket{0}
-\frac{\ii\,g\,\delta t}{\hbar}
\bra{k}\hat{H}_\text{int}\ket{0}\\
&+\frac{1}{2}\left(\frac{\ii\,\delta t}{\hbar}\right)^2
\bra{k}\hat{H}_0^2\ket{0}
+\frac{g}{2}\left(\frac{\ii\,\delta t}{\hbar}\right)^2
\bra{k}\{\hat{H}_\text{int},\hat{H}_0\}\ket{0}
+\frac{g^2}{2}\left(\frac{\ii\,\delta t}{\hbar}\right)^2
\bra{k}\hat{H}_\text{int}^2\ket{0}\\
&+\dots \, .
\end{align}
At this point each of these Kraus operators could be of either the first or second kind depending on how $g$ depends on $\delta t$. However for $k=0$ we can see that $\hat{A}_0(\delta t)$ must be a Kraus operator of the first kind since $\hat{A}_0(0)=\openone\neq0$. All Kraus operators of the second kind vanish at $\delta t=0$.

In order for the continuum dynamics to be non-unitary we need at least one of the other Kraus operators (with $k\neq0$) to be of the second kind. Moreover we need it to scale as $\sqrt{\delta t}$ as $\delta t\to0$. The obvious way to do this is to take $g=\sqrt{\kappa/\delta t}$ for small $\delta t$ such that, $g\,\delta t=\sqrt{\kappa\,\delta t}$. Taking the interaction strength to diverge in this specific way as the interaction shortens will cause the continuum limit dynamics to be non-unitary. Note the similarity of this argument to that of the example presented in section \ref{CavesMilburn}.

As in this previous example, taking this balanced limit causes collateral damage that must be cancelled out. Recall that $\hat{A}_0$ must be a Kraus operator of the first kind, and so it must be analytic at $\delta t=0$. That is, $\hat{A}_0(\delta t)$ must have a series expansion in $\delta t$ with only integer powers and no half-integers. Taking $g=\sqrt{\kappa/\delta t}$ introduced a $g\,\delta t=\sqrt{\kappa\,\delta t}$ term into $\hat{A}_0$ which must be hidden somehow. Indeed we must take $\bra{0}\hat{H}_\text{int}\ket{0}=0$ or equivalently,
\be
\text{Tr}\left(\hat{H}_\text{int}(\hat{\rho}_\text{S}\otimes\ket{0}\bra{0})\right)=0,
\ee
for every $\hat{\rho}_\text{S}$. That is, the interaction Hamiltonian must vanish on average given the initial ancilla state regardless of the system state. This is equivalent to the $\langle\hat{p}_\text{A}\rangle=0$ condition in our earlier example.

Moreover, the Kraus operators with $k\neq0$ must also be either Kraus operators of the first kind or of the second kind. We cannot have both the $\delta t$ and $g\,\delta t\sim\sqrt{\delta t}$ terms  present within any given Kraus operator. This implies further restrictions: for each $k\neq0$ we must have either,
\begin{align}
\bra{k}\hat{H}_0\ket{0}=0
\quad\text{or}\quad
\bra{k}\hat{H}_\text{int}\ket{0}=0.
\end{align}
In general the $g^2\delta t=\kappa$ approach to generating non-unitary dynamics in the continuum limit requires a fine-tuning of the interaction Hamiltonian and the ancilla state on top of the fine tuning of the interaction strength.

Finally, it is worth considering in general what scenarios we might find the $\delta t$-dependence necessary for non-unitary dynamics in the continuum limit. Are there any cases in which this duration dependence arises naturally? For the Hamiltonian or the ancilla's initial state to depend \textit{directly} on the duration of interaction seems unlikely. The ancilla would need to know the duration of the interaction from its start to adjust its state and/or how it couples to the system. However, ostensibly the duration of the interaction is not established until the end of the interaction. Explanations which involve the interaction duration directly causing the ancilla to have some initial state or causing the Hamiltonian to be a certain way are highly unlikely.

A more natural way to achieve this behavior is to take the interaction Hamiltonian or the ancilla's initial state to match the interaction duration via some indirect means, for instance through a common cause. These sorts of common cause explanations are certainly possible in principle. For instance, the experimenter may have many dials at their disposal and may choose to turn some of them in unison. For instance they could manually increase the interaction strength as they manually decrease the interaction duration. 

For a less ad-hoc attempt at a common cause explanation consider the case where the interaction strength and the duration of the interaction are both determined by the velocity, $v$, of the system as $g(v)$ and $\delta t(v)=L/v$ for some length scale $L$. One may hope that a coupling strength proportional to the system's velocity/momentum\footnote{We are ignoring relativity here.}, $g\sim v$, will do the trick. However, this is not so. The necessary relation between $g(v)$ and $\delta t(v)$ to see non-unitary dynamics in the continuum limit is, $g(v)^2\,\delta t(v)\to \kappa$ a constant as $v\to\infty$. This implies that $g(v)\sim\sqrt{v}$ as $v\to\infty$. Any other scaling behavior will either 1) not produce unitary dynamics in the continuum limit, or 2) cause the continuum limit to fail to exist.

The results of this Chapter have shown that finding non-unitary in the continuum limit requires a certain non-ananytic behavior in the update map's Kraus operators. To achieve this in the context of ancillary bombardment requires either the Hamiltonian or the ancilla state to depend on the interaction duration. As I have argued this sort of duration dependence is difficult to explain in a natural way without resorting to unmotivated fine-tuning. Taken together with the results of Chapters \ref{Ch2} and \ref{Ch2.1} and the applications described in the coming chapters this constitutes a strong argument for a shift away from continuum limit based master equation.

\chapter{Thermalizing and Purifying Dynamics in Ancillary Bombardment}\label{Ch4}
One may have the intuition that collisional models can be used as a simple model for the process of thermalization. Imagine a thermal atomic gas with one atom far from equilibrium. One might expect this atom to interact with those near to it sequentially moving towards some sort of local equilibrium. If, by diffusion the local environment is continually ``refreshed'' with new atoms one might expect the atom to eventually equilibrate with the whole gas. This picturesque description is reminiscent of the ancillary bombardment scenario discussed in Sec \ref{AncillaryBombardmentExample} suggesting that ancillary bombardment may be able to model thermalization. Indeed, use is often made of \textit{Collision Models} to model thermalization \cite{PhysRevLett.115.120403,PhysRev.129.1880,PhysRevA.91.020502,PhysRevA.79.022105,PhysRevA.77.052106} in quantum thermodynamics.

In this chapter I investigate under what circumstances ancillary bombardment can be used as a model for thermalization. I will do so by looking at two necessary conditions for dynamics to yield thermalization: being able to cause purification (see Sec. \ref{Purification} -  Sec. \ref{PureExamples}) and being sensitive to the systems' free energy scales (see Sec. \ref{Sensitivity} - Sec \ref{ThermalAB}).

As we will see, we can only have purification at leading possible order (in $\mathcal{L}_1$) if the interaction between the system and ancilla is ``complex enough''. If the way that the system and ancilla pass information back and forth (i.e., the interaction Hamiltonian) is too simple then they will not be able to significantly purify each other in the brief time they are interacting. More generally this shows that more complicated interaction Hamiltonians will yield a richer set of dynamics in the rapid bombardment regime. A full analysis of which Hamiltonians yield which types of dynamics at which orders in $\delta t$ is given for Gaussian systems in Chapter \ref{InterpolateGQM}.

In Sec. \ref{Sensitivity} we will see that in order for the system and ancilla to exchange information about their respective temperatures each of their reduced dynamics must depend on the free energy scale of the other system. As we will see in Sec. \ref{ThermalAB} this dependence generically does not show up until second order in $\delta t$, that is in $\mathcal{L}_2$. This shows that the exchange of thermal information is a relatively slow process. Indeed, this dependence on free energy scales cannot be found in the continuum limit, that is in $\mathcal{L}_0$. Moreover, these effects are not present in $\mathcal{L}_1$ either such that our earlier trick of promoting dynamics in $\mathcal{L}_1$ into the continuum limit (see Chapter \ref{Ch3}) will not help us here. This all shows us that thermalization via collision models is necessarily a finite duration effect. This provides another compelling reason to study collisional models outside of the continuum limit.

\section{Thermalizing requires Purification which requires Non-unital dynamics}\label{Purification}

Consider dynamics which causes some system, $S$, to thermalize to some fixed temperature, $T$. If the system is initially thermal and slightly colder than this temperature then its temperature will increase under such dynamics. Likewise if the system is initially slightly above this temperature the thermalizing dynamics will need to cool the system down. Since purity is a monotone of temperature, this cooling implies that the purity of the system increases. Thus in order for dynamics to thermalize a system to some temperature it must be able to increase the purity of at least some state. Dynamics which decreases (or maintains) the purity of every state cannot yield thermalization. 

Throughout this chapter when we say that dynamics \textit{can purify/cause purification} we mean that there exists some system state, $\hat{\rho}_\text{S}$, whose purity, $\mathcal{P}[\hat{\rho}_\text{S}]\coloneqq\text{Tr}(\hat{\rho}_\text{S}{}^2)$, increases under the dynamics. That is, a CPTP map $\phi$ can purify if there exists some state $\hat{\rho}_\text{S}$ such that,
\begin{align}
\mathcal{P}[\phi[\hat{\rho}_\text{S}]]
>\mathcal{P}[\hat{\rho}_\text{S}].
\end{align}
Note that this is a relatively weak notion of causing purification; the dynamics just need to slightly increase the purity of one state for us to say it can cause purification. However, as we will see achieving even this weak notion of purification is non-trivial.

More generally, we may be interested in dynamics which result in some isolated attractive fixed point for the system, $\hat{\rho}_\text{S}(\infty)$. For instance, in the case of thermalization discussed above we would have $\hat{\rho}_\text{S}(\infty)=\hat{\rho}_\beta$ for some fixed inverse temperature $\beta$. We can extend the above argument as follows: Consider a small set of states around this fixed point, this set will almost always contain states with lower purity than $\hat{\rho}_\text{S}(\infty)$. As these lower purity states are attracted to the fixed point their purity will necessarily increase. Thus it appears that dynamics being able to purify is a necessary condition for having an isolated attractive fixed point. However, there is an exception to this argument. 

Our assumption that the set of small states neighborhood around $\hat{\rho}_\text{S}(\infty)$ contains a lower purity state does not hold if $\hat{\rho}_\text{S}(\infty)$ is a local minimum of purity. In fact this is the only case where this assumption does not hold. Moreover, since purity, $\mathcal{P}[\hat{\rho}_\text{S}]$, is a convex function of the state, $\hat{\rho}_\text{S}$, there can be at most one such local minimum. For finite dimensional systems the maximally mixed state (i.e., $\hat{\rho}_\text{S}=\hat{\openone}/D$ where $D$ is state's dimension) is the unique minimum purity state. For infinite dimensional quantum systems the minimum purity state is ill-defined. Taking into account the above exception we can amend our above conclusion to be: In order for some dynamics to have an isolated attractive fixed point (other than the maximally mixed state) then this dynamics must be able to increase the purity of at least some state.

In this section, we will discuss a necessary and sufficient condition for when a CPTP map $\phi$ can cause purification of a finite dimensional system. We will consider an infinite dimensional system in Chapter \ref{Ch5} and give its purification conditions there.

For finite dimensional quantum systems, a necessary and sufficient condition for a CPTP map to be able to purify is that it is non-unital (see Theorem 4.27 of Ref. \cite{Watrous}). That is, that $\phi[\hat{\openone}/N]\neq\hat{\openone}/N$ or equivalently $\phi[\hat{\openone}]\neq\hat{\openone}$. Non-unital dynamics are those which ``displace'' the maximally mixed state. The sufficiency of this condition is hopefully clear, if the maximally mixed state is mapped to some other state then its purity has surely increased. The necessity of this condition is non-trivial and implies that if the dynamics purifies any state then it also purifies the maximally mixed state. Thus we only need to know how the dynamics acts on this one state to tell whether it can purify in general.

\section{Non-unital dynamics in Ancillary Bombardment}
As we saw in the previous section, in order for dynamics to have a isolated attractive fixed point (other than the maximally mixed state) it is necessary that the dynamics is non-unital, $\phi[\hat{\openone}]\neq\hat{\openone}$. We also discussed how this is a necessary condition for the dynamics to thermalize the system to any inverse temperature $\beta$. In this section we apply this criteria to the ancillary bombardment example from Chapter \ref{AncillaryBombardmentExample} to establish under what conditions such a model can describe thermalizing dynamics.

To begin, note that the composition of two CPTP maps $\phi_1$ and $\phi_2$ which each cannot purify yields a map $\phi_2\circ\phi_1$ which cannot purify. Thus for the dynamics, $\hat{\rho}(n\,\delta t)=\phi(\delta t)^n[\hat{\rho}(0)]$ to purify we need that $\phi(\delta t)$ itself can purify, that is, $\phi(\delta t)[\hat{\openone}]\neq\hat{\openone}$. What does this condition mean in terms of the interpolation generator $\mathcal{L}_{\delta t}$? Note that if the interpolated dynamics is unital (i.e., $\mathcal{L}_{\delta t}[\hat{\openone}]=0$) then evolving under this dynamics for a duration $\delta t$ would yield a map,
\begin{align}
\phi(\delta t)[\hat{\openone}]
&=\exp(\delta t\mathcal{L}_{\delta t})[\hat{\openone}]\\
&=\left(\hat{\openone}+\delta t\mathcal{L}_{\delta t}+\frac{1}{2}\delta t^2\mathcal{L}_{\delta t}^2+\dots\right)[\hat{\openone}]\\
&=\openone+\delta t\mathcal{L}_{\delta t}[\hat{\openone}]+\frac{1}{2}\delta t^2\mathcal{L}_{\delta t}^2[\hat{\openone}]+\dots\\
&=\openone.
\end{align}
Thus if the interpolation generator, $\mathcal{L}_{\delta t}$, is unital, then the corresponding update map, $\phi(\delta t)$, is unital as well. Thus if the dynamics to be able to purify over long time scales (a requirement for thermalization) we need the interpolation generator be non-unital, \mbox{$\mathcal{L}_{\delta t}[\openone]\neq0$}. 

\begin{comment}
Analogous results to the above discussion for CPTP maps $\phi$ hold for Markovian time-independent master equations. Namely, the dynamics generated by
\begin{align}
\frac{\dd}{\dd t}\hat{\rho}_\text{S}
=\mathcal{L}[\hat{\rho}_\text{S}]
\end{align}
will decrease (or maintain) the purity of every initial state if and only if $\mathcal{L}[\openone]=0$. See \cite{Lidar2006} for proof. In fact, it is shown in \cite{Lidar2006} that
\bel{PurityRateBound}
\frac{\dd}{\dd t}\mathcal{P}(\hat{\rho})
=\frac{\dd}{\dd t}\text{Tr} \, (\hat{\rho}{}^2)
\leq\text{Tr} \, \big(\mathcal{L}[\openone] \, \hat{\rho}^2\big).
\ee
Thus to increase the purity of some state it is necessary and sufficient that the dynamics is non-unital, $\mathcal{L}[\openone]\neq0$.
\end{comment}

It is interesting to consider at which order in $\delta t$ this happens. Expanding the interpolation generator as,
\begin{align}
\mathcal{L}_{\delta t}
=\mathcal{L}_0
+\delta t\mathcal{L}_1
+\delta t^2\mathcal{L}_2
+\dots \, ,
\end{align}
we can take $m$ to be the smallest non-negative integer such
that $\mathcal{L}_m[\openone]\neq0$. If the dynamics is only non-unital past some high order in $\delta t$ how pure can the isolated fixed point be? To establish our intuition and some basic facts about this question, let us consider a qubit example along the lines of Sec. \ref{BlochExample}. Specifically, consider the following hypothetical dynamics for the Bloch vector, $\bm{a}(t)=(a_x(t),a_y(t),a_z(t))^\intercal$, of a two-level system:
\begin{align}
\frac{\dd}{\dd t}
\begin{pmatrix}
a_x(t)\\
a_y(t)\\
a_z(t)
\end{pmatrix}
=\begin{pmatrix}
-\delta t \, \gamma & \omega_\text{S} & 0\\
-\omega_\text{S} & -\delta t \, \gamma & 0\\
0 & 0 & -\delta t^3 \, \Gamma
\end{pmatrix}
\begin{pmatrix}
a_x(t)\\
a_y(t)\\
a_z(t)
\end{pmatrix}
+\begin{pmatrix}
0\\
0\\
\delta t^5 \, b 
\end{pmatrix}.
\end{align}
This dynamics describes a qubit with free Hamiltonian, \mbox{$\hat{H}_\text{S}=\hbar\omega_\text{S}\,\hat\sigma_z/2$}, undergoing amplitude damping. Recall that the maximally mixed state is represented by the Bloch vector $\bm{a}=0$ such that, 
\begin{align}
\mathcal{L}[\hat{\openone}]\equiv\left[\frac{\dd}{\dd t}\bm{a}(t)\right]_{\bm{a}=0}
=(0,0,b\ \delta t^5)^\intercal
\neq0.
\end{align}
Thus, in this example, the dynamics is non-unital beginning at fifth order in $\delta t$.

At zeroth order (in the continuum limit) this dynamics is simply the free dynamics of the system, i.e., rotation around the $z$-axis. At first order in $\delta t$ we have phase damping in the $x$ and $y$ Bloch coordinates at a rate $\delta t\,\gamma$. Note that at first order there is a 1-dimensional set of attractive fixed points, those along the $z$-axis. Note that because these fixed points form a continuum they are not isolated fixed points.  At third order there is dynamics induced within this fixed space, leading to an isolated fixed point at $\bm{a}(\infty)=0$, that is, at the maximally mixed state. Note that since the dynamics is still unital at third order this is the only place that its isolated fixed point could be. At fifth order the dynamics becomes non-unital. The isolated fixed point of the fifth order dynamics is,
\begin{align}
\bm{a}(\infty)=\delta t^2\left(0,0,b/\gamma_z\right)^\intercal
\neq0.
\end{align}
Note that the fifth order dynamics has only made a perturbative correction to the third order fixed point $\bm{a}(\infty)=0$. Any higher order corrections to the dynamics, say $\mathcal{O}(\delta t^6)$, would only make further perturbative corrections to this fixed point. The reason that $\delta t^2$ appears in the above expression appears because the non-unital dynamics is two orders behind the order at which the dynamics achieves an isolated fixed point (fifth versus third order).

This example shows that in order to have an isolated fixed point which is not perturbatively near the maximally mixed state (as it is in the above Bloch example) we would need the dynamics to be non-unital at the same order as the dynamics first has an isolated fixed point. This is necessary to have thermalizing dynamics with a fixed point which is not at an extremely high temperature.

As we saw in Chapter \ref{Ch3}, ancillary bombardment had unitary dynamics at zeroth order (in $\mathcal{L}_0$) unless some fine tuning is done. Unitary dynamics does not have isolated fixed points. Thus, in general the dynamics will develop an isolated fixed point at or after first order in $\delta t$. For instance, in the examples discussed in Sec. \ref{BlochExample} and \ref{ZenoEffectExample} an isolated fixed point is found at first order in $\delta t$. To match this possibility we are therefore interested in finding under what conditions we have non-unital dynamics at first order. That is under what conditions do we have $\mathcal{L}_1[\hat{\openone}]\neq0$.

Recall that $\mathcal{L}_1
=\phi_2-\frac{1}{2}\phi_1{}^2$. Taking the ancillary bombardment example (see Sec. \ref{AncillaryBombardmentExample}) with a system-ancilla Hamiltonian,
\be\label{HamK2}
\hat{H}=\sum_{k} \hat{Q}_k\otimes\hat{R}_k,
\ee
we have (see equation \eqref{Phi1AB}),
\begin{align}
\phi_1[\hat{\rho}_\text{S}]
=\frac{-\ii}{\hbar}\sum_{k}
\langle \hat{R}_k\rangle \ [\hat{Q}_k,\hat{\rho}_\text{S}].
\end{align}
where $\langle\hat{R}_n\rangle = \text{Tr}_\text{A}(\hat{R}_k\,\hat{\rho}_\text{A})$. Since $\phi_1[\hat{\openone}]=0$ the $\phi_1^2$ term will not contribute to $\mathcal{L}_1[\hat{\openone}]$. Next we have (see equation \eqref{Phi2AB}),
\begin{align}
\phi_2[\hat{\rho}_\text{S}]
&=\frac{1}{2}\left(\frac{-\ii}{\hbar}\right)^2\sum_{n,m}
\text{Tr}_\text{A}\left([\hat{Q}_n\otimes\hat{R}_n,[\hat{Q}_m\otimes\hat{R}_m,\hat{\rho}_\text{S}\otimes\hat{\rho}_\text{A}]]\right).
\end{align}
Using the identity\footnote{Note that this identity requires that the operator $\hat{B}\hat{D}\hat{\rho}_2$ and all its permutations be trace class. If system $2$ is finite dimensional then this is automatic, otherwise it must be checked. Since each of these permutations contains $\hat{\rho}_2$, this amounts to checking that certain expectation values (i.e., $\langle\hat{B}\hat{D}\rangle$ and its permutations) are finite.},
\begin{align}
&\text{Tr}_\text{2}\left([\hat{A}\otimes\hat{B},[\hat{C}\otimes\hat{D},\hat{\rho}_\text{1}\otimes\hat{\rho}_\text{2}]]\right)
=\frac{1}{2}\langle\{\hat{B},\hat{D}\}\rangle \  [\hat{A},[\hat{C},\hat{\rho}_\text{1}]]
+\frac{1}{2}\langle[\hat{B},\hat{D}]\rangle \, [\hat{A},\{\hat{C},\hat{\rho}_\text{1}\}]
\end{align}
we have
\begin{align}
\phi_2[\hat{\rho}_\text{S}]
&=\frac{1}{4}\Big(\frac{-\ii}{\hbar}\Big)^2\sum_{n,m}
\langle\{\hat{R}_n,\hat{R}_m\}\rangle \, [\hat{Q}_n,[\hat{Q}_m,\hat{\rho}_\text{S}]]\\
&+\frac{1}{4}\Big(\frac{-\ii}{\hbar}\Big)^2\sum_{n,m}
\langle\ii[\hat{R}_n,\hat{R}_m]\rangle \, [\hat{Q}_n,\{\hat{Q}_m,\hat{\rho}_\text{S}\}].
\end{align}
The first of these terms vanishes when we take $\hat{\rho}_\text{S}=\hat{\openone}$ such that,
\begin{align}\label{SumProduct}
\mathcal{L}_1[\hat{\openone}]
=\frac{1}{2}\Big(\frac{-\ii}{\hbar}\Big)^2\sum_{n,m}
\langle\ii[\hat{R}_n,\hat{R}_m]\rangle \, [\hat{Q}_n,\hat{Q}_m].
\end{align}
Thus a necessary and sufficient condition for a Hamiltonian of the form \eqref{HamK2} to be able to purify at first order is,
\bel{GenCond}
\sum_{n,m}
\langle\ii[\hat{R}_n,\hat{R}_m]\rangle \, [\hat{Q}_n,\hat{Q}_m].
\neq0
\ee
In order for the above expression to be non-zero, the Hamiltonian (when written in the form \eqref{HamK2}) must have a pair of terms whose system parts do not commute and whose ancilla parts do not commute on average.

\section{Example Hamiltonians}\label{PureExamples}
In this section, we investigate a range of common Hamiltonians in light of the necessary and sufficient condition to purify at first order which we described in the previous section.

\subsection*{Tensor Product Interaction}
We begin by analyzing perhaps the simplest class of Hamiltonians, namely one where the interaction Hamiltonian is the tensor product of a scalar operator of the system with a scalar operator of the ancilla. In this case the full Hamiltonian is,
\bel{HsaJsJa}
\hat{H}=\hat{H}_\text{S}\otimes\hat{\openone}_\text{A}
+\hat{\openone}_\text{S}\otimes \hat{H}_\text{A}
+\hat{Q}_\text{S}\otimes \hat{R}_\text{A},
\ee
where $\hat{Q}_\text{S}$ and $\hat{R}_\text{A}$ are observables of the system and ancilla respectively.  This type of Hamiltonian is common in the literature of rapid repeated interaction \cite{CM,ACMZ}, as well as being the Hamiltonian used in so-called pointer measurements.

Checking the condition \eqref{GenCond} we see that this Hamiltonian cannot purify at first order. Every pair of terms in \eqref{HsaJsJa} have that either their system operators commute or their ancilla operators commute. In fact, the free Hamiltonian terms, $\hat{H}_\text{S}\otimes\openone_\text{A}$ and $\openone_\text{S}\otimes\hat{H}_\text{A}$, can never contribute to \eqref{GenCond} being non-zero since $\openone_\text{S}$ and $\openone_\text{A}$ commute with everything. In order to have purification at first order the interaction Hamiltonian must contain these pairs of non-commuting operators itself. This means that the interaction Hamiltonian must be at least Schmidt rank-2.

In \cite{Grimmer2017a} it was shown that the dynamics coming from \eqref{HsaJsJa} also cannot purify at second order (i.e., $\mathcal{L}_2[\hat{\openone}]=0$). It is shown however that it is possible for this dynamics to purify at third order since,
\be
\mathcal{L}_3[\hat{\openone}]
=\frac{1}{12\hbar^4}
\text{Tr}_\text{A}\Big([\hat{R}_\text{A},[\hat{H}_\text{A},\hat{R}_\text{A}]]\hat{\rho}_\text{A}\Big) \, 
[\hat{Q}_\text{S},[\hat{H}_\text{S},\hat{Q}_\text{S}]]
\ee
is generally non-zero.

\subsection*{Qubit-Harmonic Oscillator coupling}
As we have just seen, purification at first order requires that the system-ancilla interaction Hamiltonian be the sum of multiple tensor products. We can find an example of an interaction Hamiltonian that can purify by considering a qubit, $S$, which repeatedly interacts with sequence of harmonic oscillators, $A$,  via the interaction Hamiltonian,
\be
\hat{H}_\text{SA}=
\hbar\omega \, (\hat{\sigma}_x\otimes\hat{q}
+\hat{\sigma}_y\otimes\hat{p}),
\ee
where $\hat{q}$ and $\hat{p}$ are dimensionless quadrature operators satisfying $[\hat{q},\hat{p}]=\ii\openone$. From \eqref{SumProduct} we can compute the effect of $\mathcal{L}_1$ on the maximally mixed state as,
\begin{align}
\mathcal{L}_1[\hat{\openone}]
&\nonumber
=\Big(\frac{-\ii}{\hbar}\Big)^2 (\hbar \omega)^2 \ 
\big\langle[\hat{q},\hat{p}]\big\rangle \ [\hat{\sigma}_x,\hat{\sigma}_y]\\
&=2 \, \omega^2 \ \hat{\sigma}_z.
\end{align}
Thus the qubit system's maximally mixed state is moved in the $z$ direction under this interaction regardless of the state of the harmonic oscillator ancillas. This type of interaction can, in principle, be implemented in superconducting circuits \cite{sufluxqubit}, achieving fast switching times in the ultra strong switchable coupling regime \cite{PeroPadre}.

\subsection*{Isotropic spin coupling (\texorpdfstring{$\bm{\hat{\sigma}}_\text{S}\cdot\bm{\hat{\sigma}}_\text{A}$}{text})}
Another natural coupling which is the sum of multiple tensor products is the isotropic spin coupling, 
\begin{align}
\hat{H}_\text{SA}
&=\hbar \, J \, \bm{\hat{\sigma}}_\text{S}\cdot\bm{\hat{\sigma}}_\text{A}
=\hbar \, J \, \sum_{j=x,y,z}\hat{\sigma}_\text{S}{}_j\otimes\hat{\sigma}_\text{A}{}_j.
\end{align}
From \eqref{SumProduct} we can compute the effect of $\mathcal{L}_1$ on the maximally mixed state as,
\begin{align}\label{sig-sig}
\mathcal{L}_1[\hat{\openone}]
&\nonumber
=\frac{1}{2}\sum_{i,j=x,y,z}\Big(\frac{-\ii}{\hbar}\Big)^2 (\hbar J)^2 \ 
\big\langle[\hat{\sigma}_\text{A}{}_i,\hat{\sigma}_\text{A}{}_j]\big\rangle \ [\hat{\sigma}_\text{S}{}_i,\hat{\sigma}_\text{S}{}_j]\\
&=2 \, J^2 \ 
\langle \bm{\hat{\sigma}}_\text{A}\rangle \,\cdot\, \bm{\hat{\sigma}}_\text{S}.
\end{align}
In the above equation we can identify the ancilla's Bloch vector as $\bm{a}_\text{A}=\langle \bm{\hat{\sigma}}_\text{A}\rangle$. Thus we see that the system's maximally mixed state is moved in the direction of the ancilla's Bloch vector. The only case in which the system's maximally mixed state is fixed is if $\bm{a}_\text{A}=0$, that is if the ancillas are maximally mixed.

\subsection*{Vector-vector coupling}
We can generalize this example by considering a interaction Hamiltonian which is a product of two vector observables as
\bel{VectorCoupling}
\hat{H}_\text{SA}
=\bm{\hat{V}}_\text{S}\cdot\bm{\hat{W}}_\text{A}
\coloneqq\sum_j\hat{V}_\text{S}{}_j\otimes\hat{W}_\text{A}{}_j.
\ee
for some operator valued vectors $\bm{\hat{V}}_\text{S}$ and $\bm{\hat{W}}_\text{A}$. From \eqref{SumProduct}, the effect of $\mathcal{L}_1$ on the maximally mixed state is
\be
\mathcal{L}_1[\hat{\openone}]
=\sum_{ij}\frac{1}{2}\Big(\frac{-\ii}{\hbar}\Big)^2 \ 
\big\langle[\hat{W}_\text{A}{}_i,\hat{W}_\text{A}{}_j]\big\rangle \ [\hat{V}_\text{S}{}_i,\hat{V}_\text{S}{}_j].
\ee
Thus, for repeated interactions under \eqref{VectorCoupling} to purify efficiently, the components of $\bm{\hat{V}}$ must not commute amongst themselves, and the components of $\bm{\hat{W}}$ must not either. Many common vector observables such as $\bm{\hat{x}}$, $\bm{\hat{p}}$, $\bm{\hat{E}}(\bm{x}_0)$, and $\bm{\hat{B}}(\bm{x}_0)$, do not pass this test, while others such as $\bm{\hat{L}}$ and $\bm{\hat{\sigma}}$ do. Thus vector-vector couplings involving any of $\bm{\hat{x}}$, $\bm{\hat{p}}$, $\bm{\hat{E}}(\bm{x}_0)$, or $\bm{\hat{B}}(\bm{x}_0)$ can not purify efficiently whereas couplings involving $\bm{\hat{L}}$ or $\bm{\hat{\sigma}}$ potentially can depending on what they are coupled to. 

From this we can generalize further to the case of two vector fields coupled component-wise throughout all of space as,
\begin{align}\label{VectorFieldCoupling}
H_\text{SA}
=\int \dd\bm{x} \ \bm{\hat{V}}_\text{S}(\bm{x})\cdot\bm{\hat{W}}_\text{A}(\bm{x})
=\int\! \dd\bm{x} \ \sum_j \hat{V}_\text{S}{}_j(\bm{x})\otimes\hat{W}_\text{A}{}_j(\bm{x}).
\end{align}
we compute the effect of $\mathcal{L}_1$ on the maximally mixed state from \eqref{SumProduct} as,
\begin{align}\label{L1oIVW}
\mathcal{L}_1[\hat{\openone}]
=\Big(\frac{-\ii}{\hbar}\Big)^2\!\!\!\int\!\!\! \dd\bm{x}\!\!\int\!\!\! \dd\bm{x'} \, 
\sum_{ij} \big\langle [\hat{W}_i(\bm{x}),\hat{W}_j(\bm{x'})]\big\rangle \, 
[\hat{V}_i(\bm{x}),\hat{V}_j(\bm{x'})].
\end{align}
For this to be non-zero it is necessary that the commutators between the vector field components at different locations, $[\hat{W}_i(\bm{x}),\hat{W}_j(\bm{x'})]$ and $[\hat{V}_i(\bm{x}),\hat{V}_j(\bm{x'})]$, be non-zero.

Even if these commutators are not zero everywhere this integral may vanish. For instance many common vector fields, $\hat{\Phi}_j(x)$, are \textit{microcausal}, meaning that the equal time commutator, $[\hat{\Phi}_i(\bm{x}),\hat{\Phi}_j(\bm{x'})]$, has support only on $\bm{x}=\bm{x}'$. For instance the electric and magnetic field operators, $\bm{\hat{E}}(\bm{x})$ and  $\bm{\hat{B}}(\bm{x})$, are microcausal. If either $\hat{\bm{W}}(\bm{x})$ or $\hat{\bm{V}}(\bm{x})$ is microcausal then we have,
\begin{align}
\mathcal{L}_1[\hat{\openone}]
=\Big(\frac{-\ii}{\hbar}\Big)^2\!\!\!\int\!\!\! \dd\bm{x} \, 
\sum_{ij} \big\langle [\hat{W}_i(\bm{x}),\hat{W}_j(\bm{x})]\big\rangle \, 
[\hat{V}_i(\bm{x}),\hat{V}_j(\bm{x})].
\end{align}
If either $\hat{\bm{W}}(\bm{x})$ or $\hat{\bm{V}}(\bm{x})$ has its components commute commute with each other then this integral will vanish. This is the case for instance for the electric and magnetic field operators, $\bm{\hat{E}}(\bm{x})$ and  $\bm{\hat{B}}(\bm{x})$.

In \cite{Grimmer2017a} I have investigated the ability of the light-matter interaction to purify in the context of rapid repeated interactions. I found there that the electric dipole coupling, the electric quadrupole coupling, the magnetic dipole coupling, and the sum of all these cannot purify at first order in $\delta t$.

As we have seen in the last few sections, we can only see purification at leading possible order (in $\mathcal{L}_1$) if the interaction between the system and ancilla is ``complex enough''. In particular the interaction Hamiltonian $\hat{H}_\text{int}=\sum_k\hat{Q}_k\otimes\hat{R}_k$ must be at least Schmidt rank-2 and moreover it must have some non-trivial commutation structure. 

For instance, it is necessary for the interaction Hamiltonian to have at least two $\hat{R}$ ancilla operators which do not commute with each other. If this were not the case then all of the $\hat{R}$ operators could be diagonalized in the same basis $\{\ket{\ell}\}$ such that we have
\begin{align}
\hat{H}_\text{int} 
=\sum_\ell \hat{A}_\ell
\otimes 
\ket{\ell}\bra{\ell}    
\end{align}
for some Hermitian system operators $\hat{A}_\ell$. This is a controlled Hamiltonian in that its matrix exponential is a controlled unitary,
\begin{align}
\exp(\ii \, t \, \hat{H}_\text{int} /\hbar)
=\sum_\ell 
\exp(\ii \, t \, \hat{A}_\ell /\hbar)
\otimes 
\ket{\ell}\bra{\ell}.    
\end{align}
How this unitary acts on the system state, $\hat{\rho}_\text{S}$, only depends on the ancilla state, $\hat\rho_\text{A}$, projected onto the $\{\ket{\ell}\}$ basis. Any coherences that the ancilla might have in this basis are irrelevant. The ancilla is effectively treated as a classical control variable with probabilities $p_\ell=\bra{\ell}\rho_\text{A}\ket{\ell}$. 

If on the other hand there were two or more ancilla $\hat{R}$ operators that did not commute with each other then they would each be diagonal in a different basis. Thus ancilla coherences in any one of these bases will be able to affect the system's dynamics as they will in general be on-diagonal in another basis. We can thus see this non-commutation condition \eqref{GenCond} in effect allows for quantum information to flow between the system and ancilla in the early stage of their interaction.

\section{Thermalizing Requires Sensitivity to Free Energy Scales}\label{Sensitivity} 
In \cite{Grimmer_2019} it was shown that a necessary condition for dynamics to yield thermalization is that the dynamics somehow knows the energy scale of the systems' free Hamiltonians. The argument presented in \cite{Grimmer_2019} is quite involved, I will provide only a summary of it here. Consider two quantum systems $A$ and $B$ evolving together from initial thermal states, 
\begin{align}
\hat{\rho}_\text{A}(0)
=\frac{e^{-\beta_\text{A}(0)\hat{H}_\text{A}}}{Z_\text{A}(0)},\qquad
\hat{\rho}_\text{B}(0)
=\frac{e^{-\beta_\text{B}(0)\hat{H}_\text{B}}}{Z_\text{B}(0)}.
\end{align}
with inverse temperatures $\beta_\text{A}(0)$ and $\beta_\text{B}(0)$ and with free Hamiltonians $\hat{H}_\text{A}$ and $\hat{H}_\text{B}$. Note that both of these states are invariant under the transformations,
\begin{align}\label{LambdaDef}
\Lambda_X:
\beta_X &\to\lambda \, \beta_X, \quad
\hat{H}_X \to \hat{H}_X/\lambda;
\quad X=A,\,B.
\end{align}
That is, they are invariant under a simultaneous rescaling of the temperature and free Hamiltonian. These transformations change the temperatures of the systems but do not change their density matrix. This is a reflection of the fact that temperature has units whereas the density matrix is unitless.

Suppose that these two systems interact and evolve to a have final reduced states which are themselves thermal,
\begin{align}\label{ThermStateA}
\hat{\rho}_\text{A}(\infty)
=\frac{e^{-\beta_\text{A}(\infty)\hat{H}_\text{A}}}{Z_\text{A}(\infty)},\qquad
\hat{\rho}_\text{B}(\infty)
=\frac{e^{-\beta_\text{B}(\infty)\hat{H}_\text{B}}}{Z_\text{B}(\infty)},
\end{align}
with identical temperatures, $\beta_\text{A}(\infty)=\beta_\text{B}(\infty)$. In particular one may expect that if the systems start at the same temperature then there will be no heat flow between them (recall the zeroth law of thermodynamics) such that their final temperatures are the same as their initial temperatures. That is, if $\beta_\text{A}(0)=\beta_\text{B}(0)$ then nothing should happen and we should end up with \mbox{$\beta_\text{A}(\infty)=\beta_\text{B}(\infty)=\beta_\text{A}(0)=\beta_\text{B}(0)$}. In fact, this is the \textit{only} case where nothing should happen. If the two systems are at different temperatures then there should be a non-zero heat flow between them as they thermalize with each other. Indeed, this is essentially the definition of ``different temperatures'' by the zeroth law of thermodynamics.

Note that transformations of the form \eqref{LambdaDef} do change the system's initial temperatures. Thus for any dynamics to describe thermal contact between $A$ and $B$ it must be able to tell if such a transformation has been performed. By construction, every effect of such transformations within the system's density matrices cancels out. If the dynamics does not depend explicitly on the the system's initial temperatures (that is, no dependence outside of the system's initial density matrices) then the dynamics only changes under \eqref{LambdaDef} if it depends on the system's free Hamiltonians explicitly, outside of the system's initial density matrices. Concretely if system A and B's reduced dynamics are given by, 
\begin{align}
\frac{\dd}{\dd t}\rho_\text{A}(t)
=\mathcal{L}_\text{A}[\rho_\text{A}(t)]
\quad
\frac{\dd}{\dd t}\rho_\text{B}(t)
=\mathcal{L}_\text{B}[\rho_\text{B}(t)]
\end{align}
then $\mathcal{L}_\text{A}$ and $\mathcal{L}_\text{B}$ must depend on both $\hat{H}_\text{S}$ and $\hat{H}_\text{B}$ (outside of dependence through $\rho_\text{A}(0)$ and $\rho_\text{B}(0)$) to yield thermal contact.
    
\section{Sensitivity to Free Energy Scales in Ancillary Bombardment}\label{ThermalAB}
Let us now apply the above criteria to an ancillary bombardment scenario. In particular let us consider the rapid bombardment regime where we a can expand the interpolation generator as,
\begin{align}
\mathcal{L}_{\delta t}
=\mathcal{L}_0
+\delta t\mathcal{L}_1
+\delta t^2\mathcal{L}_2
+\dots.
\end{align}
At what order in the above expansion does the dynamics depend on both the system and ancilla's free Hamiltonians?

To help us interpret the coming discussion consider an example along the lines of the qubit scenario discussed in Sec. \ref{BlochExample}. Specifically, consider the following hypothetical dynamics for the Bloch vector, $\bm{a}_\text{S}$, of a qubit interaction with a series of ancillas, $A$, as
\begin{align}
\frac{\dd}{\dd t}
\begin{pmatrix}
a_x(t)\\
a_y(t)\\
a_z(t)
\end{pmatrix}
=\begin{pmatrix}
-\delta t \, \gamma & \omega_\text{S} & 0\\
-\omega_\text{S} & -\delta t \, \gamma & 0\\
0 & 0 & -\delta t \, \gamma
\end{pmatrix}
\begin{pmatrix}
a_x(t)\\
a_y(t)\\
a_z(t)
\end{pmatrix}
+\begin{pmatrix}
0\\
0\\
\delta t \, b_1+\delta t^2 \, b_2
\end{pmatrix},
\end{align}
where $\hat{H}_\text{S}=\hbar\omega_\text{S}\hat\sigma_\text{S,z}/2$ and where $\gamma$ and $b_1$ do not depend on $\beta_\text{A}(0)$ or $\hat{H}_\text{A}$ outside of $\hat{\rho}_\text{A}(0)$, but where $b_2$ does depend directly on $\hat{H}_\text{A}$.

At zeroth order (in the continuum limit) this dynamics describes free evolution around the $z$-axis, that is, evolution by the system's free Hamiltonian $\hat{H}_\text{S}$. At first order the dynamics is attracted to the fixed point $\bm{a}_\text{S}=\left(0,0,b_1/\gamma\right)$. By assumption this first order fixed point does not depend on $\beta_\text{A}(0)$ or $\hat{H}_\text{A}$ outside of $\hat{\rho}_\text{A}(0)$. This means that the first order fixed point is unchanged if we transform system $A$ as \eqref{LambdaDef} prior to the interaction. Since such a transformation changes the initial temperature of system $A$ this fixed point cannot represent thermal equillibrium with the environment.

At second order, the $b_2$ term may move this fixed point slightly. Since $b_2$ depends on $\hat{H}_\text{A}$ outside of $\hat{\rho}_\text{A}$ its contribution is affected by transformations of the form \eqref{LambdaDef}. However, its contribution is only perturbative in the rapid interacting regime and so 1) cannot move the fixed point far and 2) even if we only need a certain small adjustment, this term could only provide it for a certain fixed $\delta t$. 

The lessons of the above qubit example largely transfer to a generic collision model scenario. In general, in order for dynamics to describe thermalization to the temperature of the environment, the dynamics must depend on the ancilla Hamiltonian at the same order in $\delta t$ that an isolated fixed point is established. As discussed in the previous section, for a generic ancillary bombardment scenario, this can be as early as first order, that is in $\mathcal{L}_1$. Under what conditions can $\mathcal{L}_1$ depend on $\hat{H}_\text{S}$ and  $\hat{H}_\text{A}$ outside of $\hat{\rho}_\text{A}(0)$?

Let us now investigate a generic ancillary bombardment scenario with update map,
\begin{align}
\phi(\delta t)[\hat{\rho}_\text{S}]
=\text{Tr}_\text{A}\big(e^{-\ii \, \delta t \, \hat{H}/\hbar} \ \hat{\rho}_\text{S}\otimes \hat{\rho}_\text{A}(0) \ e^{\ii \, \delta t \, \hat{H}/\hbar}\big),
\end{align}
where $\hat{H}
=\hat{H}_\text{S}\otimes\hat{\openone}_\text{A}
+\hat{\openone}_\text{S}\otimes\hat{H}_\text{A}
+\hat{H}_\text{SA}$. We can expand this update map as a series in $\delta t$, we have,
\begin{align}
\phi(\delta t)
=\openone
+\delta t \, \phi_1
+\delta t^2 \, \phi_2
+\delta t^3 \, \phi_3
+\dots    
\end{align}
where,
\begin{align}\label{phinDef}
&\phi_1[\hat{\rho}_\text{S}]
=\frac{-\ii}{\hbar} \, 
\text{Tr}_\text{A}\Big(
[\hat{H},\hat{\rho}_\text{S}\otimes \hat{\rho}_\text{A}(0)]\Big),\\
\nonumber
&\phi_2[\hat{\rho}_\text{S}]
=\frac{1}{2!}\Big(\frac{-\ii}{\hbar}\Big)^2
\text{Tr}_\text{A}\Big(
[\hat{H},[\hat{H},\hat{\rho}_\text{S}\otimes \hat{\rho}_\text{A}(0)]]\Big),\\
\nonumber
&\phi_3[\hat{\rho}_\text{S}]
=\frac{1}{3!}\Big(\frac{-\ii}{\hbar}\Big)^3
\text{Tr}_\text{A}\Big(
[\hat{H},[\hat{H},[\hat{H},\hat{\rho}_\text{S}\otimes \hat{\rho}_\text{A}(0)]]]\Big),
\end{align}
etc. From this expansion we can expand $\mathcal{L}_{\delta t}$ as a series as,
\begin{align}
\mathcal{L}_{\delta t}
=\mathcal{L}_0
+\delta t \, \mathcal{L}_1
+\delta t^2 \, \mathcal{L}_2
+\delta t^3 \, \mathcal{L}_3
+\dots \, ,
\end{align}
where,
\begin{align}
\mathcal{L}_0
&=\phi_1,\\
\mathcal{L}_1
\nonumber
&=\phi_2-\frac{1}{2}\phi_1{}^2,\\
\mathcal{L}_2
\nonumber
&=\phi_3
-\frac{1}{2}(\phi_1\phi_2+\phi_2\phi_1)
+\frac{1}{3}\phi_1{}^3,
\end{align}
etc. By using the linearity of the partial trace and the commutator, we can see that the $n^{th}$ term in \eqref{phinDef} involves all the ways of picking one of $\hat{H}_\text{S}$, $\hat{H}_\text{A}$, or $\hat{H}_\text{SA}$ for each of the $n$ copies of $\hat{H}$ appearing in the expressions given by \eqref{phinDef}. In \cite{Grimmer_2019} it was found that if $\hat{\rho}_\text{A}(0)$ is thermal then the first non-vanishing term which contains $\hat{H}_\text{A}$ appears in $\phi_3$ and therefore in $\mathcal{L}_2$. Specifically this term is, 
\begin{align}\label{L2HA}
\text{Tr}_\text{A}\Big([\hat{H}_\text{SA},[\hat{H}_\text{A},[\hat{H}_\text{SA},\hat{\rho}_\text{S}\otimes \hat{\rho}_\text{A}(0)]]]\Big).
\end{align}
Similarly the first dependence on $\hat{H}_\text{S}$ (outside of the ``trivial'' dependence in $\mathcal{L}_0$) occurs in $\mathcal{L}_2$ and is given by,
\begin{align}\label{L2HS}
\text{Tr}_\text{A}\Big([\hat{H}_\text{SA},[\hat{H}_\text{S},[\hat{H}_\text{SA},\hat{\rho}_\text{S}\otimes \hat{\rho}_\text{A}(0)]]]\Big).  
\end{align}

A natural explanation for why $\hat{H}_\text{A}$ does not show up until $\mathcal{L}_2$ is provided in \cite{Grimmer_2019} and will be summarized here. By interpreting $[\hat{H}_X,\cdot]$ as a small amount of evolution with respect to $\hat{H}_X$ we can interpret \eqref{L2HA} as follows. The simplest/shortest process carrying information about the ancilla's local Hamiltonian (and therefore its temperature) is to:
\begin{itemize}
\item [1)] Interact with it (so it is not thermal anymore)
\item [2)] Let it evolve freely (bringing in its energy scale)
\item [3)] Interact with it again (to get the information out).
\end{itemize}
In the rapid bombardment regime this process ``takes too long'' and is therefore highly suppressed.

In the last two sections we have seen how dependence on the ancilla's free energy scale is necessary for thermalizing dynamics. We then showed that in an ancillary bombardment scenario this dependence doesn't show up until second order in $\delta t$, that is in $\mathcal{L}_2$. Thus any model of thermalization built from collision models must include finite duration effects. We cannot even use the $g^2\delta t$ trick to ``promote'' these $\mathcal{L}_2$ effects into the continuum limit, since this trick only applies to $\mathcal{L}_1$ dynamics. This provides yet another reason to consider collisional models outside of the continuum limit.

\chapter{Review of Gaussian Quantum Mechanics}\label{Ch5}
In this chapter I will give a broad overview of Gaussian Quantum Mechanics (GQM) in order to familiarize the reader and to establish our notation. In particular this chapter will follow the summary and characterization of GQM presented in my prior publications, \cite{Dan2018,PhysRevA.97.052120}.  In these papers many of the following claims are fleshed out and demonstrated with examples. For other introductions see \cite{Weedbrook2012, adesso1, lami,GQMRev}.

GQM is a subtheory of quantum mechanics dealing with states of continuous variable systems which have Gaussian Wigner functions (termed Gaussian states) and with transformations which preserve this Gaussianity. Such states are theoretically and experimentally relevant, including coherent states, thermal states, squeezed states and the vacuum state. Since Gaussians are characterized by relatively few parameters (their means and covariances) GQM offers a large (or even infinite) decrease in the overhead for describing quantum states and transformations.

GQM has been applied in areas including open quantum systems \cite{koga, nicacio, nicacio2}, quantum information processing \cite{Weedbrook2012,Eisler2015,Eliska2018,Kraus2009}, quantum computing \cite{D2005,Bravyi2000,B2005,serge2011,PhysRevB.88.035121,PhysRevB.94.045316}, quantum entanglement \cite{Abotero2003,botero2004,isert2018,Richter2017}, thermodynamics \cite{PhysRevLett.120.190501,PhysRevA.90.020302,PhysRevA.90.062329,Melo_2013} and quantum thermodynamics \cite{eric2016,marvy2018,campbell2015}.

\section{Phase Space Structure }\label{ReviewGQM}
Let us consider a system of $N$ coupled bosonic modes with the $n^\text{th}$ mode fully characterized by its creation and annihilation operators, \mbox{$\hat{a}_n^\dagger$ and $\hat{a}_n$,} which obey the canonical Bosonic commutation relations,
\begin{align}\label{CanonicalCommsAA}
[\hat{a}_n,\hat{a}_m]
=[\hat{a}_n^\dagger,\hat{a}_m^\dagger]
=0
\quad\text{and}\quad
[\hat{a}_n,\hat{a}_m^\dagger]
=\delta_{nm} \, \hat{\openone},
\end{align}
where  $\delta_{nm}$ is the Kronecker delta and $\hat{\openone}$ is the identity operator on the system's Hilbert space. For our purposes it is convenient to instead characterize the system in terms of its quadrature operators
\bel{qpDef}
\hat{q}_n=\frac{1}{\sqrt{2}}(\hat{a}_n^\dagger+\hat{a}_n)
\quad\text{and}\quad
\hat{p}_n=\frac{\ii}{\sqrt{2}}(\hat{a}_n^\dagger-\hat{a}_n).
\ee
From \eqref{CanonicalCommsAA}, these quadrature operators obey the canonical commutation relations, 
\begin{align}\label{CanonicalCommsQP}
[\hat{q}_n,\hat{q}_m]
=[\hat{p}_n,\hat{p}_m]
=0
\quad\text{and}\quad
[\hat{q}_n,\hat{p}_m]
=\ii \, \delta_{nm} \, \hat{\openone}.
\end{align}
Such systems can be fully characterized in terms of a pseudo-probability distribution defined on the system's phase space \cite{Groenewold,Moyal}. In particular, a state with density matrix $\hat{\rho}$ can be equivalently represented by its Wigner function,
\be
W(\bm{q},\bm{p})=\frac{1}{\pi^N}\!\int_{-\infty}^\infty \dd^N \bm{s}
\bra{\bm{q}+\bm{s}}\hat{\rho}\ket{\bm{q}-\bm{s}}\exp(-2\ii \, \bm{p}\cdot\bm{s}).
\ee
Our goal will be to translate our description of the system from Hilbert space to phase space, in particular, to matrices and vectors in phase space. To facilitate this description it is convenient to collect these $2N$ quadrature operators into the following operator-valued phase space vector,
\bel{XhatDef}
\hat{\bm{X}}
\coloneqq
(\hat{q}_1,\hat{p}_1,\hat{q}_2,\hat{p}_2,\dots,\hat{q}_N,\hat{p}_N)^\intercal.
\ee
Note that every pair of these quadrature operators, say $\hat{X}_j$ and $\hat{X}_k$, commute to a (potentially zero) multiple of the identity operator on the system's Hilbert space. In particular any pair will commute to either $\pm\ii \,  \hat{\openone}$ or to $0$. Thus we can fully capture the system's commutation relations with the phase space matrix, $\Omega$, defined by,
\begin{align}\label{OmegaDef}
[\hat{X}_j,\hat{X}_k]
&=\ii \ \Omega_{jk} \, \hat{\openone}.
\end{align}
This matrix, called the symplectic form, is given explicitly as,
\bel{OmegaExplicit}
\Omega
=\bigoplus_{n=1}^N \omega
=\openone_N\otimes\omega; \ \ \ \ \omega
=\begin{pmatrix}
0 & 1\\
-1 & 0
\end{pmatrix},
\ee
in the same representation as \eqref{XhatDef}. Note that $\Omega$ is real-valued, antisymmetric, and invertible with \mbox{$\Omega^{-1}=\Omega^T=-\Omega$}.  

It should be noted that an alternate operator ordering,
\bel{AltXhatDef}
\hat{\bm{X}}_\text{alt}=(\hat{q}_1,\dots\hat{q}_N,\hat{p}_1,\dots,\hat{p}_N)^\intercal
\ee
is also common in the literature and would yield an alternate expression for the symplectic form,
\bel{AltOmegaExplicit}
\Omega_\text{alt}
=\omega\otimes\openone_N
=\begin{pmatrix}
0 & \openone_N\\
-\openone_N & 0
\end{pmatrix}.
\ee
We prefer the ordering given by \eqref{XhatDef} as it has the conjugate pairs of observables adjacent. This ordering is helpful in addressing individual modes and in characterizing dynamics as either single-mode or multi-mode (as we will see in Sec. \ref{Characterizing}).

\section{Gaussian States}
Having captured the algebraic structure of our system's Hilbert space in terms of the phase space matrix, $\Omega$. We now discuss the class of quantum states considered in Gaussian Quantum Mechanics, those with Gaussian Wigner functions. 

The main benefit of this restriction to Gaussian states is that it allows for a significantly simplified description of quantum states and transformations while still describing a wide variety of theoretically and experimentally relevant situations. In particular, a Gaussian distribution is completely determined by its first and second statistical moments.

The system's first moments are captured by the mean of each of these operators,
\bel{XDef}
\bm{X}
\coloneqq\langle\hat{\bm{X}}\rangle
=\big(\langle\hat{q}_1\rangle,\langle\hat{p}_1\rangle,\dots,\langle\hat{q}_N\rangle,\langle\hat{p}_N\rangle\big)^\intercal.
\ee
The system's centered second moments are given by the matrix,
\be
\Lambda_{jk}
\coloneqq
\big\langle
\hat{X}_j^\dagger \, \hat{X}_k
\big\rangle
-\langle\hat{X}_j^\dagger\rangle\,
\langle\hat{X}_k\rangle.
\ee
Note that these second moments are in general complex-valued since $\hat{X}_j^\dagger$ and $\hat{X}_k$ may not commute. Nonetheless this second moment matrix is positive semi-definite, $\Lambda\geq0$. That is, for any complex phase space vector, $\bm{c}=(c_1,\dots,c_{2N})$, we have $\bm{c}^\dagger\Lambda\bm{c}\geq0$. We can see this through the following calculation,
\begin{align}
\bm{c}^\dagger\Lambda\bm{c}
&=\sum_{jk} c_j^* \Lambda_{jk} c_k\\
&=\sum_{jk} \big\langle c_j^* 
\hat{X}_j^\dagger \, \hat{X}_k
 c_k\big\rangle
-\big\langle c_j^* 
\hat{X}_j^\dagger\big\rangle
\big\langle\hat{X}_k
 c_k\big\rangle\\
&=\left\langle
\hat{C}^\dagger \,
\hat{C}
\right\rangle
-\left\langle
\hat{C}\right\rangle^*
\left\langle
\hat{C}\right\rangle\\
&\geq0,
\end{align}
where $\hat{C}=\sum_k c_k\hat{X}_k$. The final inequality follows from the density matrix being positive semi-definite, $\hat{\rho}\geq0$, in the system's Hilbert space. Thus requiring that $\hat{\rho}$ is a bona-fide quantum state places a restriction on the system's matrix of second moments. Moreover, this condition $\Lambda\geq0$ is both necessary and sufficient for the corresponding $\hat{\rho}$ to be a valid density matrix (positive semi-definite and trace one). We will return to this point later after, after removing some redundancy from these second moments.

Using the symplectic form we can see that the anti-symmetric part of $\Lambda$ is fixed, independent of the system's state. In particular,
\begin{align}
\Gamma_{jk}
&\coloneqq\Lambda_{jk}-\Lambda_{kj}\\
&=\langle\hat{X}_j \, \hat{X}_k
- \hat{X}_k \, \hat{X}_j\rangle\\
&=\langle[\hat{X}_j,\hat{X}_k]\rangle\\
&=\langle\ii \ \Omega_{jk} \, \hat{\openone}\rangle\\
&=\ii \ \Omega_{jk}.
\end{align}
Thus the anti-symmetric part of $\Lambda$ contains no information about the system's state; All of the information about the state is encoded in the symmetric part of $\Lambda$. We can collect this information by defining a symmetric $2N$ by $2N$ called the covariance matrix as,
\bel{Vdef2}
\sigma_{jk}
\coloneqq
\Lambda_{jk}+\Lambda_{kj}
=\big\langle
\hat{X}_j \, \hat{X}_k
+ \hat{X}_k \, \hat{X}_j
\big\rangle
-2\big\langle\hat{X}_j\big\rangle
\big\langle\hat{X}_k\big\rangle.
\ee
We note that an alternate definition for the covariance matrix, \mbox{$\sigma_\text{alt}=\sigma/2$} is common. We prefer the notation defined by \eqref{Vdef2} as it removes many factors of two from our equations. 

The Wigner function of a Gaussian state can be written directly in terms of its mean, $\bm{X}$, and covariance matrix, $\sigma$, as,
\be
W(\bm{Y})=\frac{1}{\pi^N\sqrt{\text{det}(\sigma)}}
\exp\big(-
(\bm{Y}-\bm{X})^\intercal
\sigma^{-1}
(\bm{Y}-\bm{X})
\big).
\ee
It is useful to visualize Gaussian states as hyperellipsoids in phases space corresponding to the all the points within one deviation of the mean, $\bm{X}$, with respects to $\sigma$, that is,
\begin{align}
\bm{Y}=\bm{X}+\bm{R}
\quad\text{for every }\bm{R}\text{ such that}\quad
\bm{R}^\intercal
\sigma^{-1}\bm{R}\leq1.
\end{align}
In the case of a single mode ($N=1$), one can think of a Gaussian state as an ellipse in a 2-dimensional phase space centered at $\bm{X}=(\langle\hat{q}\rangle,\langle\hat{p}\rangle)^\intercal$ and some major and minor axes whose size and orientation are given by the eigensystem of $\sigma$.

For this visualization to make sense $\sigma$ must have non-negative eigenvalues, that is, that it must be positive semi-definite, $\sigma\geq0$. Indeed this is exactly the condition for $W(\bm{Y})$ to be a valid probability distribution. However, this clearly cannot be the only restriction on $\sigma$ as then a particle could have an arbitrarily well defined position and momentum, violating the uncertainty principle.

As mentioned above, a necessary and sufficient condition for a Gaussian state to correspond to a valid quantum state is that its matrix of second moments is positive semi-definite, $\Lambda\geq0$. We can translate this condition into a condition on the system's covariance matrix $\sigma$, as follows. First note that 
$\Lambda=\frac{1}{2}(\sigma+\Gamma)=\frac{1}{2}(\sigma+\ii\Omega)$ such that\footnote{Throughout this text, the notation \mbox{$P\geq Q$} is used to mean that \mbox{$P-Q$} is positive semi-definite (i.e., \mbox{$P-Q\geq0$}).}
\be
\Lambda\geq0\Rightarrow\sigma\geq-\ii \, \Omega.
\ee
Next note that,
\be
\sigma\geq-\ii \, \Omega 
\Rightarrow 
\sigma^*\geq(-\ii \, \Omega)^*
\Rightarrow 
\sigma=\ii \, \Omega,
\ee
since entry-wise complex conjugation maintains the eigenvalues of a Hermitian matrix and since $\sigma$ and $\Omega$ are both real-valued. Thus for all valid Gaussian states we have both $\sigma\geq\ii \, \Omega$ and $\sigma\geq-\ii \, \Omega$. For details see \cite{Simon1994}. This condition is stronger\footnote{Indeed, since $\sigma$ is greater than both $\ii \, \Omega$ and $-\ii \, \Omega$, it is greater than their average as well, thus $\sigma\geq0$.} than the condition that $W(Y)$ merely be a valid probability distribution, $\sigma\geq0$. As one of the following examples will demonstrate the new stronger condition enforces the uncertainty principle.

Many relevant states for both theory and experiment are Gaussian states. For instance, taking a single mode ($N=1$) to be a harmonic oscillator, its thermal states (with respect to its free Hamiltonian, $\hat{H}=\hat{q}^2+\hat{p}^2$) are described by,
\bel{ThermalStateDef}
\bm{X}=0
\quad \text{and} \quad
\sigma=\begin{pmatrix}
\nu & 0 \\
0 & \nu \\
\end{pmatrix},
\ee
where $\nu\geq1$ is a monotone function of the temperature. In the ellipsoid picture, described above, this state corresponds to a circle of radius $\sqrt{\nu}$ centered at $\bm{X}=0$.

Pure coherent states are described by,
\bel{CoherentStateDef}
\bm{X}=\begin{pmatrix}
\langle \hat{q} \rangle \\
\langle \hat{p} \rangle \\
\end{pmatrix}
\quad \text{and} \quad
\sigma
=\begin{pmatrix}
1 & 0 \\
0 & 1 \\
\end{pmatrix}.
\ee
Visualized as an ellipsoid, this state corresponds to  a circle of unit radius centered at $\bm{X}$.

A family of single-mode squeezed states are described by,
\bel{N1SqueezedStateDef}
\bm{X}=0
\quad \text{and} \quad
\sigma=\begin{pmatrix}
\sigma_{qq} & 0 \\
0 & \sigma_{pp} \\
\end{pmatrix},
\ee
obeying the uncertainty principle $\sigma_{qq} \, \sigma_{pp}\geq1$. This state corresponds to an ellipsoid centered at $\bm{X}=0$ with its major and minor axes in the $q$ and $p$ directions with lengths $\sqrt{\sigma_{qq}}$ and $\sqrt{\sigma_{pp}}$ respectively. More generally, any single-mode squeezed state is Gaussian.

\section{Unitary Gaussian Transformations}
We now turn our attention toward unitary transformations that preserve the Gaussianity of the states they act on. That is to those unitary transformations which take Gaussian states to Gaussian states. Such transformations are called Gaussian unitary transformations.

Differential Gaussian unitary transformations are generated by Hamiltonians that  are at most quadratic in the system's quadrature operators \cite{Schumaker1986}. Any such Hamiltonian can be converted into the standard form\footnote{See Appendix A of \cite{Dan2018}
},
\bel{QuadHamForm}
\hat{H}=\frac{1}{2}\hat{\bm{X}}^\intercal  \, F \, \hat{\bm{X}}
+\bm{\alpha}^\intercal\hat{\bm{X}},
\ee
where $F$ is a real-valued $2N\times 2N$ symmetric matrix and $\bm{\alpha}$ is a real-valued vector of length $2N$.

In the Heisenberg picture, evolution under \eqref{QuadHamForm} yields\footnote{See Appendix A of \cite{Dan2018}},
\bel{HeisenbergEqsGaussian}
\frac{\dd}{\dd t}\bm{\hat{X}}
=\ii[\hat{H},\bm{\hat{X}}]
=\Omega (F \bm{\hat{X}}+\bm{\alpha} \, \hat{\openone}).
\ee
Note that in the above expression $\hat{H}$ is a linear map on the system's Hilbert space and acts on $\bm{\hat{X}}$ componentwise. On the other hand, $F$ is a linear map on the system's phase space and acts on $\bm{\hat{X}}$ as a phase space vector, yielding linear combinations of its (operator-valued) components. Differential unitary evolution in Hilbert space (with respects to a quadratic Hamiltonian) is differential linear-affine evolution in phase space.

For a time independent Hamiltonian, integrating \eqref{HeisenbergEqsGaussian} for a time interval $[0,t]$ yields,\footnote{See Appendix A of \cite{Dan2018}.} 
\bel{GaussianUnitaryUpdate}
\bm{\hat{X}}(t)
=\hat{U}_\text{G}(t)\hat{\bm{X}}(0)\hat{U}_\text{G}(t)^\dagger
=S(t)\hat{\bm{X}}(0)+\bm{d} \, \hat{\openone},
\ee
where $\hat{U}_\text{G}(t)=\exp(\ii \, \hat{H} \, t)$ and,
\begin{align}
\label{SHamDef}
S(t)&=\text{exp}(\Omega F \, t),\\
\label{dHamDef}
\bm{d}(t)&=\frac{\text{exp}(\Omega F \, t)-\openone_{2N}}{\Omega F} \, \Omega\bm{\alpha}.
\end{align}
In the above expression $\hat{U}_\text{G}(t)$ is a linear map on the system's Hilbert space and acts on $\bm{\hat{X}}$ componentwise. On the other hand, $S(t)$ is a linear map on the system's phase space and acts on $\bm{\hat{X}}$ as a phase space vector, yielding linear combinations of its (operator-valued) components. Finally note that $\Omega F$ does not need to be invertible to make sense of \eqref{dHamDef}, if one understands it in terms of the following definition:
\bel{(ExpX-1)byXDef}
\frac{\text{exp}(M \, t)-\openone}{M}
\coloneqq\sum_{m=0}^\infty \frac{t^{m+1}}{(m+1)!}M^m,
\ee
for a general square matrix $M$.

From \eqref{HeisenbergEqsGaussian}, the differential evolution of $\bm{X}$ and $\sigma$ under a quadratic Hamiltonian, \eqref{QuadHamForm}, can be straightforwardly computed as,
\begin{align}
\label{SymplecticDiffXUpHam}
\frac{\dd}{\dd t}\bm{X}
&=\Omega (F \bm{X}+\bm{\alpha}),\\
\label{SymplecticDiffVUpHam}
\frac{\dd}{\dd t}\sigma
&=(\Omega F) \, \sigma
+\sigma \, (\Omega F)^\intercal.
\end{align}
From \eqref{GaussianUnitaryUpdate} the finite time evolution of the system's mean and covariance matrix is given by
\begin{align}
\label{SymplecticXUp1}
\bm{X}(0)&\longrightarrow \bm{X}(t)=S(t) \, \bm{X}(0)+\bm{d}(t),\\
\label{SymplecticVUp1}
\sigma(0)&\longrightarrow \sigma(t)=S(t) \, \sigma(0) \, S^\intercal(t).
\end{align}
Note that $S(t)\Omega S(t)^\intercal=\Omega$. This is a reflection of the fact that generic unitary evolution, $U$, preserves the commutation relations as,
\begin{align}
[\hat{X}_j,\hat{X}_k]=\ii\,\Omega_{jk}\hat{\openone}
\Rightarrow
[U\hat{X}_j U^\dagger,U\hat{X}_k U^\dagger]
=U[\hat{X}_j,\hat{X}_k ]U^\dagger
=U(\ii\Omega_{jk}\hat{\openone})U^\dagger
=\ii\,\Omega_{jk}\hat{\openone}.
\end{align}

More generally, if we allow for a time-dependent Hamiltonian\footnote{Notice that in order to implement a general symplectic transformation a time dependent generator is generally needed. This follows from the exponential in the symplectic group not being surjective.}, any transformation of the form,
\bel{GeneralUnitaryUpdate}
\bm{\hat{X}}
\longrightarrow
S\hat{\bm{X}}+\bm{d} \, \hat{\openone},
\ee
and therefore,
\begin{align}
\label{SymplecticXUp2}
\bm{X}&\longrightarrow S \bm{X}+\bm{d},\\
\label{SymplecticVUp2}
\sigma&\longrightarrow S\, \sigma \, S^\intercal,
\end{align}
with generic real-valued $S$ and $\bm{d}$ can be implemented as long as it preserves the commutation relation (i.e., the symplectic form) as,
\bel{SympTranDef}
S\Omega S^\intercal=\Omega.
\ee
Such a matrix $S$ is said to be symplectic and to implement a transformation. Together with $\bm{d}$, the update \eqref{SymplecticXUp2} and \eqref{SymplecticVUp2} constitutes a symplectic-affine transformation. Thus Gaussian unitary transformations on the system's Hilbert space correspond to symplectic\footnote{For convenience we will often drop the ``-affine'' suffix, referring to these transformations as simply symplectic.} transformations on the system's phase space.

\section{Gaussian Channels}\label{GaussianChannel}
The unitary transformations described in the previous section are not the most general class of transformations that  preserve the Gaussian nature of the state. In addition to the Gaussian unitary transformations described above, one can implement non-unitary Gaussian transformations by allowing the system to interact with an environment. In direct analogy with the Stinespring dilation theorem \cite{Stinespring:1955}, one can implement any completely positive trace preserving (CPTP) Gaussian transformation as a Gaussian unitary transformation in some larger Hilbert space (or equivalently as a symplectic-affine transformation in a larger phase space) \cite{GaussianDilation}.

To see concretely how such a dilation works let us consider a system, $\text{S}$, to be a Gaussian system composed of $N_\text{S}$ bosonic modes. Likewise  consider an ancilla, $\text{A}$ to  be a Gaussian system composed of $N_\text{A}$ bosonic modes. Together they form a joint system, $\text{SA}$, which is Gaussian and is composed of $N_\text{S}+N_\text{A}$ modes. Note that the dimensions of $\text{S}$'s, $\text{A}$'s and $\text{SA}$'s phase spaces are $2N_\text{S}$, $2N_\text{A}$, and $2N_\text{S}+2N_\text{A}$ respectively.

The system and ancilla's quadrature operators are collected together into the joint operator vector,
\be
\hat{\bm{X}}_\text{SA}
=(\hat{\bm{X}}_\text{S}^\intercal,\hat{\bm{X}}_\text{A}^\intercal)^\intercal
=\hat{\bm{X}}_\text{S}\oplus\hat{\bm{X}}_\text{A}.
\ee
Since the system's and ancilla's observables act on different Hilbert spaces, all pairs of their observables commute with each other. Thus they have the joint symplectic form,
\be
\Omega_\text{SA}=
\begin{pmatrix}
\Omega_\text{S} & 0\\
0 & \Omega_\text{A}
\end{pmatrix}
=\Omega_\text{S}\oplus\Omega_\text{A}
,
\ee
where $\Omega_\text{S}$ and $\Omega_\text{A}$ are the symplectic forms in the phase space of S and A respectively. 

We assume that the system and ancilla are initially uncorrelated, having the initial joint mean vector,
\be
\bm{X}_\text{SA}(0)
=(\bm{X}_\text{S}(0)^\intercal,\bm{X}_\text{A}(0)^\intercal)^\intercal
=\bm{X}_\text{S}(0)\oplus\bm{X}_\text{A}(0),
\ee
and the initial joint covariance matrix,
\be
\sigma_\text{SA}(0)=
\begin{pmatrix}
\sigma_\text{S}(0) & 0\\
0 & \sigma_\text{A}(0)
\end{pmatrix}
=\sigma_\text{S}(0)\oplus\sigma_\text{A}(0).
\ee
Further we assume that they evolve under a quadratic time-independent Hamiltonian,
\bel{HSADef1}
\hat{H}_\text{SA}
=\frac{1}{2}\hat{\bm{X}}_\text{SA}^\intercal  \, F_\text{SA} \, \hat{\bm{X}}_\text{SA}
+\bm{\alpha}^\intercal_\text{SA}\hat{\bm{X}}_\text{SA},
\ee
where $F_\text{SA}$ is real and symmetric and $\bm{\alpha}_\text{SA}$ is real. 

It is useful to divide this Hamiltonian into subblocks corresponding to the system and ancilla phase spaces as,
\be
F_\text{SA}=
\begin{pmatrix}
F_\text{S} & G\\
G^\intercal & F_\text{A}
\end{pmatrix},
\quad \quad
\bm{\alpha}_\text{SA}=
\begin{pmatrix}
\bm{\alpha}_\text{S}\\
\bm{\alpha}_\text{A}
\end{pmatrix}.
\ee
Note that $F_\text{S}$ and $F_\text{A}$ are symmetric and that $G$ is not generally square, having dimensions $2 N_\text{S}$ by $2 N_\text{A}$. 

Divided this way we can see that $F_\text{S}$ and $\bm{\alpha}_\text{S}$ correspond to the system's free Hamiltonian,
\be
\hat{H}_\text{S}
=\frac{1}{2}\hat{\bm{X}}_\text{S}^\intercal  \, F_\text{S} \, \hat{\bm{X}}_\text{S}
+\bm{\alpha}^\intercal_\text{S}\hat{\bm{X}}_\text{S}.
\ee
Similarly $F_\text{A}$ and $\bm{\alpha}_\text{A}$ correspond to the ancilla's free Hamiltonian,
\be
\hat{H}_\text{A}
=\frac{1}{2}\hat{\bm{X}}_\text{A}^\intercal  \, F_\text{A} \, \hat{\bm{X}}_\text{A}
+\bm{\alpha}^\intercal_\text{A}\hat{\bm{X}}_\text{A}.
\ee
Finally, we can see that the $G$ matrix contains all of the couplings between the system and the ancilla, corresponding to the interaction Hamiltonian, 
\be
\hat{H}_\text{I}
=\frac{1}{2}\hat{\bm{X}}_\text{S}^\intercal  \, G \, \hat{\bm{X}}_\text{A}
+\frac{1}{2}\hat{\bm{X}}_\text{A}^\intercal  \, G^ \intercal\, \hat{\bm{X}}_\text{S}.
\ee

Next we compute the effect that evolving for a time $t$ under this Hamiltonian has on system $S$. In order to do this we compute the evolution of the joint system, $SA$, then isolate the effect on the system $S$. The joint evolution is unitary and therefore given by a symplectic-affine transformation in the joint phase space. Specifically,
\begin{align}
\label{SXUpdt1}
\bm{X}_\text{SA}(t)
&=S_\text{SA}(t) \, \bm{X}_\text{SA}(0)+\bm{d}_\text{SA}(t),\\
\label{SVUpdt1}
\sigma_\text{SA}(t)
&=S_\text{SA}(t) \, \sigma_\text{SA}(0) \, S^\intercal_\text{SA}(t),
\end{align}
where,
\begin{align}
\label{AppSHamDef1}
S_\text{SA}(t)&=\text{exp}(\Omega_\text{SA} F_\text{SA} \, t),\\
\label{AppdHamDef1}
\bm{d}_\text{SA}(t)&=\frac{\text{exp}(\Omega_\text{SA} F_\text{SA} \, t)-\openone_{2 N_\text{S}+2 N_\text{A}}}{\Omega_\text{SA} F_\text{SA}} \, \Omega_\text{SA} \, \bm{\alpha}_\text{SA}.
\end{align}
In order to find the effective update on the system's state we can divide these into blocks as,
\be
S_\text{SA}(t)
=\begin{pmatrix}
M_\text{SS}(t) & M_\text{SA}(t) \\
M_\text{AS}(t) & M_\text{AA}(t) \\
\end{pmatrix}
\ \ \text{and} \ \ 
\bm{d}_\text{SA}(t)
=\begin{pmatrix}
\bm{d}_\text{S}(t) \\ \bm{d}_\text{A}(t) \\
\end{pmatrix}.
\ee
Expanding \eqref{SXUpdt1} and \eqref{SVUpdt1} over the direct sum between the system and ancilla's phase spaces, one can identify that the reduced state of the system ($\bm{X}_\text{S}$ and $\sigma_\text{S}$) is updated as,
\begin{align}\label{5161}
\bm{X}_\text{S}(t)
&=T(t) \, \bm{X}_\text{S}(0)
+\bm{d}(t),\\
\label{5162}
\sigma_\text{S}(t)
&=T(t) \, \sigma_\text{S}( 0) \, T^\intercal(t)
+R(t),
\end{align}
where 
\begin{align}\label{TSMdSDef1}
T(t)&=M_\text{SS}(t),\\
\bm{d}(t)&=M_\text{SA}(t) \ \bm{X}_\text{A}(0)+\bm{d}_\text{S}(t),\\
R(t)&=M_\text{SA}(t) \, \sigma_\text{A}(0) \, M^\intercal_\text{SA}(t).
\end{align}
Equations \eqref{5161} and \eqref{5162} generalize the unitary (symplectic) evolution given by \eqref{SymplecticXUp1} and \eqref{SymplecticVUp1} in two ways: 1) $T(t)$ unlike $S(t)$ is not necessarily symplectic and 2) the matrix $R(t)$ does not appear in the symplectic evolution.

More generally, allowing the Hamiltonians to be time-dependent, we can implement any transformation of the form,
\begin{align}
\label{GeneralUpdateX}
\bm{X}&\to T\bm{X}+\bm{d},\\
\label{GeneralUpdateV}
\sigma&\to T \, \sigma \, T^\intercal+R,
\end{align}
where $\bm{d}$ is a real $2N$-dimensional vector, $T$ and $\bm{R}$ are $2N$ by $2N$ real matrices, $R$ is symmetric so long as it takes valid Gaussian states to valid Gaussian states. That is, so long as for every covariance matrix $\sigma$,
\be
\sigma+\ii\Omega\geq0
\Rightarrow
T(\sigma+\ii\Omega)T^\intercal+R\geq0.
\ee
This is the Gaussian analogue of the complete positivity condition for quantum channels. An equivalent statement of this condition is that \cite{GQMRev},
\bel{FiniteCPCond1}
R\geq\ii \, (T \, \Omega \,  T^\intercal-\Omega).
\ee
Note that this implies that $R\geq 0$.

To see what sort of master equations are possible for open Gaussian systems, we can take the update given by \eqref{GeneralUpdateX} and \eqref{GeneralUpdateV} to be differential, as,
\begin{align}
T(\dd t)&=\openone_{2N}+\dd t \ \Omega  \, A,\\
\bm{d}(\dd t)&=\dd t \ \Omega \, \bm{b},\\
R(\dd t)&=\dd t \ C,
\end{align}
where $\bm{b}$ is a real $2N$-dimensional vector, $A$ and $C$ are $2N$ by $2N$ real matrices, $C$ is symmetric. Since $\Omega$ is invertible, and since $A$ and $\bm{b}$ are arbitrary, assuming that a factor of $\Omega$ precedes $A$ and $\bm{b}$ is justified.

From this differential update one can find that the general form of the Gaussian master equations is,
\begin{align}
\label{GeneralDiffXUp}
\frac{\dd}{\dd t}\bm{X}(t)
&=\Omega(A \bm{X}(t)+\bm{b}),\\
\label{GeneralDiffVUp}
\frac{\dd}{\dd t}\sigma(t)
&=(\Omega A) \, \sigma(t)
+\sigma(t) \, (\Omega A)^\intercal
+C.
\end{align}
This generalized the differential symplectic evolution given by \eqref{SymplecticDiffXUpHam} and \eqref{SymplecticDiffVUpHam} in two ways: in two ways: 1) $A$, unlike $F$ is not necessarily symmetric and 2) the matrix $C$ does not appear in the symplectic evolution.

The differential version of the complete positivity condition \eqref{FiniteCPCond1} is,
\bel{DiffCPCond}
C\geq\ii \, \Omega (A-A^\intercal)\Omega,
\ee
from which it follows that $C\geq0$.

\section{Characterizing Gaussian Master Equations - A 4-way partition}\label{Characterizing}
Before applying the interpolated collision model formalism to Gaussian systems, allow me to briefly overview the types of dynamics that can arise from generic Gaussian master equations as well as a scheme for classifying these different types of dynamics presented. In particular I will summarize the classification presented in my prior publication \cite{Dan2018}.

I will classify Gaussian dynamics according to the following four dichotomies:
\begin{itemize}
    \item Symplectic vs. Unsymplectic (i.e., Unitary vs. Non-unitary) 
    \item Passive vs. Active
    \item Single-Mode vs. Multi-Mode
    \item State-Dependent vs. State-Independent.
\end{itemize}

\subsubsection{Symplectic vs. Unsymplectic}
As discussed above, Gaussian unitary transformations are represented in phase space by symplectic transformations, that is, by transformations which preserve the sympectic form $S\Omega S^\intercal=\Omega$. Likewise we can identify any part of the dynamics which does not preserve the symplectic form (i.e. which is unsymplectic) as non-unitary.

Comparing general differential evolution \eqref{GeneralDiffXUp} and \eqref{GeneralDiffVUp} with unitary (symplectic) evolution \eqref{SymplecticDiffXUpHam} and \eqref{SymplecticDiffVUpHam} we can identify that $\bm{b}$ and the symmetric part of $A$ correspond to symplectic evolution whereas $C$ and the anti-symmetric part of $A$ correspond to unsymplectic evolution.

\subsubsection{Passive vs. Active}
Next we can characterize the dynamics by their effect on the average total excitation number,
\begin{align}
\langle \hat{n}\rangle
&=\sum_{k=1}^N\left\langle \hat{q}_k^2+\hat{p}_k^2-\frac{1}{2}\right\rangle
=\text{Tr}(\sigma/2+\bm{X}\bm{X}^\intercal)-N/2.
\end{align}
Note that the trace in the above equation is over the system's phase space. The rate of change of the expected particle number can be computed from \eqref{GeneralDiffXUp} and \eqref{GeneralDiffVUp} as,
\begin{align}\label{Activity}
\frac{\dd}{\dd t}\langle \hat{n}\rangle
&=\text{Tr}\Big(
\Omega A \big(\sigma/2+\bm{X}\bm{X}^\intercal\big)
+\big(\sigma/2+\bm{X}\bm{X}^\intercal
\big)(\Omega A)^\intercal
+\bm{X}(\Omega\bm{b})^\intercal
+\Omega\bm{b}\bm{X}^\intercal
+C\Big),\\
&=\text{Tr}\Big(\big(\Omega A+(\Omega A)^\intercal\big)\big(\sigma/2+\bm{X}\bm{X}^\intercal\big)\Big)
+2\bm{X}^\intercal\Omega\bm{b}
+\text{Tr}\big(C\big).
\end{align}
We can define passive dynamics as that which preserves the average excitation number. Likewise we can define active dynamics as that which is capable of changing the average excitation number. Note that any change in the average excitation number must be attributed to at least one of the following conditions: either $\Omega A+(\Omega A)^\intercal\neq0$, or $\bm{b}\neq0$, or $\text{Tr}(C)\neq0$. Thus, $\bm{b}$, the ``traceful'' part\footnote{There are many ways to divide a matrix into a part that contributes to the trace and a part that  does not. One cannot in general uniquely specify which part of a part of a matrix is ``traceful''. However, as discussed in \cite{Dan2018}, in the context that considers all four partitions there is natural way to do this.} of $C$, and the part of $A$ which is symmetric when multiplied by $\Omega$ on the left,
\begin{align}\label{AaDef}
A_\text{A}
&\coloneqq \frac{1}{2}\Omega^{-1}\big(\Omega A+(\Omega A)^\intercal\big)
=\frac{1}{2}\big(A+\Omega^{-1} A^\intercal \Omega^\intercal\big),
\end{align}
all generate active dynamics. 

Similarly one can see that the ``traceless'' part of $C$ and the part of $A$ which is anti-symmetric when multiplied by $\Omega$ on the left,
\begin{align}\label{ApDef}
A_\textsc{p}
&\coloneqq \frac{1}{2}\Omega^{-1}\big(\Omega A-(\Omega A)^\intercal\big)
=\frac{1}{2}\big(A-\Omega^{-1} A^\intercal \Omega^\intercal \big),
\end{align}
are passive.

Note that the $C$ term will always be in total active: since $C\geq0$ is positive semi-definite: $C\neq0$ implies $\text{Tr}(C)>0$. However, different parts of $C$ can be considered either active or passive depending on their trace. At the moment (and we will revisit this later) there is not a natural way to decompose $C$ into a ``traceful'' and ``traceless'' part.

\subsubsection{State-Dependent vs State-Independent}
We can further classify dynamics by whether it enters \eqref{GeneralDiffXUp} and \eqref{GeneralDiffVUp} via a linear or an affine term. The linear terms ($\Omega A \bm{X}(t)$ and $\Omega A \,  \sigma(t)$ and $\sigma(t) (\Omega A)^\intercal$) depend on the system's current state ($\bm{X}(t)$ and $\sigma(t)$). Contrast this with the affine terms ($\Omega\bm{b}$ and $C$) which are independent of the system's current state. Thus we can identify $A$ as generating state dependent dynamics and $\bm{b}$ and $C$ as generating state-independent dynamics.

\subsubsection{Single-Mode vs. Multi-Mode}
Finally we can classify the dynamics as either acting on one or several modes. The $n^{th}$ mode is characterized by the pair of quadrature operators $\hat{q}_n$ and $\hat{p}_n$. Due to the operator ordering we chose in \eqref{XhatDef} these canonical pairs of observables are adjacent. Thus, dividing $A$ and $C$ into $2$ by $2$ blocks, we can identify dynamics as being either single-mode or multi-mode depending on whether they are block on- or off-diagonal respectively. Since $\bm{b}$ can be decomposed into a sum of terms each acting within a sector without mixing, it can be identified as entirely single-mode.

In \cite{Dan2018} the above four binary partitions were carried out simultaneously. For instance, it was identified exactly which part of the dynamics was simultaneously 1) symplectic and 2) active and 3) state-dependent and 4) single-mode. It was found that the only dynamics with all four of these properties is a sort of single-mode squeezing. Similarly the 15 other combinations of these four properties were investigated and named in \cite{Dan2018}.

The 4-way characterization carried out in \cite{Dan2018} was quite involved and I merely summarize the results here. To outline the procedure, $A$ and $C$ are first divided into $2\times2$ blocks. As discussed above these on-diagonal blocks correspond to single-mode dynamics and the off-diagonal blocks correspond to multi-mode dynamics. Each of these $2\times2$ blocks are then expanded in terms of the basis:
\bel{2by2basis}
\openone_2
=\begin{pmatrix}
1 & 0 \\
0 & 1 \\
\end{pmatrix},
\,
\omega=\begin{pmatrix}
0 & 1 \\
-1 & 0 \\
\end{pmatrix},
\, 
X=\begin{pmatrix}
0 & 1 \\
1 & 0 \\
\end{pmatrix},
\,
Z=\begin{pmatrix}
1 & 0 \\
0 & -1 \\
\end{pmatrix}.
\ee
The result of these expansions are
\begin{align}
A&=A_I \otimes\openone_2
+A_w \otimes\omega
+A_x \otimes X
+A_z \otimes Z\\
C&=C_I \otimes\openone_2
+C_w \otimes\omega
+C_x \otimes X
+C_z \otimes Z
\end{align}
for some $N\times N$ matrices $A_\mu$ and $C_\mu$ for $\mu\in\{I,w,x,z\}$. Applying the four partitions to the above expansion allows us to identify the various parts of the dynamics. For instance, the 1) symplectic and 2) active and 3) state-dependent and 4) single-mode dynamics (single mode squeezing) can be identified with the diagonal-entries of $A_x$ and $A_z$. Table \ref{SUAPTable} shows the 
results of the partition.

\begin{table}[h]
\centering
\begin{tabular}{|r|c|c|c|c|} 
\hline
& \multicolumn{2}{c|}{\bf Active} & \multicolumn{2}{c|}{\bf Passive}\\ 
\hline
{\bf Symplectic} 
& \ $A_\textsc{sa}$  ${}^{(s/m)}$
& \ $\bm{b}$ \ \ \, ${}^{(s/\,\ )}$
& \ $A_\textsc{sp}$  ${}^{(s/m)}$
& \quad\quad\quad \\
\hline
{ \bf Unsymplectic} 
& \ $A_\textsc{ua}$  ${}^{(s/m)}$
& \ $C_\textsc{ua}$  ${}^{(s/\,\ )}$
& \ $A_\textsc{up}$  ${}^{(\,\ /m)}$
& \ $C_\textsc{up}$  ${}^{(s/m)}$ \\ 
\hline
\quad  
& \quad \bf S.D. \quad
& \quad \bf S.I. \quad
& \quad \bf S.D. \quad
& \quad \bf S.I. \quad\\ 
\hline
\end{tabular}
\caption{The result of the partition described above. Note that the horizontal division within each cell indicates state dependence (S.D.) or independence (S.I.). The parenthetical note indicates if the dynamics can be either single-mode (s) or multi-mode (m). Note that there is no symplectic, passive, and state-independent dynamics, either single- or multi-mode. \textit{Reproduced with permission from} \cite{Dan2018}.}
\label{SUAPTable}
\end{table}
Note that in the above table for organizational convenience we define the following labels: symplectic passive (SP), symplectic active (SA), unsymplectic active (UA), and unsymplectic passive (UP). The distinction between state-dependent and state-independent dynamics ($A$ versus $\bm{b}$ or $C$) is obvious and is thus left unlabeled. The distinction between single-mode and multi-mode dynamics is also left unlabeled although it is indicated in the Table \ref{SUAPTable}.

One may note that the above table is not full, there appear to be missing types of dynamics. For instance there is no symplectic, passive, and state-independent dynamics, either single- or multi-mode. This phenomena is discussed in \cite{Dan2018} in detail. In total there are only 11 types of dynamics for Gaussian systems. Each of these dynamics is given in Table \ref{Table2} according to its effect on an arbitrary Gaussian state (see \cite{Dan2018}).

\afterpage{%
    %\clearpage% Flush earlier floats (otherwise order might not be correct)
    %\thispagestyle{empty}% empty page style (?)
    \begin{sidewaysfigure}% Landscape page
        \centering % Center table
        \begin{tabular}{||cccc||c||}
        \hline 
        \quad Single-mode? \quad & \quad Symplectic? \quad  &  \quad Passive? \quad  &  \quad State-Dependent? \quad  &            \\
        \quad (else Multi-mode) \quad  &  \quad (else Unsymplectic) \quad  &  \quad (else Active) \quad  &  \quad (else Independent) \quad  &       Name \\
        \hline Yes &        Yes &        Yes &        Yes & Single-mode Rotation \\
        \hline Yes &        Yes &        Yes &         No &        Not Possible \\
        \hline Yes &        Yes &         No &        Yes & Single-mode Squeezing \\
        \hline Yes &        Yes &         No &         No & Displacement \\
        \hline Yes &         No &        Yes &        Yes &        Not Possible \\
        \hline Yes &         No &        Yes &         No & Single-mode Squeezed Noise \\
        \hline Yes &         No &         No &        Yes & Amplification/Relaxation \\
        \hline Yes &         No &         No &         No & Free Thermal Noise \\
        \hline No &        Yes &        Yes &        Yes & Multi-mode Rotation \\
        \hline No &        Yes &        Yes &         No &        Not Possible \\
        \hline No &        Yes &         No &        Yes & Multi-mode Squeezing \\
        \hline No &        Yes &         No &         No &        Not Possible \\
        \hline No &         No &        Yes &        Yes & Multi-mode Counter-Rotation \\
         \hline No &         No &        Yes &         No & Multi-mode Squeezed Noise \\
         \hline No &         No &         No &        Yes & Multi-mode Counter-Squeezing \\
         \hline No &         No &         No &         No &        Not Possible \\
        \hline
        \end{tabular}
        \captionof{table}{The result of our division. Eleven of the possible sixteen types of dynamics are logically possible. These are named in this table. See Sec.IV. of \cite{Dan2018} for phase space plots of each type of dynamics. \textit{Reproduced with permission from} \cite{Dan2018}.}\label{Table2}
    \end{sidewaysfigure}
    %\clearpage% Flush page
}

Another important thing to note about the above partition is that many of the above-named dynamics are not completely positive in isolation. For instance if we want to isolate the dynamics which are single-mode, unsymplectic, active and state dependent (amplification/relaxation) then forced to have $A=\gamma\,\omega$, $\bm{b}=0$ and $C=0$ for some $\gamma\in\mathbb{R}$. The master equation for the this dynamics is,
\begin{align}
\frac{\dd}{\dd t}\bm{X}(t)
&=-\gamma\bm{X}(t)\\
\frac{\dd}{\dd t}\sigma(t)
&=-2\gamma\,\sigma(t).
\end{align}
Under this dynamics the covariance matrix converges exponentially to $\sigma=0$, a state with no uncertainty. This is not a valid Gaussian state as it violates the uncertainty principle. Recall that any dynamics which is completely positive (satisfying \eqref{DiffCPCond}) will always produce valid Gaussian states from valid initial states. This dynamics must not be completely positive, indeed it does not satisfy \eqref{DiffCPCond}.

In \cite{Dan2018} an analysis of the complete positivity of each of the 16 possible types of dynamics was conducted. There it was shown that all of the symplectic types of dynamics were completely positive in isolation. Of the unsymplectic dynamics, only free thermal noise (unsymplectic, active, state independent, single mode, see Table \ref{Table2}) is completely positive in isolation. All other types of unsymplectic dynamics require a certain amount of free thermal noise in order to be completely positive.

Given our discussion of thermalization and purification for finite dimensional systems in Chapter \ref{Ch4}, it is interesting to ask what dynamics can purify Gaussian systems (an example of an infinite dimensional system). The condition for purification given in Chapter \ref{Ch4} (that the dynamics be non-unital) only applies for finite dimensional systems. For Gaussian systems we need a new condition.

The purity of a Gaussian state is given (in our notation, \eqref{Vdef2}) by \cite{GPurity},
\be
\mathcal{P}=\text{Tr}(\hat{\rho}^2)
=\frac{1}{\text{det}(\sigma)}.
\ee
Recall that the eigenvalues of the covariance matrix, $\sigma$, determine the length squared of the 1-deviation hyper-ellipse in each direction in phase space. Thus the ``volume'' of the Gaussian state (the measure of the points inside this 1-deviation hyper-ellipse) is,
\be
\text{Vol}(\bm{X},\sigma)
\coloneqq
\sqrt{\text{det}(\sigma)}.
\ee
That is, the purity is the inverse of the state's ``volume'' squared.

The time-derivative of the determinant of $\sigma$ is worked out in \cite{Dan2018} as,
\begin{align}
\frac{\dd}{\dd t}\text{det}\big(\sigma(t)\big)
&=\text{det}\big(\sigma(t)\big) \, \text{Tr}\big(2 \, \Omega A+\sigma^{-1}(t) \, C\big).
\end{align}
Note that $\sigma\geq0$ implies that $\text{det}(\sigma)\geq0$. Thus the condition that the dynamics preserves the purity/volume of all states is thus $\text{Tr}\big(\, \Omega A\big)=0$ and $C=0$. Thus a condition that the dynamics increases the purity of at least one state is that there exists a valid state $\sigma$ such that,
\bel{GaussianPurificationInequality}
\text{Tr}\big(\Omega A\big)
<-\frac{1}{2}\text{Tr}\big(\sigma^{-1} \, C\big)
\leq0.
\ee
By taking $\sigma$ to be arbitrarily hot/mixed ($\sigma=\nu\openone$ as $\nu\to\infty$) we can satisfy this inequality as long as $\text{Tr}\big(\Omega A\big)<0$. That is, if $\text{Tr}\big(\Omega A\big)<0$ then a sufficiently hot state will be purified by the dynamics. In summary, $\text{Tr}\big(\Omega A\big)<0$ is a necessary and sufficient condition for differential Gaussian dynamics to increase the purity of at least one state. Within the partition outlined above the only Gaussian dynamics with $\text{Tr}\big(\Omega A\big)\neq0$ is amplification/purification.

Now that we have introduced Gaussian Quantum Mechanics we can apply the Interpolated Collision Model Formalism developed in this thesis. In particular in the next chapter we will write derive the interpolation generator from the quadratic Hamiltonian \eqref{HSADef1}. We will then expand the interpolation generator as a series in $\delta t$ and establish exactly how each of the terms is built from the various parts of the Hamiltonian. Comparing this expansion of the interpolation generator with the above outlined partition we will be able to determine exactly which Hamiltonians correspond to which dynamics at which orders in $\delta t$.

\chapter{Gaussian Interpolated Collision Model Formalism}\label{InterpolateGQM}
Now that we have reviewed Gaussian Quantum Mechanics and established some basic facts about the sort of master equations that Gaussian systems can obey, we are ready to apply the interpolated collision model formalism.

\section{Constructing the Interpolation Schemes}\label{AppGQMInterpolate}
In this section we will consider a Gaussian system composed of $N$ modes and characterized by its mean vector, $\bm{X}$, and its covariance matrix, $\sigma$, and with a symplectic form $\Omega$. Beginning from an initial state $\bm{X}(0)$ and $\sigma(0)$ the system will be repeatedly updated by a Gaussian channel as
\begin{align}
\label{UpdateSchemeXX}
\bm{X}((n+1)\delta t)
&=T(\delta t) \, \bm{X}(n \, \delta t)
+\bm{d}(\delta t),\\
\label{UpdateSchemeVV}
\sigma((n+1)\delta t)
&=T(\delta t) \, \sigma(n \, \delta t) \, T(\delta t)^\intercal
+R(\delta t).
\end{align}
for some real-valued $2N\times2N$ matrices $T(\delta t)$ and $R(\delta t)$ and a real valued vector $\bm{d}(\delta t)$ with $R(\delta t)$ symmetric. This update scheme defines the system state at every discrete time points $t=n\,\delta t$.

For example, this update map could come from the system interacting with another Gaussian system as described in Sec. \ref{GaussianChannel}. We will consider this case in detail in Sec. \ref{AncillaryBombardmentGQM}. For now let us continue with a generic update map which is completely positive, that is satisfying,
\bel{FiniteCPCond2}
R(\delta t)\geq\ii \, (T(\delta t) \, \Omega \,  T(\delta t)^\intercal-\Omega).
\ee

Our goal will be to construct a Gaussian master equation of the general form 
\begin{align}
\label{GQMInterpMasterEqsX}
\bm{X}'(t)
&=\Omega(A_{\delta t} \, \bm{X}(t)
+\bm{b}_{\delta t}),\\
\label{GQMInterpMasterEqsV}
\sigma'(t)
&=(\Omega A_{\delta t}) \, \sigma(t)
+\sigma(t) \, (\Omega A_{\delta t})^\intercal
+C_{\delta t},
\end{align}
for some generators $A_{\delta t}$, $\bm{b}_{\delta t}$, and $C_{\delta t}$ which exactly matches the evolution given by the above discrete dynamics. To do this we will adapt the interpolated collision model formalism described in Chapter \ref{Ch2}. 

The main complication in attempting to apply this methods of Chapter \ref{Ch2} directly is that 1) the state is given by a vector $\bm{X}$ and a matrix, $\sigma$, instead of a single vector and 2) the update equations in this case are linear-affine instead of just being linear. These complications can be overcome with the following two linear isomorphisms.

First we use the vectorization map, $\vec$, which maps outer products to tensor products as
\bel{OuterToTensor}
\vec(\lambda \ \bm{u}\bm{v}^\intercal)
\coloneqq\lambda \ \bm{u}\otimes\bm{v}
\ee
for some scalar $\lambda$ and real vectors $\bm{u}$ and $\bm{v}$. By linearity this defines its action on any matrix. From this it follows that for any matrices $X$, $Y$ and $Z$,
\bel{VecIdentity}
\vec(X \, Y \, Z^\intercal)=(X\otimes Z)\,\vec(Y).
\ee
This operation can be represented by the vector formed by taking the entries of a matrix in order as follows,
\be
\vec\begin{pmatrix}
a & b \\
c & d
\end{pmatrix}
=(a,b,c,d)^\intercal.
\ee
Note that $\text{vec}^{-1}$ is trivially defined by ``restacking'' the matrices entries. Using this map we can turn $\bm{X}$ and $\sigma$ into one big vector as
\begin{align}
\{\bm{X},\sigma\} 
\Longleftrightarrow
(\bm{X},\text{vec}(\sigma))^\intercal
\end{align}

The second isomorphism embeds the state in an affine space (a hyperplane offset from the origin) as,
\begin{align}
(\bm{X},\text{vec}(\sigma))^\intercal
\Longleftrightarrow
(1,\bm{X},\text{vec}(\sigma))^\intercal.
\end{align}
This has the effect of converting our linear affine update equations to merely linear ones. This is completely analogous to the Bloch example worked out is Sec. \ref{BlochExample}.

Applying these linear isomorphisms, applying the methods of Chapter \ref{Ch2}, and then reversing the isomorphisms is rather involved and is carried out in Appendix A of \cite{PhysRevA.97.052120}. The result is the interpolation generators,
\begin{align}
\label{AdtDef}
\Omega A_{\delta t}
&=\frac{1}{\delta t}\text{Log}(T(\delta t)),\\
\label{bdtDef}
\Omega \, \bm{b}_{\delta t}
&=\frac{1}{\delta t}
\frac{\text{Log}(T(\delta t))}{T(\delta t)-\openone_{2N}}\bm{d}(\delta t),\\
\label{CdtDef}
C_{\delta t}
&=\vec^{-1}\Big(\frac{1}{\delta t}\frac{\text{Log}(T(\delta t) \otimes T(\delta t))}{T(\delta t) \otimes T(\delta t)-\openone_{4N^2}} \, \vec\big(R(\delta t)\big)\Big).
\end{align}
It is worth noting that the expressions for $\Omega\bm{b}_{\delta t}$, and $C_{\delta t}$ are to be understood via the series expansion,
\bel{LogSeries2}
\frac{\text{Log}(X)}{X-\openone}
=\sum_{m=0}^\infty\frac{(-1)^m}{m+1}(X-\openone)^m.
\ee
Understood this way \mbox{$T(\delta t)-\openone_{2N}$} and \mbox{$T(\delta t)\, \otimes \, T(\delta t)-\openone_{4N^2}$} do not need to be invertible. Finally, we should note that, as in Chapter \ref{Ch2}, we take the logarithm's principal branch cut, such that $\text{Log}(\openone)=0$. This along with the assumptions 
\bel{NothingNoTime}
T(0)=\openone_{2N}, \ \ \  \bm{d}(0)=0, \ \ \ \text{and} \ \ \ R(0)=0
\ee
(nothing happens in no time) and that,
\bel{FiniteRate}
T'(0), \ \ \  \bm{d}'(0), \ \ \ \text{and} \ \ \ R'(0) \ \ \ \text{exist}
\ee
(things happen at a finite rate) ensures that the interpolation generators converge as $\delta t\to0$.

If in addition to the minimal regularity assumed above --- \eqref{NothingNoTime} and \eqref{FiniteRate} --- we have that $T(\delta t)$, $\bm{d}(\delta t)$, and $R(\delta t)$ are analytic at $\delta t=0$, then we can then expand them as a series in $\delta t$ as,
\begin{align}
\label{TSeries}
T(\delta t)
&=\openone_{2N}
+\delta t \, T_1
+\delta t^2 \, T_2
+\delta t^3 \, T_3
+\delta t^4 \, T_4
+\dots,\\
\label{dSeries}
\bm{d}(\delta t)
&=0
+\delta t \, \bm{d}_1
+\delta t^2 \, \bm{d}_2
+\delta t^3 \, \bm{d}_3
+\delta t^4 \, \bm{d}_4
+\dots,\\
\label{RSeries}
R(\delta t)
&=0
+\delta t \, R_1
+\delta t^2 \, R_2
+\delta t^3 \, R_3
+\delta t^4 \, R_4
+\dots \, .
\end{align}
Using these series expansions, through \eqref{AdtDef}, \eqref{bdtDef}, and \eqref{CdtDef}, we can expand each interpolation generator as a series in $\delta t$ as well,
\begin{align}
\label{ASeries}
A_{\delta t}
&=A_0
+\delta t \, A_1
+\delta t^2 \, A_2
+\delta t^3 \, A_3
+\dots,\\
\label{bSeries}
\bm{b}_{\delta t}
&=\bm{b}_0
+\delta t \, \bm{b}_1
+\delta t^2 \, \bm{b}_2
+\delta t^3 \, \bm{b}_3
+\dots,\\
\label{CSeries}
C_{\delta t}
&=C_0
+\delta t \, C_1
+\delta t^2 \, C_2
+\delta t^3 \, C_3
+\dots,
\end{align}
where the first few terms of the expansion of $A_{\delta t}$ are given by
\begin{align}
\label{A0def}
\Omega A_0=T_1,&\\
\label{A1def}
\Omega A_1=T_2
&-\frac{1}{2}T_1{}^2,\\
\label{A2def}
\Omega A_2=T_3
&-\frac{1}{2}(T_1 T_2+T_2 T_1)
+\frac{1}{3}T_1{}^3.
%\label{A3def}
%\Omega A_3=T_4
%&-\frac{1}{2}(T_1 T_3+T_2 T_2+T_3 T_1)\\
%\nonumber
%&+\frac{1}{3}(T_1{}^2 T_2+T_2 T_1{}^2+T_1 T_2 T_1)
%-\frac{1}{4}T_1{}^4.
\end{align}
The first few terms of the expansion of $\bm{b}_{\delta t}$ are given by,
\begin{align}
\Omega \, \bm{b}_0
=\bm{d}_1&,\\
\Omega \, \bm{b}_1
=\bm{d}_2
&-\frac{1}{2}T_1\bm{d}_1,\\
\Omega \, \bm{b}_2
=\bm{d}_3
&-\frac{1}{2}(T_1\bm{d}_2+T_2\bm{d}_1)
+\frac{1}{3}T_1^2\bm{d}_1.
%\Omega \, \bm{b}_3
%=\bm{d}_4
%&-\frac{1}{2}(T_1\bm{d}_3+T_2\bm{d}_2+T_3\bm{d}_1)\\
%\nonumber
%&+\frac{1}{3}(T_1^2\bm{d}_2+T_2 T_1\bm{d}_1+T_1 T_2\bm{d}_1)
%-\frac{1}{4}T_1^3\bm{d}_1.
\end{align}
Finally, the first few terms of the expansion of $C_{\delta t}$ are given by,
\begin{align}
C_0&=R_1,\\
C_1&=R_2 
-\frac{1}{2}(T_1 R_1+R_1 T_1^\intercal),\\
C_2&=R_3 
-\frac{1}{2}(T_2 R_1+R_1 T_2^\intercal+T_1 R_2+R_2 T_1^\intercal)
+\frac{1}{3}(T_1{}^2 R_1+R_1 T_1^\intercal{}^2)
+\frac{1}{6} T_1 R_1 T_1^\intercal.
%C_3&=R_4\\
%\nonumber
%&-\frac{1}{2}
%(T_3 R_1+R_1 T_3^\intercal
%+T_2 R_2+R_2 T_2^\intercal
%+T_1 R_3+R_3 T_1^\intercal)\\
%\nonumber
%&+\frac{1}{3}
%(T_1 T_2 R_1+R_1 T_2^\intercal T_1^\intercal
%+T_2 T_1 R_1+R_1 T_1^\intercal T_2^\intercal\\
%\nonumber
%&+T_1^2 R_2+R_2 T_1^\intercal{}^2)
%+\frac{1}{6}
%(T_1 R_1 T_2^\intercal
%+T_2 R_1 T_1^\intercal
%+T_1 R_2 T_1^\intercal)\\
%\nonumber
%&-\frac{1}{4}
%(T_1^3 R_1+R_1 T_1^\intercal{}^3)
%-\frac{1}{12}
%(T_1^2 R_1 T_1^\intercal
%+T_1 R_1 T_1^\intercal{}^2).    
\end{align}
Higher order terms in these series can be calculated easily.

\section{Gaussian Ancillary Bombardment}\label{AncillaryBombardmentGQM}

In this section we construct the Gaussian channel corresponding to a specific physically motivated situation that we refer to as \textit{Gaussian ancillary bombardment}, in analogy with the ancillary bombardment introduced in Sec. \ref{AncillaryBombardmentExample}. Much of the groundwork for this scenario has already been covered in the thesis in Sec. \ref{GaussianChannel}. For the reader's convenience we restate the results of that section here. 

Let us assume that the system and ancilla are initially uncorrelated, having the initial joint mean vector,
\be
\bm{X}_\text{SA}(0)
=(\bm{X}_\text{S}(0)^\intercal,\bm{X}_\text{A}(0)^\intercal)^\intercal
=\bm{X}_\text{S}(0)\oplus\bm{X}_\text{A}(0)
\ee
and the initial joint covariance matrix,
\be
\sigma_\text{SA}(0)=
\begin{pmatrix}
\sigma_\text{S}(0) & 0\\
0 & \sigma_\text{A}(0)
\end{pmatrix}
=\sigma_\text{S}(0)\oplus
\sigma_\text{A}(0).
\ee
Further we assume that they evolve under a quadratic time-independent Hamiltonian,
\bel{HSADef2}
\hat{H}_\text{SA}
=\frac{1}{2}\hat{\bm{X}}_\text{SA}^\intercal  \, F_\text{SA} \, \hat{\bm{X}}_\text{SA}
+\bm{\alpha}^\intercal_\text{SA}\hat{\bm{X}}_\text{SA},
\ee
where $F_\text{SA}$ is real and symmetric and $\bm{\alpha}_\text{SA}$ is real. Dividing this Hamiltonian into blocks as,
\be
F_\text{SA}=
\begin{pmatrix}
F_\text{S} & G\\
G^\intercal & F_\text{A}
\end{pmatrix},
\quad \quad
\bm{\alpha}_\text{SA}=
\begin{pmatrix}
\bm{\alpha}_\text{S}\\
\bm{\alpha}_\text{A}
\end{pmatrix},
\ee
we can identify the system's free Hamiltonian,
\be
\hat{H}_\text{S}
=\frac{1}{2}\hat{\bm{X}}_\text{S}^\intercal  \, F_\text{S} \, \hat{\bm{X}}_\text{S}
+\bm{\alpha}^\intercal_\text{S}\hat{\bm{X}}_\text{S},
\ee
and the ancilla's free Hamiltonian,
\be
\hat{H}_\text{A}
=\frac{1}{2}\hat{\bm{X}}_\text{A}^\intercal  \, F_\text{A} \, \hat{\bm{X}}_\text{A}
+\bm{\alpha}^\intercal_\text{A}\hat{\bm{X}}_\text{A}.
\ee
Note that $F_\text{S}$ and $F_\text{A}$ are symmetric. The interaction Hamiltonian between the system and ancilla is given by, 
\be
\hat{H}_\text{I}
=\frac{1}{2}\hat{\bm{X}}_\text{S}^\intercal  \, G \, \hat{\bm{X}}_\text{A}
+\frac{1}{2}\hat{\bm{X}}_\text{A}^\intercal  \, G^ \intercal\, \hat{\bm{X}}_\text{S}.
\ee

The joint system evolving under these Hamiltonians undergoes a symplectic transformation given by, 
\begin{align}
\label{SXUpdt2}
\bm{X}_\text{SA}(\delta t)
&=S_\text{SA}(\delta t) \, \bm{X}_\text{SA}(0)+\bm{d}_\text{SA}(\delta t),\\
\label{SVUpdt2}
\sigma_\text{SA}(\delta t)
&=S_\text{SA}(\delta t) \, \sigma_\text{SA}(0) \, S^\intercal_\text{SA}(\delta t),
\end{align}
where,
\begin{align}
\label{AppSHamDef2}
S_\text{SA}(\delta t)&=\text{exp}(\Omega_\text{SA} F_\text{SA} \, \delta t),\\
\label{AppdHamDef2}
\bm{d}_\text{SA}(\delta t)&=\frac{\text{exp}(\Omega_\text{SA} F_\text{SA} \, \delta t)-\openone_{2 N_\text{S}+2 N_\text{A}}}{\Omega_\text{SA} F_\text{SA}} \, \Omega_\text{SA} \, \bm{\alpha}_\text{SA}.
\end{align}
To find the effect of this transformation on the system's reduced state we divide $S_\text{SA}(\delta t)$ and $\bm{d}_\text{SA}(\delta t)$ into blocks as,
\be
S_\text{SA}(\delta t)
=\begin{pmatrix}
M_\text{SS}(\delta t) & M_\text{SA}(\delta t) \\
M_\text{AS}(\delta t) & M_\text{AA}(\delta t) \\
\end{pmatrix}
\ \ \text{and} \ \ 
\bm{d}_\text{SA}(\delta t)
=\begin{pmatrix}
\bm{d}_\text{S}(\delta t) \\ \bm{d}_\text{A}(\delta t) \\
\end{pmatrix}.
\ee
From this we can identify $T(\delta t)$, $\bm{d}(\delta t)$ and $R(\delta t)$ as,
\begin{align}\label{TSMdSDef2}
T(\delta t)&=M_\text{SS}(\delta t),\\
\bm{d}(\delta t)&=M_\text{SA}(\delta t) \ \bm{X}_\text{A}(0)+\bm{d}_\text{S}(\delta t),\\
R(\delta t)&=M_\text{SA}(\delta t) \, \sigma_\text{A}(0) \, M^\intercal_\text{SA}(\delta t).
\end{align}

With some effort, these can be expanded as a series in $\delta t$ (as in \eqref{TSeries}, \eqref{dSeries}, and \eqref{RSeries}) but now with coefficients ($T_k$, $\bm{d}_k$, and $R_k$) constructed from the Hamiltonians (i.e., $F_\text{S}$, $F_\text{A}$, and $G$). Using the results of the previous section, we can then write the interpolation generators $A_{\delta t}$, $\bm{b}_{\delta t}$, and $C_{\delta t}$ as a series in $\delta t$ (as in \eqref{ASeries}, \eqref{bSeries}, and \eqref{CSeries}) but now with coefficients ($A_k$, $\bm{b}_k$, and $C_k$) written explicitly in terms of $F_\text{S}$, $F_\text{A}$, and $G$.

This calculation is quite involved but is ultimately straightforward. For the first few terms of the expansion of $A_{\delta t}$ it yields,
\begin{align}
A_0=&F_\text{S},\\
\label{A1DefHam}
A_1=&\frac{1}{2}G \, \Omega_\text{A} G^\intercal,\\
A_2=&-\frac{1}{12} G \, \Omega_\text{A} G^\intercal \Omega_\text{S} F_\text{S}
-\frac{1}{12} F_\text{S} \Omega_\text{S} G \, \Omega_\text{A}  G^\intercal
+\frac{1}{6} G \, \Omega_\text{A} F_\text{A} \Omega_\text{A} G^\intercal.
%\nonumber
%A_3=&-\frac{1}{24} G \, \Omega_\text{A}  F_\text{A} \Omega_\text{A}  G^\intercal \Omega_\text{S} F_\text{S}
%-\frac{1}{24} F_\text{S}\Omega_\text{S} G \, \Omega_\text{A}  F_\text{A} \Omega_\text{A}  G^\intercal\\
%\nonumber
%&+\frac{1}{24} F_\text{S}\Omega_\text{S} G \, \Omega_\text{A}  G^\intercal \Omega_\text{S} F_\text{S}
%+\frac{1}{24} G \, \Omega_\text{A}  F_\text{A} \Omega_\text{A}  F_\text{A} \Omega_\text{A}  G^\intercal\\
%&-\frac{1}{12} G \, \Omega_\text{A}  G^\intercal \Omega_\text{S} G \, \Omega_\text{A}  G^\intercal.
\end{align}
For the first few terms of the expansion of $\bm{b}_{\delta t}$ we find,
\begin{align}
\bm{b}_0&=
\bm{\alpha}_\text{S}
+G\bm{X}_\text{A}(0),\\
\bm{b}_1&=
\frac{1}{2} G \, \Omega_\text{A} F_\text{A}\bm{X}_\text{A}(0)
+\frac{1}{2} G \, \Omega_\text{A}\bm{\alpha}_\text{A},\\ 
\bm{b}_2&=-\frac{1}{12} F_\text{S} \Omega_\text{S} G \, \Omega_\text{A} \bm{\alpha}_\text{A}
+\frac{1}{6} \Omega_\text{S} G \, \Omega_\text{A} F_\text{A} \Omega_\text{A} \bm{\alpha}_\text{A}-\frac{1}{12} F_\text{S} \Omega_\text{S} G \, \Omega_\text{A} F_\text{A} \bm{X}_\text{A}(0)\\
\nonumber
&+\frac{1}{6} G \, \Omega_\text{A} F_\text{A} \Omega_\text{A} F_\text{A} \bm{X}_\text{A}(0)
-\frac{1}{12} G \, \Omega_\text{A} G^\intercal \Omega_\text{S} \bm{\alpha}_\text{S}
-\frac{1}{12} G \, \Omega_\text{A} G^\intercal \Omega_\text{S} G \bm{X}_\text{A}(0).
%\bm{b}_3&=\frac{1}{24} G \, \Omega_\text{A} F_\text{A} \Omega_\text{A} F_\text{A} \Omega_\text{A} \bm{\alpha}_\text{A}
%-\frac{1}{24} F_\text{S}\Omega_\text{S} G \, \Omega_\text{A} F_\text{A} \Omega_\text{A} \bm{\alpha}_\text{A}\\
%\nonumber
%&-\frac{1}{24} G \, \Omega_\text{A} F_\text{A} \Omega_\text{A} G^\intercal \Omega_\text{S} \bm{\alpha}_\text{S}
%+\frac{1}{24} F_\text{S}\Omega_\text{S} G \, \Omega_\text{A} G^\intercal \Omega_\text{S} \bm{\alpha}_\text{S}\\
%\nonumber
%&-\frac{1}{24} G \, \Omega_\text{A} F_\text{A} \Omega_\text{A} G^\intercal \Omega_\text{S} G \bm{X}_\text{A}(0)
%-\frac{1}{12} G \, \Omega_\text{A} G^\intercal \Omega_\text{S} G \, \Omega_\text{A} F_\text{A} \bm{X}_\text{A}(0)\\
%\nonumber
%&+\frac{1}{24} F_\text{S}\Omega_\text{S} G \, \Omega_\text{A} G^\intercal \Omega_\text{S} G \bm{X}_\text{A}(0)
%-\frac{1}{24} F_\text{S}\Omega_\text{S} G \, \Omega_\text{A} F_\text{A} \Omega_\text{A} F_\text{A} \bm{X}_\text{A}(0)\\
%\nonumber
%&+\frac{1}{24} G \, \Omega_\text{A} F_\text{A} \Omega_\text{A} F_\text{A} \Omega_\text{A} F_\text{A} \bm{X}_\text{A}(0)
%-\frac{1}{12} G \, \Omega_\text{A} G^\intercal \Omega_\text{S} G \, \Omega_\text{A} \bm{\alpha}_\text{A}.
\end{align}
Finally, the first few terms of the expansion of $C_{\delta t}$ are,
\begin{align}
C_0&=0,\\
\label{C1DefHam}
C_1&=\Omega_\text{S} G \sigma_\text{A}(0) G^\intercal \Omega^\intercal_\text{S},\\
\label{C2DefHam}
C_2&=\frac{1}{2} \Omega_\text{S} G \big(\Omega_\text{A} F_\text{A} \sigma_\text{A}(0)+\sigma_\text{A}(0) (\Omega_\text{A} F_\text{A})^\intercal\big) G^\intercal \Omega^\intercal_\text{S}.
%C_3&=\frac{1}{6} \Omega_\text{S} G\sigma_\text{A}(0) F_\text{A} \Omega_\text{A}{}^\intercal F_\text{A} \Omega_\text{A}{}^\intercal G^\intercal \Omega_\text{S}{}^\intercal\\
%\nonumber
%&+\frac{1}{3} \Omega_\text{S} G \, \Omega_\text{A} F_\text{A} \sigma_\text{A}(0) F_\text{A} \Omega_\text{A}{}^\intercal G^\intercal \Omega_\text{S}{}^\intercal\\
%\nonumber
%&+\frac{1}{6} \Omega_\text{S} G \, \Omega_\text{A} F_\text{A} \Omega_\text{A} F_\text{A} \sigma_\text{A}(0) G^\intercal \Omega_\text{S}{}^\intercal\\
%\nonumber
%&-\frac{1}{12} \Omega_\text{S} G \Omega_\text{A} G^\intercal \Omega_\text{S} G\sigma_\text{A}(0) G^\intercal \Omega_\text{S}{}^\intercal\\
%\nonumber
%&-\frac{1}{12} \Omega_\text{S} G\sigma_\text{A}(0) G^\intercal \Omega_\text{S}{}^\intercal G \Omega_\text{A}{}^\intercal G^\intercal \Omega_\text{S}{}^\intercal\\
%\nonumber
%&-\frac{1}{12} \Omega_\text{S} G\sigma_\text{A}(0) F_\text{A} \Omega_\text{A}{}^\intercal G^\intercal \Omega_\text{S}{}^\intercal F_\text{S} \Omega_\text{S}{}^\intercal\\
%\nonumber
%&-\frac{1}{12} \Omega_\text{S} F_\text{S} \Omega_\text{S} G \Omega_\text{A} F_\text{A} \sigma_\text{A}(0) G^\intercal \Omega_\text{S}{}^\intercal\\
%\nonumber
%&-\frac{1}{12} \Omega_\text{S} G \, \Omega_\text{A} F_\text{A} \sigma_\text{A}(0) F_\text{A} \Omega_\text{A}{}^\intercal G^\intercal \Omega_\text{S}{}^\intercal\\
%\nonumber
%&-\frac{1}{12} \Omega_\text{S} F_\text{S} \Omega_\text{S} G\sigma_\text{A}(0) G^\intercal \Omega_\text{S}{}^\intercal F_\text{S} \Omega_\text{S}{}^\intercal.
\end{align}

Now that we have the interpolation generators expanded in terms of the system-ancilla Hamiltonian we can ask what type of dynamics shows up at what orders in $\delta t$ and under what Hamiltonians. Such a study was conducted in \cite{PhysRevA.97.052120} and is summarized in Table \ref{Table22}. We now will outline the results  presented in Table \ref{Table22} before discussing them in more detail below.

At zeroth order (in the continuum limit) the dynamics is given by the system's free Hamiltonian ($F_\text{S}$ and $\bm{\alpha}_\text{S}$) plus an additional induced displacement (coming from the $G \bm{X}_\text{A}(0)$ term in $\bm{b}_0$). At higher orders in $\delta t$ the rotation, squeezing and amplification effects (coming from $A$) that are available to the system alternate between symplectic and unsymplectic. That is, $A_k$ is alternatingly symmetric and anti-symmetric. Past zeroth order in $\delta t$ the dynamics will generically be able to access all types of displacement ($\bm{b}$) and noise ($C$) terms.

\afterpage{
    \begin{sidewaysfigure}
        \centering % Center table
        \begin{tabular}{||c|c|c|c|c|c|c||}
        \hline Type of Dynamics & \quad 0th (Free) \quad &  \quad 0th (Induced) \quad & \quad Odd ( $\geq$ 1st) \quad & \quad Even ( $\geq$ 2nd) \quad \\

        \hline Single-mode Rotation &        Yes & No &         No &        Yes \\

        \hline Single-mode Squeezing &      Yes &    No &         No &        Yes \\

        \hline Displacement & Yes &  Yes &        Yes &        Yes \\

        \hline Single-mode Squeezed Noise &   No &       No &        Yes &        Yes \\

        \hline Amplification/Relaxation &   No &       No &        Yes &         No \\

        \hline Thermal Noise &      No &    No &        Yes &        Yes \\

        \hline Multi Mode Rotation &   Yes &       No &         No &        Yes \\

        \hline Multi Mode Squeezing &   Yes &       No &         No &        Yes \\

        \hline Multi Mode Counter-Rotation &   No &       No &        Yes &         No \\

        \hline Multi Mode Noise &   No &       No &        Yes &        Yes \\

        \hline Multi Mode Counter-Squeezing &   No &       No &        Yes &         No \\
        \hline
        \end{tabular}
        \captionof{table}{The dynamics available to a bombarded Gaussian system at each order in $\delta t$. The eleven types of dynamics listed in this table are described in detail in \cite{Dan2018}. The zeroth order effects are further divided into those available through the system's free Hamiltonian and those which can be induced through the interaction. \textit{Reproduced with permission from} \cite{PhysRevA.97.052120}.
    }\label{Table22}
    \end{sidewaysfigure}
}

\section{Continuum Limit Dynamics}
The zeroth order dynamics (i.e, in the continuum limit, as $\delta t\to 0$) is unitary, since $A_0$ is symmetric and $C_0$ vanishes. At zeroth order we have the dynamics,
\begin{align}\label{GQMInterpMasterEqs}
\bm{X}_\text{S}'(t)
&=\Omega(F_\text{S} \, \bm{X}_\text{S}(t)
+\alpha_\text{S} +G \, \bm{X}_\text{S}(0)),\\
\sigma_\text{S}'(t)
&=(\Omega F_\text{S}) \, \sigma_\text{S}(t)
+\sigma_\text{S}(t) \, (\Omega F_\text{S})^\intercal.
\end{align}
Comparing this to \eqref{SymplecticDiffXUpHam} and \eqref{SymplecticDiffVUpHam} we can see that this is just evolution under the effective Hamiltonian,
\begin{align}\label{Heff0Gaussian}
\hat{H}_\text{eff}^{(0)}
&=\frac{1}{2}\hat{\bm{X}}_\text{S}^\intercal  \, F_\text{S} \, \hat{\bm{X}}_\text{S}
+\hat{\bm{X}}_\text{S}^\intercal(\bm{\alpha}_\text{S}
+ G\bm{X}_\text{A}(0))\\
\nonumber
&=\hat{H}_\text{S}+\hat{\bm{X}}_\text{S}{}^\intercal G\bm{X}_\text{A}(0).
\end{align}
This is in line with the general result discussed in Chapter \ref{Ch3} showing that (under some minor assumptions) all collision models produce unitary dynamics in the continuum limit. 

This Hamiltonian is the system's free Hamiltonian $\hat{H}_\text{S}$ plus a new induced Hamiltonian $\hat{\bm{X}}_\text{S}{}^\intercal G\bm{X}_\text{A}(0)$. Note that this induced Hamiltonian is only able to be linear in the system's quadrature operators $\hat{\bm{X}}_\text{S}$. This means that the induced dynamics by the rapid collisions can only displace the system state in phase space (not rotate it or squeeze it). It is for this reason that the zeroth order dynamics in Table \ref{Table22} is divided into free and induced dynamics; there may be rotation and squeezing dynamics at zeroth order, but only if these types of dynamics were already present in the system's free Hamiltonian. No new squeezing or rotations are possible at zeroth order.

\section{First Order Dynamics}\label{FirstOrderGaussian}
At first order, we can see a new displacement term (from $b_1$), the first noise in the dynamics (from $C_1$) and several types of unsymplectic rotations and squeezings (from $A_1$). 

At this order the dynamics coming from both $A_1$ and $C_1$ is non-unitary ($A_1$ is antisymmetric, and noise is always non-unitary), thus the only unitary effects at first order come from $\bm{b}_1$. These effects give a first order correction to the effective Hamiltonian,
\be
\hat{H}_\text{eff}
=\hat{H}_\text{eff}^{(0)}
+\delta t \, \hat{H}_\text{eff}^{(1)}
+\mathcal{O}(\delta t^2),
\ee
of,
\be
\hat{H}_\text{eff}^{(1)}
=\hat{\bm{X}}_\text{S}^\intercal \, \bm{b}_1
=\frac{1}{2}\hat{\bm{X}}^\intercal_\text{S} 
G \, \Omega_\text{A} 
\big(F_\text{A}\bm{X}_\text{A}(0)
+\bm{\alpha}_\text{A}\big).
\ee
This correction can be understood as accounting for the ancilla freely evolving during the interaction.

The first order noise term is given by,
\be
C_1=\Omega_\text{S} G \, \sigma_\text{A}(0) \, G^\intercal \Omega_\text{S}^\intercal,
\ee
which we note is positive semi-definite \mbox{($C_1\geq 0$)}, since \mbox{$\sigma_\text{A}(0)\geq0$}. This noise vanishes only if $G=0$ (there is no interaction) or if $\sigma_\text{A}(0)$ is singular (i.e., infinitely squeezed) and  $G^\intercal\Omega_\text{S}^\intercal$ maps entirely into the kernel of $\sigma_\text{A}(0)$. 

As discussed in the previous section, a necessary and sufficient condition for Gaussian dynamics to cause purification is $\text{Tr}(\Omega A)<0$. Since the zeroth order dynamics is unitary the first opportunity for purification is at first order. We have purification at first order if and only if,
\bel{FirstOrderPurifyNandS}
0>\text{Tr}\big(\Omega_\text{S} A_1\big)
= \frac{1}{2} \text{Tr}\big(\Omega_\text{S} \, G \,  \Omega_\text{A} \,  G^\intercal\big).
\ee

In Chapter \ref{Ch4} we were able to understand the first-order purification condition for finite dimensional (non-Gaussian) systems as ruling out tensor product interaction Hamiltonians. That is, we saw that interaction Hamiltonians of the form,
\be
H_\text{int}=\hat{Q}_\text{S}\otimes \hat{R}_\text{A},
\ee
will not purify at first order. Does this result apply for Gaussian systems as well?

In order for the above interaction Hamiltonian to be quadratic (bi-linear) in $\hat{\bm{X}}_\text{S}$ and $\hat{\bm{X}}_\text{A}$ then $\hat{Q}_\text{S}$ and $\hat{R}_\text{A}$ must each be linear in their respective quadrature operators as
\be
\hat{Q}_\text{S}
=\bm{u}^\intercal\hat{\bm{X}}_\text{S}
=\hat{\bm{X}}_\text{S}^\intercal\bm{u}
\quad\text{and}\quad
\hat{R}_\text{A}
=\bm{v}^\intercal\hat{\bm{X}}_\text{A}
=\hat{\bm{X}}_\text{A}^\intercal\bm{v},
\ee
for some real vectors $\bm{u}$ and $\bm{v}$. Thus we can write the interaction Hamiltonian as,
\be
\hat{H}_\text{int}
=\hat{Q}_\text{S}\otimes \hat{R}_\text{A}
=\frac{1}{2}\hat{\bm{X}}_\text{S}^\intercal  \, G \, \hat{\bm{X}}_\text{A}
+\frac{1}{2}\hat{\bm{X}}_\text{A}^\intercal  \, G^ \intercal\, \hat{\bm{X}}_\text{S},
\ee
with,
\be
G=\bm{u}\bm{v}^\intercal.
\ee
Thus, in Gaussian quantum mechanics, tensor product interaction Hamiltonians correspond to rank one interaction matrices. By linearity, an quadratic interaction Hamiltonian consisting of a sum of $k$ tensor products corresponds to an interaction matrix consisting of a sum of $k$ outer products.

From \eqref{FirstOrderPurifyNandS} we can quickly see that a rank one interaction cannot purify at leading order since,
\begin{align}
\text{Tr}\big(\Omega_\text{S} G \Omega_\text{A} G^\intercal\big)
&=\text{Tr}\big(\Omega_\text{S} \bm{u}\bm{v}^\intercal \Omega_\text{A} \bm{v}\bm{u}^\intercal\big)\\
\nonumber
&= \bm{u}^\intercal\Omega_\text{S} \bm{u} \ \bm{v}^\intercal \Omega_\text{A} \bm{v}\\
\nonumber
&=0,
\end{align}
since $\Omega_\text{S}$ and $\Omega_\text{A}$ are antisymmetric. Thus we can extended the result of Chapter \ref{Ch4} that tensor product interaction Hamiltonians cannot cause purification at leading order in $\delta t$ for Gaussian systems.

Finally we can ask if the thermalization results from Chapter \ref{Ch4} hold here. Do we only see dependence on the ancilla's free Hamiltonian (i.e., $F_\text{A}$ and $\alpha_\text{A}$) at second order in $\delta t$? At first glance it appears that we do see dependence on $F_\text{A}$ and $\alpha_\text{A}$ in $\bm{b}_1$. However if the ancilla is initially in a thermal state (or any stationary state with respects to its free dynamics) then these terms are zero. For a thermal ancilla state the first dependence on $F_\text{A}$ and $\alpha_\text{A}$ shows up in $A_2$, $b_2$ and $C_2$. As discussed in Chapter \ref{Ch4} this dependence is necessary for the system's dynamics to be sensitive to the ancilla's temperature. This is necessary for the dynamics to drive the system into thermal equillibrium with the ancillas.

\chapter{Conclusion}\label{Ch6}

In this thesis I have argued against the use of continuum limit master equations for collisional models as a model of open quantum systems. The continuum limit is unphysical (all realistic interactions are of finite duration) and thus should be interpreted as an approximation. As such it must be motivated. 

The canonical way to justify taking such a limit ($\delta t\to0$) is to claim that, while of course realistic interactions have finite duration, the duration of these interactions may be much much shorter than any other timescales in the problem ($\delta t\ll T_\text{other}$). This argument holds water except that, as I have shown in Chapter \ref{Ch3}, collision models tend to produce unitary dynamics in the continuum limit. Since our goal is to explain the flow of information (especially quantum information) between the system and the environment, this is a complete failure. Taking the continuum limit of a collision model will generally result in a non-viable model of open quantum dynamics.

This unitary-in-the-continuum-limit tendency can be avoided however. Indeed there is a standard trick for doing so. By taking the interaction strength $g\sim \sqrt{\kappa/\delta t}$ to diverge as $\delta t\to0$ such that $g^2\delta t=\kappa$ is constant we can find non-unitary dynamics in the continuum limit. In Chapter \ref{Ch3} I have analyzed this work-around in detail, providing a necessary and sufficient condition for a generic collision model to produce non-unitary dynamics in the continuum limit. The $g^2\delta t$ trick (or something very much like it) is fact necessary. 

To use this trick we must now justify not only $\delta t\ll T_\text{other}$ but also $\kappa/g^2\sim\delta t\ll T_\text{other}$. That is, there must be two time scales $\kappa/g^2$ and $\delta t$ which are both much much less than all other time scales and which are approximately equal to each other as $\delta t$ becomes small. In Chapter \ref{Ch3} I have discussed that such a coincidence of timescales is difficult to explain without resorting to some sort of fine-tuning.

In this thesis I have not only argued against using the continuum limit of a collision model to study open quantum dynamics, but has also provided an alternative. The main loss if one is to abandon taking the continuum limit of a collision models is the master equation that taking this limit naturally provides. In Chapter \ref{Ch2} I developed a new method of obtaining a master equation from a generic collision model without taking the continuum limit. The master equation is produced by this method is the unique one such that 1) the master equation is linear, time-independent, and Markovian, 2) solving the master equation we find exact agreement with the discrete dynamics at the times $t=n\,\delta t$, and 3) the master equation can be analyzed in the continuum limit.

This approach allows us to use master equations to describe the dynamics of collision models without taking $\delta t\to0$. We can thus avoid both of the issues described above that this limit causes. Moreover all of the continuum limit results are recoverable from the new approach by simply taking $\delta t\to0$.

Since this new approach captures finite duration effects, it allows us to ask and answer questions that the previous approach does not. For instance, it is often the case that we can expand the interpolating master equation as a series in the interaction duration $\delta t$. We can then ask which types of dynamics are present at which orders in $\delta t$ under what circumstances (e.g., under what interaction Hamiltonians). Three lines of questioning along these lines have been pursued in this thesis.

Firstly, motivated by a desire to model the process of thermalization using collision models, I have investigated the capacity of dynamics to cause purification. Dynamics that cannot purify the state of any system also cannot decrease the temperature of a system and so cannot be thermalizing dynamics. As I have shown in the first half of Chapter \ref{Ch4}, in order for an interaction to be able to purify at leading possible order its interaction Hamiltonian must not be  ``too simple''. In order for the system and its environment to quickly exchange quantum information the ``language'' they use to speak to each other (i.e., the interaction Hamiltonian) must be sufficiently rich. In particular the interaction Hamiltonian must be at least Schmidt rank-2 and obey a certain commutation inequality.

Secondly, still interested in models of thermalization, I have investigated under what conditions the dynamics arising from collision models is dependent on the environment's free energy scale. As I have discussed in the second half of Chapter \ref{Ch4}, in order for dynamics to thermalize a system to the temperature of its environment, it must be sensitive to changes in the free Hamiltonian of the environment. Otherwise someone could simultaneously adjust the environment's temperature and free energy scale and leave the system's dynamics unchanged. In Chapter \ref{Ch4} it was found that this dependence does not arise in general until second order in $\delta t$. The process of exchanging information about free energy scales (and therefore about temperatures) is a relatively slow process. Thus a microscopic model of thermalization is impossible without accounting for finite duration effects.

Finally, in Chapters \ref{Ch5} and \ref{InterpolateGQM} I have asked ``for Gaussian quantum systems, which types of dynamics are present at which orders in $\delta t$ under which interaction Hamiltonians?'' and provided a complete answer. This involved first in Chapter \ref{Ch5} an overview of Gaussian Quantum Mechanics (GQM) as well as a complete classification of what types of dynamics are possible in GQM. In Chapter \ref{InterpolateGQM}, the interpolation generator was then computed to all orders in $\delta t$ for a generic quadratic interaction Hamiltonian. This expansion of the interpolation generator was  then compared with the classification system for Gaussian dynamics. The result is a complete dictionary indicating what system-environment Hamiltonians correspond to what types of open dynamics.

Taken together, the results of this thesis provide compelling evidence that the continuum limit approach to deriving a master equation from a collision model is both problematic and unnecessary in the study of open quantum systems. The interpolative approach suggested in this thesis both generalizes this continuum limit approach and overcomes its numerous shortcomings. Widening our view to finite duration effects (which appear to be essential to understand thermalization) promises to be a fruitful area of future study.

Immediate future applications of the work presented in this thesis include:
\begin{enumerate}
    \item The formalism developed in Chapter \ref{Ch2} can likely be extended to include the possibility of random interaction durations. In realistic scenarios the duration of each interaction (including the gaps between these interactions) can vary between interactions in an uncontrolled and unpredictable way.
    \item The ``dictionary'' between system-environment interaction Hamiltonians and the resulting open dynamics presented in Chapter \ref{InterpolateGQM} for bosonic Gaussian systems can be straightforwardly extended to fermionic Gaussian systems. A classification of fermionic Gaussian dynamics analogous to the bosonic one presented in Chapter \ref{Ch5} has already published in \cite{Onuma_Kalu_2019}. Moreover a similar dictionary should be computable for general qubit-qubit interactions.
    \item One interesting class of collision models is those involving an atom (or collection of atoms) crossing a cavity containing a quantum field. Since the cavity with necessarily have a finite width $L$ the duration of each interaction $\delta t=L/v$ must also be finite. Indeed, if these collision models are treated relativistically we must have $v<c$ such that $\delta t > L/c$. Thus if treated relativistically the $\delta t\to0$ limit cannot be achieved even conceptually.
    
    This class of collision models has potential applications that range from the harvesting of entanglement from quantum fields \cite{PhysRevA.88.052310} to experimental verification of the Unruh effect (publication in progress). 
\end{enumerate}

\addcontentsline{toc}{chapter}{Bibliography}
\bibliographystyle{plain}
\bibliography{references}
\chapter*{APPENDICES}
\appendix
\addcontentsline{toc}{chapter}{APPENDICES}
\chapter{Kraus Representation of Analytic Update Map}\label{KrausApp}

%======================================================================
Chapter \ref{L0isUnitary} considers a generic collision model in which the update map $\phi(\delta t)$ is analytic at $\delta t=0$ and has $\phi(0)=\openone$. To facilitate such a general treatment the update map is written in Kraus form as,
\bel{KrausRepApp}
\phi(\delta t)[\hat{\rho}]
=\sum_k \hat{A}_k(\delta t)\,\hat{\rho}\,\hat{A}_k(\delta t)^\dagger
\ee
where $\hat{A}_k(\delta t)$ are operators (called Kraus operators) satisfying the trace preserving condition, 
\be
\sum_k \hat{A}_k(\delta t)^\dagger \hat{A}_k(\delta t)=\hat{\openone}.
\ee
Given that the update map is analytic at $\delta t=0$ what can we say about the Kraus operators at $\delta t=0$? 

\section{Characterizing functions whose Square is Analytic}
Note that the update map $\phi(\delta t)$ is roughly the square of the Kraus operators. This motivates the following simplified example.

Consider a function, $f:\mathbb{R}_+\to\mathbb{R}_+$ for which $f(x)^2$ is analytic at $x=0$. Given this regularity of $f(x)^2$ around $x=0$ what can we say about $f(x)$ around $x=0$? Is it analytic? If not, what types of non-analytic behavior can it have?

Since $f(x)^2$ is analytic at $x=0$ we can expand it a series
\begin{align}
f(x)^2
=a_0
+a_{1}\, x
+a_{2}\, x^2
+\dots
\end{align}
for some non-zero domain of convergence around $x=0$. Taking a square root we immediately find that,
\begin{align}
f(x)
=\sqrt{f(x)^2}
=\sqrt{a_0
+a_{1} \,x
+a_{2} \,x^2
+\dots}.
\end{align}
Note that the square root function is analytic everywhere except at $0$. Thus if $a_0\neq0$ then the square root itself can be expanded analytically yielding,
\begin{align}
f(x)=
\sqrt{a_0} \ \Big(1
+\frac{a_1}{2 \, a_0} x
+\frac{4 a_0 a_2-a_1{}^2}{8 \, a_0{}^2} x^2
+\dots\Big).
\end{align}
Thus, if $a_0=f(0)^2\neq0$ then $f(x)$ is analytic at $x=0$.

If $a_0=0$ then we can take $m\geq1$ to be the smallest integer such that $a_m\neq0$ and repeat the argument. Specifically in this case we would have,
\begin{align}
f(x)
&=\sqrt{a_m \,x^m
+a_{m+1} \,x^{m+1}
+a_{m+2} \,x^{m+2}
+\dots}\\
&=x^{m/2}
\sqrt{a_m+
a_{m+1} \,x^{1}
+a_{m+2} \,x^{2}
+\dots}\\
&=x^{m/2} \, \sqrt{a_m} \ 
\Big(1
+\frac{a_{m+1}}{2 \, a_m} x
+\frac{4 a_m a_{m+2}-a_{m+1}{}^2}{8 \, a_m{}^2} x^2
+\dots\Big).
\end{align}
Note that if $m\geq1$ is even then $f(x)$ is analytic at $x=0$. On the other hand, if $m\geq1$ is odd then $f(x)/\sqrt{x}$ is analytic at $x=0$. If $m=\infty$ then $f(x)^2=0$ such that $f(x)=0$ is analytic at $x=0$.

Thus any function, $f(x)$, whose square is analytic at $x=0$ is of one of the following two types:
\begin{enumerate}
    \item $f(x)$ is analytic at $x=0$. In this case we have,
    \be
    f(x)=f_0+
    f_1  \, x 
    +f_2  \, x^2
    +\dots
    \ee
    for some $f_0$, $f_1$, $f_2$, $\dots$.
    \item $f(x)$ is not analytic at $x=0$. In this case $f(x)/\sqrt{x}$ is analytic at $x=0$. Namely,
    \be
    f(x)=\sqrt{x} \, 
    \big(f_{1/2}+
    f_{3/2} \, x 
    +f_{5/2} \, x^2
    +\dots
    \big).
    \ee
    for some $f_{1/2}$, $f_{3/2}$, $f_{5/2}$, $\dots$.
\end{enumerate}

\section{Identifying Two Kinds of Kraus Operators}
We will now use the proof in the above section to justify the classification of Kraus operators given by \eqref{KrausType1} and \eqref{KrausType2}.

For every $\delta t\geq0$ we can compute the update map's Choi matrix as,
\be
J(\delta t)=(\phi(\delta t)\otimes\openone)[\ket{\psi}\bra{\psi}],
\ee
where $\ket{\psi}=\sum_n\ket{n}\otimes\ket{n}$ is an (unnormalized) maximally entangled state. Since $J(\delta t)$ is related to $\phi(\delta t)$ by a linear map, we know that $J(\delta t)$ is analytic at $\delta t=0$. 

This regularity at $\delta t=0$ can also be transferred to the eigenvalues and eigenvectors of $J(\delta t)$. That is, we can find an eigenvalue decomposition
\bel{EigenDecomp}
J(\delta t)=\sum_k \lambda_k(\delta t)\ket{u_k(\delta t)}\bra{u_k(\delta t)}
\ee
for which $\ket{u_k(\delta t)}$ and $\lambda_k(\delta t)\geq0$ are themselves analytic at $\delta t=0$ \cite{DeepAnalysis}.

It is possible to explicitly construct a Kraus representation of a channel directly from such an eigendecomposition of its Choi matrix. Assuming a Kraus representation of the form \eqref{KrausRepApp} we have
\begin{align}\label{ChoiKraus}
J(\delta t)
&=(\phi(\delta t)\otimes\openone)[\ket{\psi}\bra{\psi}]\\
&=\sum_k (\hat{A}_k(\delta t)\otimes\hat{\openone})\ket{\psi}\bra{\psi}(\hat{A}_k(\delta t)\otimes\hat{\openone})^\dagger.
\end{align}

To help identify \eqref{ChoiKraus} with \eqref{EigenDecomp} we use the vectorization map, $\vec$, which maps outer products to tensor products as
\bel{VecDef}
\vec(\lambda \, \ket{\psi_1}\bra{\psi_2})
\coloneqq\lambda \, \ket{\psi_1}\otimes\ket{\psi_2}.
\ee
By linearity the above expression defines the vec map's action on any matrix. For instance note that
\begin{align}
\vec(\hat{\openone})
&=\vec(\sum_n\ket{n}\bra{n})\\
&=\sum_n\vec(\ket{n}\bra{n})\\
&=\sum_n \ket{n}\otimes\ket{n}\\
&=\ket{\psi}.
\end{align}
That is, vec of the identity operator is the unnormalized maximally entangled state.

Moreover this vec operation has the identity,
\be
\vec(\hat{X} \, \hat{Y} \, \hat{Z}^\dagger)=(\hat{X}\otimes \hat{Z})\,\vec(\hat{Y})
\ee
for any matrices $X$, $Y$ and $Z$. Thus we have
\begin{align}
J(\delta t)
&=\sum_k (\hat{A}_k(\delta t)\otimes\hat{\openone})\,\vec(\hat{\openone})\,\vec(\hat{\openone})^\dagger\,(\hat{A}_k(\delta t)\otimes\hat{\openone})^\dagger\\
&=\sum_k \vec(\hat{A}_k(\delta t))\,\vec(\hat{A}_k(\delta t))^\dagger.
\end{align}
Thus we can identify \eqref{ChoiKraus} with \eqref{EigenDecomp} by taking $\vec (A_k(\delta t))=\sqrt{\lambda_k(\delta t)} \ket{u_k(\delta t)}$ or equivalently,
\be
A_k(\delta t)=\sqrt{\lambda_k(\delta t)} \ \vec^{-1}\big(\ket{u(\delta t)}\big).
\ee

Recall that we can take $\ket{u(\delta t)}$ to be analytic at $\delta t=0$. Since $\vec$ is a linear map we can take $\vec^{-1}\big(\ket{u(\delta t)}\big)$ to be analytic at $\delta t=0$ as well. Next recall that we can take $\lambda_k(\delta t)\geq0$ to be analytic at $\delta t=0$. Using the results of the previous section, we can determine that $\sqrt{\lambda_k(\delta t)}$ is either analytic at $\delta t=0$ or $\sqrt{\delta t}$ times something analytic at $\delta t=0$.

Thus in total each Kraus operator is either analytic at $\delta t=0$ or $\sqrt{\delta t}$ times something analytic at $\delta t=0$. This justifies the classification given in Sec. \ref{L0isUnitary}.

\end{document}